\documentclass[a4paper,11pt]{article}
\pdfoutput=1 

\usepackage{jcappub} 

\usepackage[T1]{fontenc} 

\usepackage{graphicx}
\usepackage[utf8]{inputenc}
\usepackage[english]{babel}   
\usepackage{amssymb}
\usepackage{amsmath}
\usepackage{slashed}
\usepackage{color}
\usepackage{mathtools}
\usepackage{hyperref}
\usepackage{floatrow}
\usepackage{makecell}
\usepackage{subfig}
\usepackage{multirow}
\usepackage{booktabs}
\usepackage{enumitem}
\usepackage{placeins}
\usepackage{arydshln}

\newcolumntype{P}[1]{>{\centering\arraybackslash}p{#1}}

\title{The Simons Observatory: impact of bandpass, 
polarization angle and calibration uncertainties on small-scale power spectrum analysis}

\author[a,b,c,1]{S. Giardiello,\note{Corresponding author.}}
\author[c]{M. Gerbino,}
\author[b,c,d]{L. Pagano,}
\author[e]{D. Alonso,}
\author[f]{B. Beringue,}
\author[g,h]{B. Bolliet,}
\author[a]{E. Calabrese,}
\author[i]{G. Coppi,}
\author[f]{J. Errard,}
\author[a]{G. Fabbian,}
\author[a]{I. Harrison,}
\author[j]{J. C. Hill,}
\author[a]{H. T. Jense,}
\author[k]{B. Keating,}
\author[e]{A. La Posta,}
\author[c]{M. Lattanzi,}
\author[l]{A. I. Lonappan,}
\author[m]{G. Puglisi,}
\author[n]{C. L. Reichardt,}
\author[o]{S. M. Simon}

\affiliation[a]{School of Physics and Astronomy, Cardiff University, The Parade, CF24 3AA Cardiff, Wales, UK}
\affiliation[b]{Dipartimento di Fisica e Scienze della Terra, Universit\`a degli Studi di Ferrara, via Saragat 1, I-44122 Ferrara, Italy}
\affiliation[c]{Istituto Nazionale di Fisica Nucleare, Sezione di Ferrara, via Saragat 1, I-44122 Ferrara, Italy}
\affiliation[d]{Institut d'Astrophysique Spatiale, CNRS, Univ. Paris-Sud, Universit\'{e} Paris-Saclay, B\^{a}t. 121, 91405 Orsay cedex, France}
\affiliation[e]{Department of Physics, University of Oxford, Denys Wilkinson Building, Keble Road,  OX1 3RH Oxford, UK}
\affiliation[f]{Universit\'e Paris Cit\'e, CNRS, Astroparticule et Cosmologie, F-75013 Paris, France}
\affiliation[g]{Kavli Institute for Cosmology, University of Cambridge, Madingley Road, CB3 0HA Cambridge, UK}
\affiliation[h]{DAMTP, Centre for Mathematical Sciences, Wilberforce Road, CB3 0WA Cambridge, UK}
\affiliation[i]{Dipartimento di Fisica, Universit\`a di Milano - Bicocca, Piazza della Scienza, 3 - 20126 Milano, Italy}
\affiliation[j]{Department of Physics, Columbia University, NY 10027, New York, USA}
\affiliation[k]{Department of Physics UC San Diego, CA 92093, La Jolla, USA}
\affiliation[l]{Dipartimento di Fisica, Universit\`a di Roma Tor Vergata, Via della Ricerca Scientifica, 1, 00133, Roma, Italy}
\affiliation[m]{Dipartimento di Fisica e Astronomia, Universit\`a degli Studi di Catania, via S. Sofia, 64, 95123, Catania, Italy; INFN - Sezione di Catania, Via S. Sofia 64, 95123 Catania, Italy; INAF - Osservatorio Astrofisico di Catania, via S. Sofia 78, 95123 Catania, Italy}
\affiliation[n]{School of Physics, The University of Melbourne, VIC 3030 Parkville, Australia}
\affiliation[o]{Astrophysics Department, Particle Physics Division, Fermi National Accelerator Laboratory, IL 60510, Batavia, USA}

\emailAdd{GiardielloS@cardiff.ac.uk}
\emailAdd{martina.gerbino@fe.infn.it}
\emailAdd{luca.pagano@unife.it}
\emailAdd{david.alonso@physics.ox.ac.uk}
\emailAdd{beringue@apc.in2p3.fr}
\emailAdd{bb667@cam.ac.uk}
\emailAdd{CalabreseE@cardiff.ac.uk}
\emailAdd{gabriele.coppi@unimib.it}
\emailAdd{josquin@apc.in2p3.fr}
\emailAdd{FabbianG@cardiff.ac.uk}
\emailAdd{HarrisonI@cardiff.ac.uk}
\emailAdd{jch2200@columbia.edu}
\emailAdd{JenseH@cardiff.ac.uk}
\emailAdd{bkeating@ucsd.edu}
\emailAdd{adrien.laposta@physics.ox.ac.uk}
\emailAdd{lattanzi@fe.infn.it}
\emailAdd{anto.lonappan@roma2.infn.it}
\emailAdd{giuseppe.puglisi2@unict.it}
\emailAdd{christian.reichardt@unimelb.edu.au}
\emailAdd{smsimon@fnal.gov}

\abstract{
We study the effects due to mismatches in passbands, polarization angles, and temperature and polarization calibrations in the context of the upcoming cosmic microwave background experiment Simons Observatory (SO). Using the SO multi-frequency likelihood, we estimate the bias and the degradation of constraining power in cosmological and astrophysical foreground parameters assuming different levels of knowledge of the instrumental effects. We find that incorrect but reasonable assumptions about the values of
all the systematics examined here can have significant effects on cosmological
analyses, hence requiring marginalization approaches at the likelihood level. 
When doing so, we find that the most relevant effect is due to bandpass shifts. When marginalizing over them, the posteriors of parameters describing astrophysical microwave foregrounds (such as radio point sources or dust) get degraded, while cosmological parameters constraints are not significantly affected. 
Marginalization over polarization angles with up to 0.25$^\circ$ uncertainty causes an irrelevant bias $\lesssim 0.05 \sigma$ in all parameters. 
Marginalization over calibration factors in polarization broadens the constraints on the effective number of relativistic degrees of freedom $N_\mathrm{eff}$ by a factor 1.2, interpreted here as a proxy parameter for non standard model physics targeted by high-resolution CMB measurements.}

\def\setsymbol#1#2{\expandafter\def\csname #1\endcsname{#2}}
\def\getsymbol#1{\csname #1\endcsname}

\def\Planck{\textit{Planck}}

\newbox\tablebox    \newdimen\tablewidth
\def\leaderfil{\leaders\hbox to 5pt{\hss.\hss}\hfil}
\def\tablenote#1 #2\par{\begingroup \parindent=0.8em
    \abovedisplayshortskip=0pt\belowdisplayshortskip=0pt
    \noindent
    $$\hss\vbox{\hsize\tablewidth \hangindent=\parindent \hangafter=1 \noindent
    \hbox to \parindent{$^#1$\hss}\strut#2\strut\par}\hss$$
    \endgroup}

\def\L2{\ifmmode L_2\else $L_2$\fi}

\def\DeltaT{\ifmmode \Delta T\else $\Delta T$\fi}
\def\deltat{\ifmmode \Delta t\else $\Delta t$\fi}
\def\fknee{\ifmmode f_{\rm knee}\else $f_{\rm knee}$\fi}
\def\Fmax{\ifmmode F_{\rm max}\else $F_{\rm max}$\fi}
\def\solar{\ifmmode{\rm M}_{\mathord\odot}\else${\rm M}_{\mathord\odot}$\fi}
\def\Msolar{\ifmmode{\rm M}_{\mathord\odot}\else${\rm M}_{\mathord\odot}$\fi}
\def\Lsolar{\ifmmode{\rm L}_{\mathord\odot}\else${\rm L}_{\mathord\odot}$\fi}
\def\inv{\ifmmode^{-1}\else$^{-1}$\fi}
\def\mo{\ifmmode^{-1}\else$^{-1}$\fi}
\def\sup#1{\ifmmode ^{\rm #1}\else $^{\rm #1}$\fi}
\def\expo#1{\ifmmode \times 10^{#1}\else $\times 10^{#1}$\fi}
\def\,{\thinspace}
\def\lsim{\mathrel{\raise .4ex\hbox{\rlap{$<$}\lower 1.2ex\hbox{$\sim$}}}}
\def\gsim{\mathrel{\raise .4ex\hbox{\rlap{$>$}\lower 1.2ex\hbox{$\sim$}}}}

\def\simprop{\mathrel{\raise .4ex\hbox{\rlap{$\propto$}\lower 1.2ex\hbox{$\sim$}}}}
\def\deg{\ifmmode^\circ\else$^\circ$\fi}
\def\pdeg{\ifmmode $\setbox0=\hbox{$^{\circ}$}\rlap{\hskip.11\wd0 .}$^{\circ}
          \else \setbox0=\hbox{$^{\circ}$}\rlap{\hskip.11\wd0 .}$^{\circ}$\fi}
\def\arcs{\ifmmode {^{\scriptstyle\prime\prime}}
          \else $^{\scriptstyle\prime\prime}$\fi}
\def\arcm{\ifmmode {^{\scriptstyle\prime}}
          \else $^{\scriptstyle\prime}$\fi}
\newdimen\sa  \newdimen\sb
\def\parcs{\sa=.07em \sb=.03em
     \ifmmode \hbox{\rlap{.}}^{\scriptstyle\prime\kern -\sb\prime}\hbox{\kern -\sa}
     \else \rlap{.}$^{\scriptstyle\prime\kern -\sb\prime}$\kern -\sa\fi}
\def\parcm{\sa=.08em \sb=.03em
     \ifmmode \hbox{\rlap{.}\kern\sa}^{\scriptstyle\prime}\hbox{\kern-\sb}
     \else \rlap{.}\kern\sa$^{\scriptstyle\prime}$\kern-\sb\fi}
\def\ra[#1 #2 #3.#4]{#1\sup{h}#2\sup{m}#3\sup{s}\llap.#4}
\def\dec[#1 #2 #3.#4]{#1\deg#2\arcm#3\arcs\llap.#4}
\def\deco[#1 #2 #3]{#1\deg#2\arcm#3\arcs}
\def\rra[#1 #2]{#1\sup{h}#2\sup{m}}

\def\dots{\relax\ifmmode \ldots\else $\ldots$\fi}
\def\WHzsr{\ifmmode $W\,Hz\mo\,sr\mo$\else W\,Hz\mo\,sr\mo\fi}
\def\mHz{\ifmmode $\,mHz$\else \,mHz\fi}
\def\GHz{\ifmmode $\,GHz$\else \,GHz\fi}
\def\mKs{\ifmmode $\,mK\,s$^{1/2}\else \,mK\,s$^{1/2}$\fi}
\def\muKs{\ifmmode \,\mu$K\,s$^{1/2}\else \,$\mu$K\,s$^{1/2}$\fi}
\def\muKRJs{\ifmmode \,\mu$K$_{\rm RJ}$\,s$^{1/2}\else \,$\mu$K$_{\rm RJ}$\,s$^{1/2}$\fi}
\def\muKHz{\ifmmode \,\mu$K\,Hz$^{-1/2}\else \,$\mu$K\,Hz$^{-1/2}$\fi}
\def\MJysr{\ifmmode \,$MJy\,sr\mo$\else \,MJy\,sr\mo\fi}
\def\MJysrmK{\ifmmode \,$MJy\,sr\mo$\,mK$_{\rm CMB}\mo\else \,MJy\,sr\mo\,mK$_{\rm CMB}\mo$\fi}
\def\microns{\ifmmode \,\mu$m$\else \,$\mu$m\fi}

\def\muK{\ifmmode \,\mu$K$\else \,$\mu$\hbox{K}\fi}
\def\microK{\ifmmode \,\mu$K$\else \,$\mu$\hbox{K}\fi}
\def\muW{\ifmmode \,\mu$W$\else \,$\mu$\hbox{W}\fi}
\def\kms{\ifmmode $\,km\,s$^{-1}\else \,km\,s$^{-1}$\fi}
\def\kmsMpc{\ifmmode $\,\kms\,Mpc\mo$\else \,\kms\,Mpc\mo\fi}
\providecommand{\sorthelp}[1]{}

\def\NHUNIT{\ifmmode {\rm \,cm^{-2}} \else $\rm \,cm^{-2}$ \fi} 

\def\muKcmb{\ifmmode \,\mu$K$_{\rm CMB}$\else \,$\mu$K$_{\rm CMB}$\fi}
\newcommand{\planck}{\Planck}

\newcommand{\lcdm}{$\Lambda$CDM}

\newcommand{\OmegaM}{\ifmmode\Omega_{\rm M}\else $\Omega_{\rm M}$\fi}

\providecommand{\Planck}{\textit{Planck}}
\providecommand{\planck}{\Planck}

\providecommand{\text}[1]{\rm{#1}}

\providecommand{\muK}{\mu\rm{K}}

\newcommand{\begm}{\begin{pmatrix}}
\newcommand{\enm}{\end{pmatrix}}

\def\pmb#1{\setbox0=\hbox{#1}%
    \kern-.025em\copy0\kern-\wd0
    \kern.05em\copy0\kern-\wd0
    \kern-.025em\raise.0433em\box0}

\def\p2Y{\;_2Y}
\def\m2Y{\;_{-2}Y}
\def\beglet{
  \addtocounter{equation}{1}%
  \setcounter{parentequation}{\value{equation}}%
  \setcounter{equation}{0}%
  \def\theequation{\arabic{parentequation}\alph{equation}}%
  \ignorespaces
}
\def\endlet{
  \setcounter{equation}{\value{parentequation}}%
  \def\theequation{\arabic{equation}}%
}
\providecommand{\beglet}{\begin{subequations}}
\providecommand{\endlet}{\end{subequations}}

\newcommand{\mksym}[1]{\ifmmode {\rm #1}\else #1\fi}

\providecommand{\text}[1]{\rm{#1}}

\providecommand{\muK}{\mu\rm{K}}

\newcommand\ba{\begin{eqnarray}}
\newcommand\ea{\end{eqnarray}}
\newcommand\bea{\begin{eqnarray}}
\newcommand\eea{\end{eqnarray}}

\newcommand\be{\begin{equation}}
\newcommand\ee{\end{equation}}

\begin{document}
\maketitle
\flushbottom

\newpage
\section{Introduction}
The Cosmic Microwave Background (CMB) temperature and polarization anisotropies at intermediate-to-small angular scales carry the signature of many phenomena that characterize the full history of the Universe. Measurements of CMB anisotropies can be used to precisely constrain the basic parameters of the standard \lcdm\, model~\cite{planck2016-l06,Aiola_2020,SPT-3G:2022hvq}, study the distribution of primordial perturbations~\cite{astro2020, Achucarro:2022qrl, Chang:2022tzj}, constrain particle and energy abundances~\cite{Brito:2022lmd, How2013, Abazajian:2013bxd}, set limits on the nature and the properties of neutrinos and other light relic components~\cite{Dvorkin:2022jyg, Kreisch:2022zxp, Dvorkin:2019jgs}, and explore complex models of dark matter and dark energy~\cite{Hill:2021yec, Li:2022mdj}. The current limits are dominated by the \Planck\, CMB measurements which anchor cosmological models at scales up to $\sim 0.1$ degree (or multipoles $\ell\sim 2000-2500$)~\cite{planck2016-l05}. The Atacama Cosmology Telescope (ACT) and the South Pole Telescope (SPT) continue to test models at smaller scales, until the CMB becomes a completely sub-dominant signal (at multipoles $\ell\sim 3000-4000$)~\cite{ACT:2020frw, SPT-3G:2022hvq}, and are expected to soon match the \planck\, constraining power. Both SPT and ACT have been extending the reach of \planck\, by making new high resolution, high
sensitivity E-mode polarization measurements. This means that both of them can provide new tests of \lcdm\ and at the same time look for new features in a multipole range not measured by \Planck. This range will then be further explored by the next-generation of ground-based experiments like the Simons Observatory (SO)~\cite{SO} and CMB-S4~\cite{CMB-S4:2016ple} which will improve sensitivity at all angular scales not yet cosmic-variance limited, especially in polarization.

With such a tremendous gain in statistical power, adequate control of instrumental systematics uncertainties will be crucial to robustly exploit the information content of CMB measurements and constrain key properties of the Universe and its components. Residual systematic effects can propagate across the data reduction pipeline and modify the reconstructed statistics of the CMB by mimicking genuine physical effects of cosmological origin. If not accounted for, instrumental systematic effects can thus be a source of bias in cosmological analyses. This work addresses a number of instrumental systematic effects important for the characterization of the SO small-scale data, targeted by the SO Large Aperture Telescope (LAT). We cover the impact of uncertainty in the frequency passbands, miscalibration and polarization angles mismatches. A rigid shift of the frequency passbands can cause significant biases in the reconstruction of the frequency-dependent signal, mostly astrophysical emissions dominating at small angular scales; miscalibration and polarization angle mismatch affect the reconstructed amplitude of the signal at all angular scales.
These three effects can be analytically modeled at the likelihood level, see e.g. the similar SO study for the large angular scale measurements in Ref. \cite{Abitbol:2020fvn}. Other systematic effects, such as beam leakages and beam uncertainties,  
are left to future studies.

We model instrumental effects at the power spectrum level in a systematics library, \texttt{syslibrary}\footnote{\href{https://github.com/simonsobs/syslibrary}{\texttt{https://github.com/simonsobs/syslibrary}, version 0.1.0}}, built into the SO power spectrum analysis pipeline. More specifically, the systematics are applied to the theory power spectra during the likelihood analysis which is performed using the SO multi-frequency likelihood {\ttfamily LAT\_MFLike}\footnote{\href{https://github.com/simonsobs/LAT_MFLike/tree/v0.9.2}{\texttt{https://github.com/simonsobs/LAT\_MFLike}, version 0.9.2}}. We then run Monte Carlo Markov Chain (MCMC) inference to explore parameter constraints in the presence of systematics, covering both the standard \lcdm\ model and the \lcdm+$N_{\rm eff}$ extension -- where the effective number of relativistic species, $N_{\rm eff}$, is free to capture small-scale cosmological behaviour -- and also exploring the full range of astrophysical emissions entering the small-scale data. We present the results for the $\Lambda$CDM+$N_{\rm eff}$ model in the main text and discuss the \lcdm~cases in Appendix \ref{app:result_lcdm}. The focus on $\Lambda$CDM+$N_{\rm eff}$ is justified by the fact that the improvement of the constraints on $N_{\rm eff}$ is one of the main science goals of the LAT. To assess the impact of systematic effects and to identify acceptable measures of mitigation, we run the analysis in two scenarios: i) setting a mismatch in one class of systematic parameters, i.e., fixing these parameters to the wrong value; ii) marginalizing over the systematic parameters assuming different levels of accuracy.

This paper is organized as follows. Section \ref{sec:LAT} gives a quick overview of the SO LAT and the sky model considered in this work; Section \ref{sec:systematics} describes how the systematic effects considered are implemented at the power spectrum level; Section \ref{sec:analysis} presents the analysis setup, including both the generation of simulated spectra and the likelihood settings; Section \ref{sec:result} presents the results for the \lcdm+$N_{\rm eff}$ cosmology with mismatches and with marginalized systematic parameters; finally, in Section \ref{sec:conclusion} we summarize the findings of the previous sections and draw conclusions.

\section{The SO Large Aperture Telescope} \label{sec:LAT}
As of March 2024, SO is in the process of deploying multiple telescopes on the Cerro Toco plateau in the Atacama desert. While an array of small-aperture telescopes will focus on the signature of primordial gravitational waves, all the observatory science goals considered here will depend on observations from a single Large Aperture Telescope (LAT).  
In the nominal SO configuration, this is a $6$-$\textrm{m}$ telescope hosting the LAT Receiver (LATR) camera with 30,000 TES bolometric detectors \cite{Zhu:2021beu} distributed among seven optics tubes that span six frequency bands from 27 to 280 GHz. Observations will cover $\sim 40 \%$ of the sky and reach $6$ $(2)$ $\mu$K-arcmin sensitivity in the baseline (goal) scenarios~\cite{SO}.
 
The angular resolution and sensitivity of each band were presented in Table 1 of Ref.~\cite{SO}. We recall the sensitivities to the CMB temperature and polarization signal (including both instrumental and atmospheric noise) for each frequency in Figs.~\ref{fig:fg_comp_tt} and \ref{fig:fg_comp_pol} of Appendix~\ref{app:result_lcdm}. The most relevant channels for our work are the frequencies $\nu = [93, 145, 225]$~GHz, which practically hold all the constraining power for cosmological models (see discussion in Ref.~\cite{SO}). We work in power spectrum space and focus on the temperature and polarization auto- and cross-spectra. The three frequency channels considered here all have arcminute resolution and therefore the spectra span scales from $\ell=30$ to $\ell=9000$. This is consistent with the approach of Ref.~\cite{SO}.\footnote{In the SO overview paper~\cite{SO} foreground cleaning was done before cosmological analyses and foreground uncertainty incorporated in the noise. Because of this, only multipoles below $\ell=4000$ were retained in the parameters runs. Here, we choose to work with the full multi-frequency spectra/noise and marginalise foregrounds at power spectrum level, and for this we need the full range of accessible multipoles. Differently from Ref.~\cite{SO}, we do not include observations from \emph{Planck} since the work done here is only applicable to the SO LAT.}

\subsection{The SO LAT sky} \label{sec:so_lat_sky}
The multi-frequency maps collected by the LAT will include several sources of emission, both in temperature and polarization. The CMB will in fact be contaminated by a number of frequency-dependent astrophysical components. Our sky model is therefore given by the sum of CMB and astrophysical foreground power spectra:
\begin{equation}\label{eq:LATsky}
C^{\mathrm{XY},\nu_i \nu_j }_{\ell, \mathrm{sky}} =  C^{\mathrm{XY}}_{\ell, \mathrm{CMB}} +  C^{\mathrm{XY},\nu_i \nu_j}_{\ell, \mathrm{FG}},
\end{equation}
where $C^{\mathrm{XY},\nu_i \nu_j}_{\ell, \mathrm{FG}}$ is the sum of the spectra of all the foreground components (see Appendix~\ref{app:fg_models}) at a given frequency combination $\nu_i, \nu_j$, and X, Y loop over the LAT temperature T and E-mode polarization generating three sets of CMB spectra: the temperature auto spectrum, TT, the E modes auto spectrum EE, and temperature-polarization correlation TE. These spectra are not delensed. We do not include B-modes since, at the scales of interest of the SO LAT, the BB spectrum is fully dominated by lensing. Therefore, the cosmological information contained in BB will be captured by the lensing reconstruction analysis and not included in the small-scale power spectrum vector, which is instead the focus of this analysis\footnote{If we were to include the lensing reconstruction likelihood in this analysis, we would need to correctly propagate the same systematic effects to the lensing reconstruction spectrum. The propagation pipeline would differ significantly from what studied in this work (see, e.g., ~\cite{Mirmelstein:2020pfk}) and is beyond the scope of the current analysis.}.  

\subsubsection{CMB}
For the CMB component of our sky model, we consider two scenarios: the basic \lcdm\ cosmology -- characterised by the sound horizon angular scale $\theta_{\rm MC}$ (or alternatively the Hubble constant $H_0$)\footnote{This choice depends on the Boltzmann code used.}, the amplitude and tilt of the scalar perturbations $A_s$ and $n_\mathrm{s}$, the baryon and dark matter densities $\Omega_\mathrm{b}h^2$ and $\Omega_\mathrm{c}h^2$, the optical depth to reionization $\tau_\mathrm{reio}$ -- and the effective number of relativistic species $N_{\rm eff}$ when \lcdm\ is extended by a single parameter. As in Ref.~\cite{SO}, we use $N_{\rm{eff}}$ for the \lcdm\ extension since it captures well the sensitivity of CMB small scales to extended cosmological models (we provide a description of its effects at the power spectrum level at the end of Section \ref{sec:systematics}). 

\subsubsection{Astrophysical Foregrounds}
For the astrophysical foregrounds, we follow the widely adopted prescription of modeling foreground signals at the power spectrum level with a parametric approach~\cite{planck2016-l05,Dunkley:2013vu,ACT:2020frw,SPT:2020psp}. Given the similarity between the SO LAT and the ACT telescope, in this work we use the same foreground model of the latest ACT power spectrum analysis~\cite{ACT:2020frw}. We note that in this work we are not assessing the completeness and/or validity of the specific foreground model, we assume the ACT model as a baseline and keep it fixed. Hence, we do not study here the effect of systematics due to modelling of astrophysical foregrounds or the coupling of instrumental systematics and foreground models -- this is left for future investigations. Given the foreground model, we still want to assess how much its free parameters (amplitudes and spectral indices, described below) are affected by systematics effects. Thus, we are sampling both cosmological and foreground parameters in all our analysis.

The full parametrization is detailed in Appendix~\ref{app:fg_models} and described in Refs.~\cite{ACT:2020frw, Dunkley:2013vu}. Here we report the relevant assumptions and parameters which we will use in the simulations of this study and that will be sampled in our analysis. 

In temperature we consider:
\begin{itemize}[noitemsep,topsep=3pt]
\item[--] thermal dust emission from the Milky Way (dust), scaling a power law with an amplitude $a_\mathrm{dust}^\mathrm{TT}$ normalized at $\ell_0 = 500$ and $\nu_0 = 150$ GHz (which is the reference frequency for all the components);
\item[--] the thermal and kinematic Sunyaev-Zel’dovich effects (tSZ and kSZ), sampling template amplitudes $a_\mathrm{kSZ}$ and $a_\mathrm{tSZ}$ (both at a pivot scale $\ell_0 = 3000$);
\item[--] terms for the Cosmic Infrared Background clustered (CIBC) and Poisson (CIBP). For the Poisson component, we scale a shot noise term sampling an amplitude $a_p$ (at $\ell_0 = 3000$) and an emissivity index $\beta_p$. For the clustered term, we scale a template with an amplitude $a_c$ (at $\ell_0 = 3000$) and an emissivity index $\beta_c$;
\item[--] power from unresolved radio point sources (radio) (below the masking threshold of 15 mJy at 150 GHz -- assuming that SO will have a flux cut similar to the deep patch of ACT~\cite{ACT:2020frw}), sampling the amplitude $a_s$ normalized at $\ell_0 = 3000$;
\item[--] and the tSZ-CIB cross-correlation (tSZxCIB), sampling the correlation parameter $\xi$ which multiplies a template normalized at $\ell_0 = 3000$.
\end{itemize} 
In polarization, for both EE and TE, we include only contributions\footnote{We note that Refs.~\cite{2019MNRAS.490.5712G,Bonavera:2017hjt,Datta:2018oae,Trombetti:2017kim} have measured upper levels of the contribution of dusty galaxy emission in polarization. We estimate this to be subdominant and therefore not include it in our baseline. However, we note that future work studying specific foreground modelling will need to consider this contribution in more detail.} from:
\begin{itemize}[noitemsep,topsep=3pt]
\item[--] polarized Galactic dust, sampling the amplitudes $a_\mathrm{dust}^\mathrm{TE}$, $a_\mathrm{dust}^\mathrm{EE}$ at $\ell_0 = 500$,
\item[--] and radio point sources, sampling the amplitudes $a_\mathrm{ps}^\mathrm{TE}$,  $a_\mathrm{ps}^\mathrm{EE}$ at normalized at $\ell_0 = 3000$.
\end{itemize}
The total foreground power as well as each component are shown in Figs.~\ref{fig:fg_comp_tt} and~\ref{fig:fg_comp_pol} in the Appendix \ref{app:fg_models}.

\section{Modelling of instrumental systematics} \label{sec:systematics}

The scientific goals of the SO LAT set strong requirements on the knowledge of the instrument (see e.g., Refs.~\cite{Hu:2002vu, Mirmelstein:2020pfk, Gallardo:2018rix, Crowley:2018eib, Gudmundsson2021ApOpt}). An imperfect modelling of the instrumental specifications may lead to systematic effects that propagate through the analysis pipeline and cause a non-negligible inflation of the error budget associated to the final scientific results as well as bias in the actual value of key parameters. While significant effort is devoted to the application of mitigation strategies and corrections prior to the likelihood analysis, modelling of residual unaccounted-for systematics is key to reduce biases and correctly account for instrumental effects and propagate data-processing uncertainties. In this work, we focus on three broad classes of instrumental systematics that are known to significantly affect observations at small angular scales in total intensity and polarization. In particular, we model the effects induced on the estimated CMB and foreground power spectra by the imperfect knowledge of the following: i) mean frequency of the transmission efficiency in a given frequency band (hereafter bandpass shift); ii) orientation of the polarizer that selects the polarization signal to be collected by the detector in a given frequency band (hereafter polarization angle mismatch); iii) overall calibration of the collected signal in intensity and polarization in a given frequency band (hereafter miscalibration).

We model the effects of bandpass shifts, polarization angles and calibration systematics at the power spectrum level. The sum of CMB and foreground power spectrum $C^{\mathrm{XY},\nu_i \nu_j}_{\ell}$ as given by Eq.~\ref{eq:LATsky} is modified by the effects of bandpass shifts ($\Delta^{\nu_i}$), calibration ($c^{\mathrm{XY}, \nu_i \nu_j}$) and polarization angle ($R^{\mathrm{XY}}(\alpha^{\nu_i},\alpha^{\nu_j})$) systematics according to the following:
\begin{align}
	\label{eq:cl_general}
	\tilde{C}^{\mathrm{XY}, \nu_i \nu_j}_{\ell} &= c^{\mathrm{XY}, \nu_i \nu_j} R^{\mathrm{XY}}(\alpha^{\nu_i},\alpha^{\nu_j}) \, \left( C^{\mathrm{XY}}_{\ell, \mathrm{CMB}} +  C^{\mathrm{XY},\nu_i \nu_j}_{\ell, \mathrm{FG}}(\Delta^{\nu_i}, \Delta^{\nu_j}) \right).
\end{align}
The individual effects are detailed in the following.

\subsection{Bandpass shifts}
We model the SO LAT passbands as top-hat functions\footnote{We do not include atmospheric opacity.}, characterized by a relative bandwidth $\delta\equiv (\nu_{\rm max} - \nu_{\rm min})/\nu_\textrm{center}$ of [0.3, 0.2, 0.13] which allow us to explore $\sim 30$~GHz for each channel, i.e., we consider ranges of [79 - 107], [130 - 159] and [210 - 240] at 93, 145 and 225~GHz respectively. The exact values of $\delta$ have been chosen to have passbands with $\sim 55 - 58$ samples, achieving a resolution of 0.5 GHz. Having more bandpass datapoints would slow down the foreground evaluation considerably\footnote{The passbands designed for SO are slightly different from the ones considered here, having a relative bandwidth $\sim 0.3$ at all frequencies. This means that our 93 GHz band is very similar to the full SO passband but we lose 34\% of the passband at 145 GHz and 50\% at 225 GHz (from the edges of the passbands). To test the impact of this approximation, we computed the total foreground spectrum assuming top-hat passbands with relative width of 0.3 and our fiducial parameters and note that the application of $\Delta^{\nu} \sim 1$ GHz shifts the total foreground power spectra in a similar way when using our passbands or the ones with 0.3 relative bandwidth (we have a maximum of 0.5$\sigma_{C_{\ell}}$ difference between the two cases).}. 

We consider a rigid shift of the bandpass profile in frequency, modelled as an offset $\Delta^{\nu}$ applied to the central frequency $\nu_\textrm{center}$.
This shift impacts the foreground signal collected by the LAT, as the foreground power in a given frequency channel is given by the integrated signal over the bandpass profile:

\begin{equation}\label{eq:cl_fg}
C^{\mathrm{XY},\nu_i \nu_j}_{\ell, \mathrm{FG}} = \sum_{k} \frac{\int d\nu_i F_{\rm CMB}(\nu_i+\Delta^{\nu_{i}}) \int d\nu_j F_{\rm CMB}(\nu_j+\Delta^{\nu_j}) f^k_{\mathrm{SED}}(\nu_i+\Delta^{\nu_i},\nu_j+\Delta^{\nu_j})}{\int d\nu_i F_{\rm CMB}(\nu_i+\Delta^{\nu_i}) \int d\nu_j F_{\rm CMB}(\nu_j+\Delta^{\nu_j})} C^{\mathrm{XY},k}_{\ell, \mathrm{FG}}. \\	
\end{equation}
In the previous equation, following the description widely adopted in the literature \cite{Dunkley:2013vu}, the frequency-dependent spectral energy distribution (SED) $f^k_{\mathrm{SED}}(\nu_i,\nu_j)$ of the foreground component $k$ and its $\ell$-dependence in harmonic space $C^{\mathrm{XY},k}_{\ell,\mathrm{FG}}$ are fully factorized. The factor
\begin{equation} \label{eq:dbbdt}
\begin{split}
F_{\rm CMB}(\nu) \propto \frac{x^2 e^x}{(e^x -1)^2} \nu^2 \tau_c(\nu), \quad \text{with} \quad x = \frac{h \nu}{k_B T_{CMB}},
\end{split}
\end{equation}
converts from CMB thermodynamic units to brightness. The first part $x^2 e^x/(e^x -1)^2$ accounts for the conversion from CMB thermodynamic to antenna temperature units. The $\nu^2$ factor comes from the fact that the passbands $\tau_c(\nu)$ are measured with respect to a Rayleigh-Jeans source, and allows conversion from antenna temperature units to intensity \cite{planck2013-p03d, Abitbol:2020fvn}. 
Finally, the denominator in Eq. \eqref{eq:cl_fg} sets the final units as $\mathrm{\mu K_{CMB}}$.\footnote{Note that the shifted frequency ranges are used also in the denominator of Eq.~\ref{eq:cl_fg}. This ensures, in absence of miscalibration, perfectly calibrated CMB spectra.}

\subsection{Polarization angles}
A mismatch $\alpha^{\nu}$ in the calibration of the polarization angle of the instrument polarimeters results in the rotation of the Stokes parameters $Q^{\nu}$ and $U^{\nu}$ collected at a given frequency~\cite{Zaldarriaga:1996xe}:
\begin{equation}
\begin{split}
    Q^{\nu}(\alpha^{\nu}) &= \cos(2\alpha^{\nu}) Q^{\nu}_0 + \sin(2\alpha^{\nu}) U^{\nu}_0\\
    U^{\nu}(\alpha^{\nu}) &= -\sin(2\alpha^{\nu}) Q^{\nu}_0 + \cos(2\alpha^{\nu}) U^{\nu}_0,
    \end{split}
\end{equation}
where $Q^{\nu}_0$ and $U^{\nu}_0$ are the Stokes parameters as they would be detected with a perfectly calibrated polarimeter.

The term $R^{\mathrm{XY}}(\alpha^{\nu_i},\alpha^{\nu_j})$ in Eq. \eqref{eq:cl_general} takes into account the subsequent modification of the EE and TE/ET\footnote{We consider the TE and ET spectra individually in the likelihood analysis, i.e., we do not compute the symmetrized TE spectrum $C^{\nu_i\nu_j, TE}_{\ell} = C^{\nu_i\nu_j, ET}_{\ell} = (C^{\nu_i\nu_j, TE}_{\ell} + C^{\nu_i\nu_j, ET}_{\ell})/2$.} spectra caused by a miscalibration in the polarization angles $\alpha^{\nu_i/\nu_j}$:

\begin{equation}\label{eq:rotmatrix}
\begin{split}
R^{\mathrm{TT}}(\alpha^{\nu_i},\alpha^{\nu_j}) &= 1 \\
R^{\mathrm{TE}}(\alpha^{\nu_i},\alpha^{\nu_j}) &= \cos(2 \alpha^{\nu_j}) \\
R^{\mathrm{ET}}(\alpha^{\nu_i},\alpha^{\nu_j}) &= \cos(2 \alpha^{\nu_i}) \\
R^{\mathrm{EE}}(\alpha^{\nu_i},\alpha^{\nu_j}) &= \cos(2 \alpha^{\nu_i})\cos(2 \alpha^{\nu_j}).
\end{split}
\end{equation}
Note that the TT spectrum is not affected by this systematic effect. 
We also do not include a contribution from the BB spectra to EE and TE/ET spectra in the analysis since for the values of polarization angles mismatch that we are considering in this paper ($\lesssim 0.5^\circ$) the B-to-E leakage is of the order of $10^{-4}$ and substantially negligible. Therefore, the effect of the polarization angle mismatch is equivalent to an effective polarization efficiency which here  
we capture with 3 free parameters ($\alpha^{93}, \alpha^{145}, \alpha^{225}$). 

\subsection{Calibration}
We calibrate the observations applying calibration factors to the $a_{\ell m}$ in each frequency channel. This leads to a total calibration factor $c^{\mathrm{XY}, \nu_i \nu_j}$, at the power spectrum level, as defined in Eq.~\ref{eq:cl_general}. The total calibration factor is the result of the combination of two per-channel calibration factors, $\mathrm{Cal}^{\nu_{i,j}}$, and a polarization efficiency, $\mathrm{Cal}_{\rm E}^{\nu_{i,j}}$:

\begin{equation} \label{eq:calfactor}
\begin{split}
c^{\mathrm{TT}, \nu_i \nu_j} &= \frac{1}{\mathrm{Cal}^{\nu_i} 
 \mathrm{Cal}^{\nu_j}}, \quad  c^{\mathrm{EE}, \nu_i \nu_j} = \frac{1}{\mathrm{Cal}^{\nu_i} 
 \mathrm{Cal}^{\nu_j} \mathrm{Cal}_{\rm E}^{\nu_i} \mathrm{Cal}_{\rm E}^{\nu_j}}, \\
 c^{\mathrm{TE}, \nu_i \nu_j} &= \frac{1}{\mathrm{Cal}^{\nu_i} 
 \mathrm{Cal}^{\nu_j}  \mathrm{Cal}_{\rm E}^{\nu_j}}, \quad c^{\mathrm{ET}, \nu_i \nu_j} = \frac{1}{\mathrm{Cal}^{\nu_i} 
 \mathrm{Cal}^{\nu_j}  \mathrm{Cal}_{\rm E}^{\nu_i}}.
\end{split}
\end{equation}

The calibration factors for each channel $\mathrm{Cal}^{\nu_{i,j}}$ are included to mimic the procedure of Ref.~\cite{ACT:2020frw}, where an overall calibration is applied to the spectra of each channel before co-adding them and its uncertainty is propagated to the covariance matrix. 
$\mathrm{Cal}^{\nu}$ acts effectively as calibration in temperature. The overall calibration error per channel is accounted for with a Gaussian prior associated to $\mathrm{Cal}^{\nu}$.  
This leaves us with a total of six free parameters in the calibration model, three $\mathrm{Cal}^{\nu_{i,j}}$ and three $\mathrm{Cal}_E^{\nu_{i,j}}$. \\

\begin{figure}[t!]
	\centering
	{\includegraphics[width=0.8\textwidth]{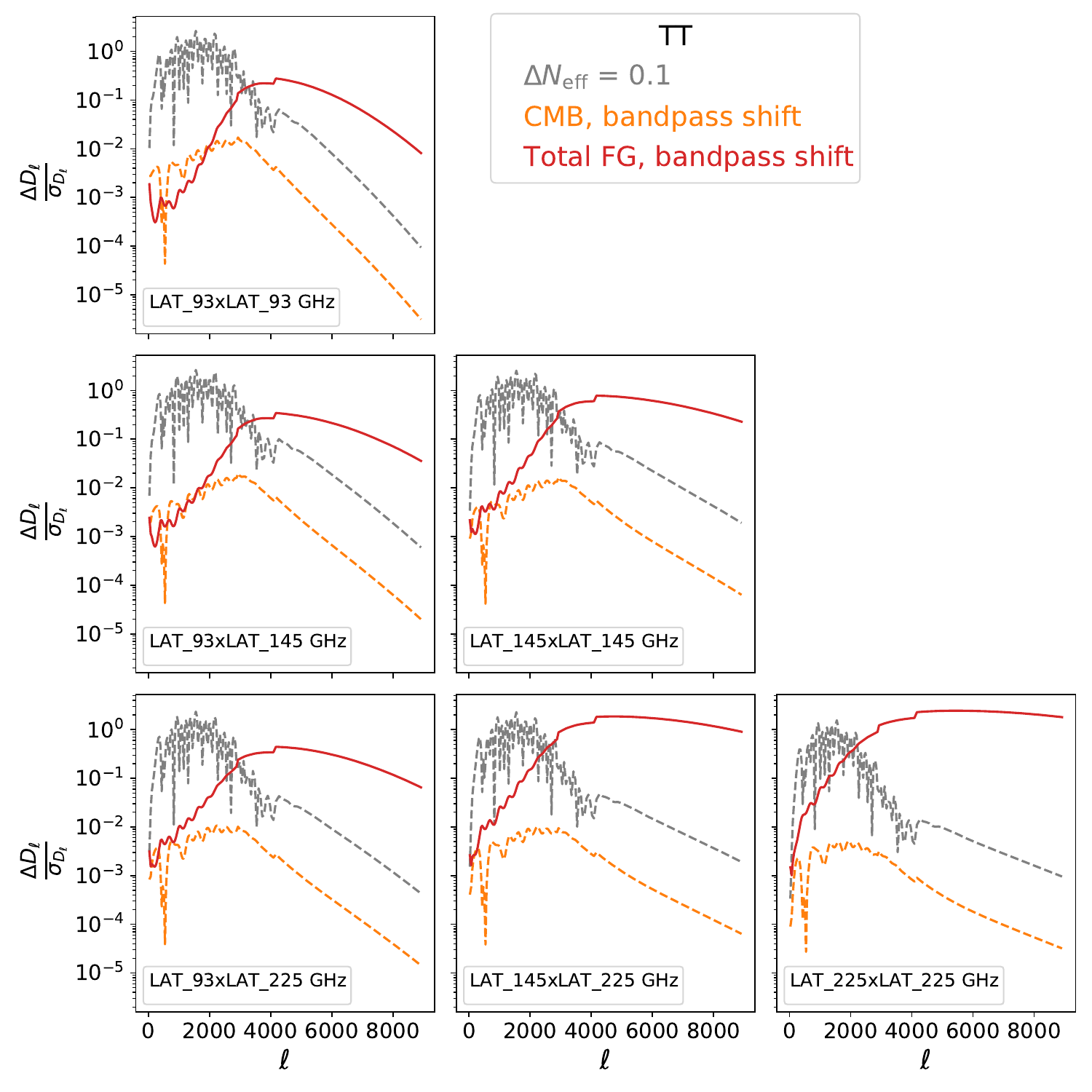}}
	{\caption{Difference in the TT multi-frequency power spectra in units of $\sigma_{D_{\ell}}$ when introducing a bandpass shift.  
 We plot residual spectra for $D^{\mathrm{TT}}_{\ell, \mathrm{CMB}}$ (in orange) and $D^{\mathrm{TT},\nu_1 \nu_2}_{\ell, \mathrm{FG}}$ (in red). Differences are computed from the bestfit parameters of two MCMC runs, assuming \lcdm: the ideal case without systematics, and the case where we marginalize over bandpass shifts with flat priors (more details in Section~\ref{sec:result} and Appendix~\ref{app:result_lcdm}). Therefore, the shape of the residuals is due to both the impact of the systematics and to the interplay of systematics, cosmological and foreground parameters in the joint fit. As expected, due to the frequency-dependent nature of the passband shift, residual for the power spectra of the foreground components are much more relevant. 
 We also plot in grey the $\Delta D_{\ell, \text{CMB}}$/$\sigma_{D_{\ell}}$, where the difference is computed between the bestfit of the ideal case without systematics and the same bestfit but with $N_{\rm eff}$ fixed to $3.144$ (corresponding to a $2\sigma$ shift with respect to the expected value $N_{\rm eff}=3.044$ in \lcdm\ ). Note the similarity in the shapes of the orange and grey curves (see details in the main text). 
 In all cases, dashed lines correspond to a negative difference. } \label{fig:cmb_fg_bsh_syst}}
\end{figure}

\begin{figure}[t!]
	\centering
	{\includegraphics[width=0.8\textwidth]{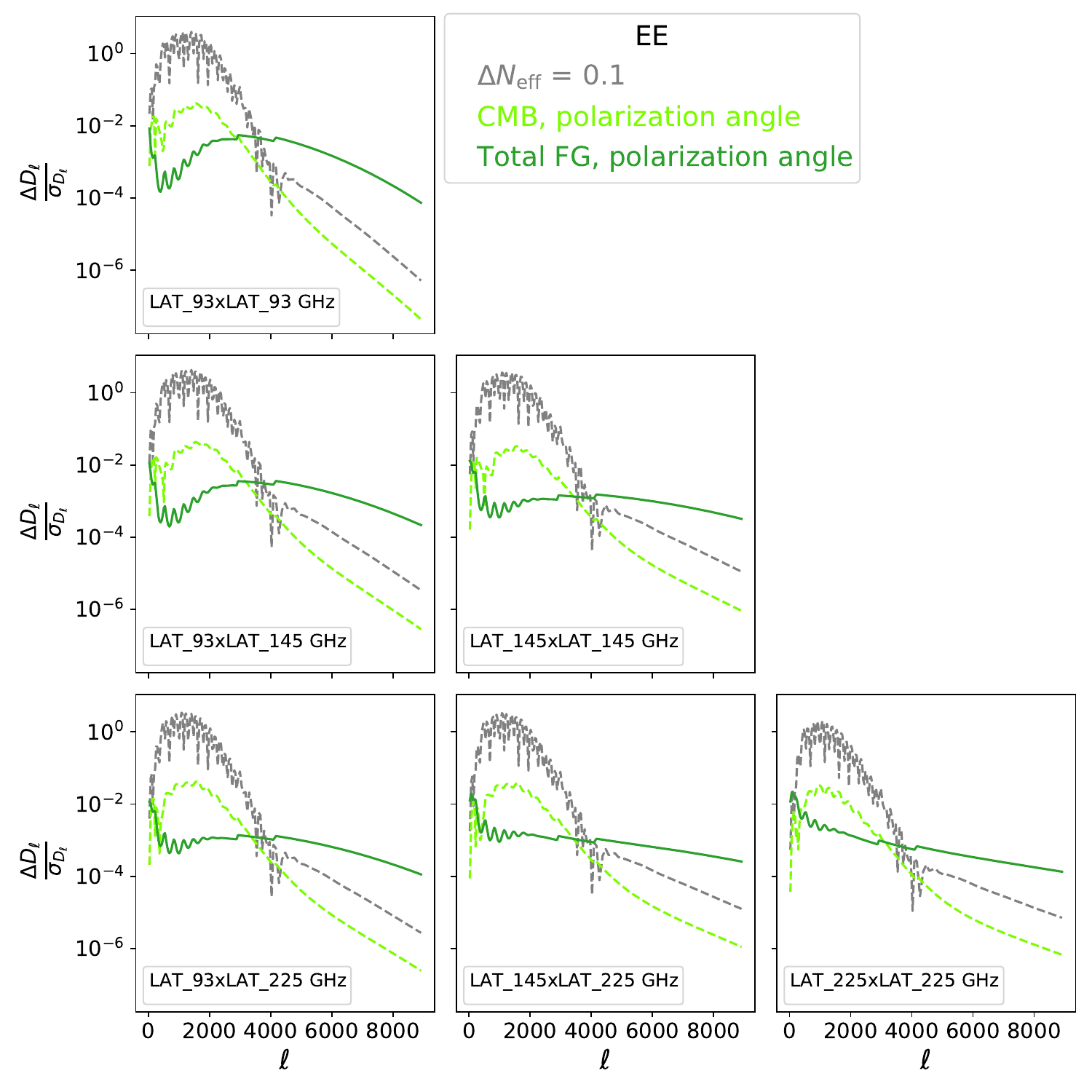}}
	{\caption{Same as Fig. \ref{fig:cmb_fg_bsh_syst}, but highlighting the effect of non-ideal polarization angles on EE power spectra. 
  The spectra affected by systematics are computed from the bestfit parameters of a MCMC run where we marginalize over polarization angles with flat priors. We plot residual spectra for $D^{\mathrm{EE}}_{\ell, \mathrm{CMB}}$ in light green and $D^{\mathrm{EE},\nu_1 \nu_2}_{\ell, \mathrm{FG}}$ in dark green. The effect of non-ideal polarization angles produces a more prominent impact on the CMB power spectrum. Note that the shape of the CMB residuals is similar to the difference arising from a $2\sigma$ change in $N_{\rm eff}$ with respect to the standard expected value (grey curve).
}\label{fig:cmb_fg_polang_syst}}
\end{figure}

\begin{figure}[ht!]
\centering
	{\includegraphics[width=0.8\textwidth]{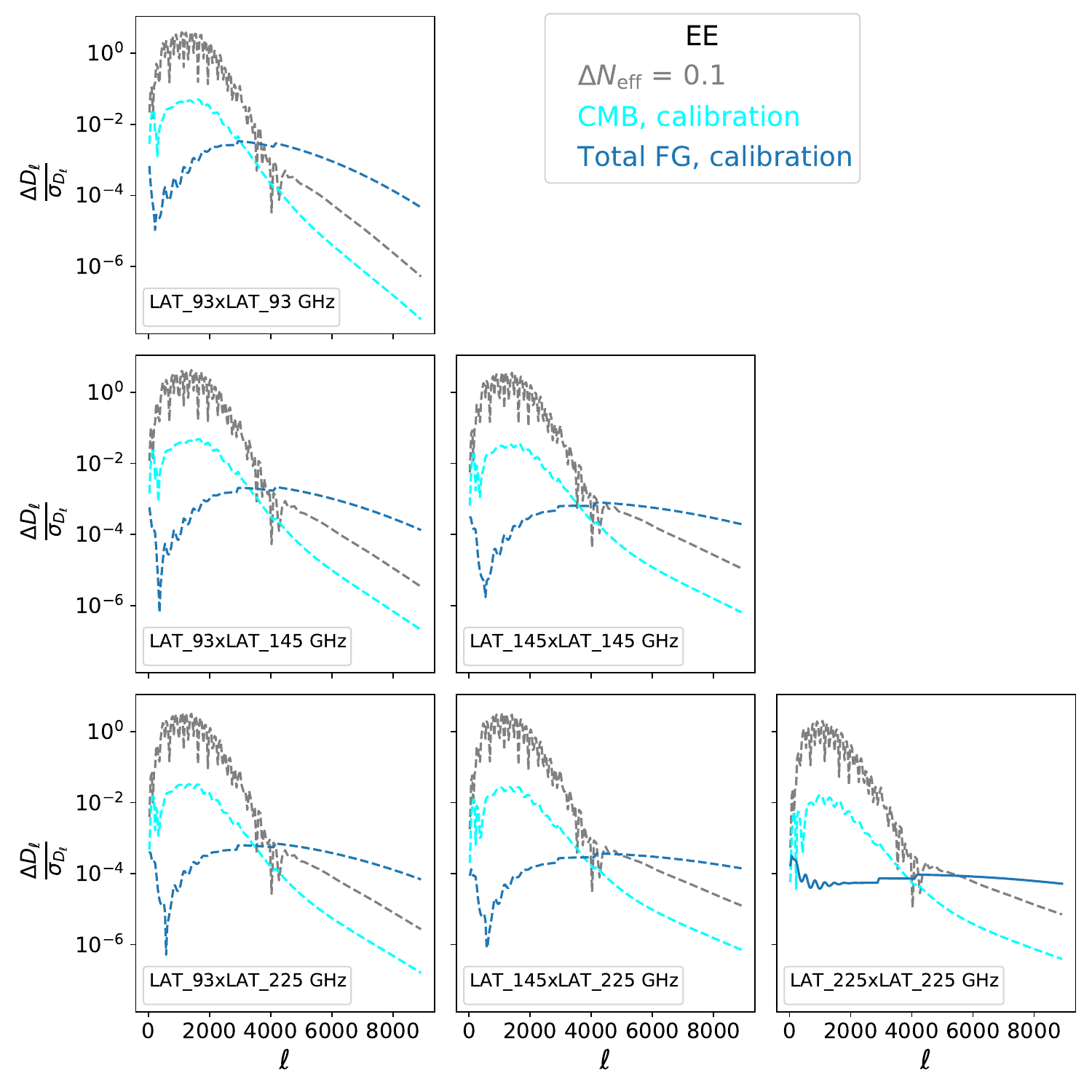}}
	{\caption{Same as Fig. \ref{fig:cmb_fg_bsh_syst}, but highlighting the effect on the EE power spectra of non-ideal calibrations. The spectra affected by systematics are computed from the bestfit parameters of a MCMC run where we marginalize over polarization efficiencies with flat priors and over calibrations per channel with a Gaussian prior. We plot residual spectra for $D^{\mathrm{EE}}_{\ell, \mathrm{CMB}}$ in cyan and $D^{\mathrm{EE},\nu_i \nu_j}_{\ell, \mathrm{FG}}$ in blue. Note that the shape of the CMB residuals is similar to the difference arising from a $2\sigma$ change in $N_{\rm eff}$ with respect to the standard expected value (grey curve).} \label{fig:cmb_fg_cal_syst}}
\end{figure}

We illustrate how the three classes of systematic effects manifest themselves in power spectrum space in Figures~\ref{fig:cmb_fg_bsh_syst}-\ref{fig:cmb_fg_cal_syst}. 
These show the difference between the case when the total CMB and foreground power is affected by a given systematic effect and a benchmark set of CMB and foreground spectra with no systematics. To obtain these figures, we take the difference between spectra computed from the bestfit parameters of the two MCMC runs with and without systematics, assuming a \lcdm\ cosmological model. As a result, the spectra affected by systematics arise from the interplay of cosmological and foreground parameters, which readjust compared to the ideal case to balance the effect of the systematics parameters. 

We note that the features exhibited by the residuals span the full range of angular scales probed by SO. They resemble the features we expect from key parameters that identify some of the main science targets of SO. For example, we recall that a change in $N_{\rm{eff}}$ and/or $\Omega_c h^2$ modifies the balance between radiation and matter density at early times -- thus the epoch of matter-radiation equality. Changes in $N_\mathrm{eff}$ and/or $\Omega_b h^2$ alter the angular scale of the diffusion damping. Modifications to $\Omega_b h^2$ also alter the photon density and velocity fields. All the aforementioned effects lead to a modification of the amplitude of the CMB acoustic peaks in all spectra \cite{Galli:2014kla, Hou:2011ec}. Similarly, a change in the properties of the primordial spectrum of perturbations captured by $A_s$ and $n_s$ as well as modifications to clustering of matter as represented by $\sigma_8$ can modify the overall amplitude or the relative power of the CMB spectra over the full range of multipoles. 

To highlight the importance of the residual due to systematics compared to the science targets of SO, we show in grey as a reference the residual power spectra caused by a shift of $\Delta N_{\rm eff} = 0.1$ with respect to the standard value 3.044, setting $N_{\rm eff} = 3.144$. This is roughly the $2\sigma$ measurement expected by SO. The grey curves are computed as the difference between the bestfit spectra of the MCMC run in the ideal scenario (i.e., no systematics applied) and the same bestfit, but increasing the value of $N_{\rm eff}$ as mentioned before. In some regions of the multipole space, residuals due to systematics are comparable in shape to the cosmological signal targeted by SO. In other regions, although subdominant, systematics exhibit a coherent trend which integrated over many multipoles will impact the cosmological results.

\section{Analysis framework} \label{sec:analysis}
In order to quantify the effects of the instrumental systematics on the estimation of parameters describing both cosmology and foregrounds, we generate a set of input spectra using the LAT noise described in Ref.~\cite{SO} and the sky model of Sec.~\ref{sec:so_lat_sky}, and process them in a full cosmological exploration analysis. 
We consider three cases: 
\begin{itemize}[noitemsep,topsep=0pt]
    \item as benchmark case we simulate a set of spectra without systematics, and we fit them with a consistent model;
    \item in a second case, we fit previously simulated spectra introducing a mismatch in the modelization of systematic effects. This case allows us to study the bias induced by uncorrected-for instrumental effects;
    \item finally, we fit previously simulated spectra with a free-to-vary instrument model, effectively marginalizing over the systematic parameters. This case allows us to verify that the extra degrees of freedom allow to recover unbiased cosmological parameters, and to study possible impacts on the final sensitivity. 
\end{itemize}

\subsection{Simulations} \label{section:simulated_data}

\begin{table}[tp!]
    \centering
\begin{tabular} {c c c | c c c}
\hline
\multicolumn{3}{c}{\makecell{Cosmology }} & \multicolumn{3}{c}{\makecell{Astrophysics}} \\
\hline
parameters & fiducials & priors & parameters & fiducials & priors \\  
\hline 
$\mathbf{\Lambda}$\bf{CDM} & & & $a_\mathrm{tSZ}$ & 3.30 &  \\
$\theta_\mathrm{MC}$ & 0.0104092 &  & $a_\mathrm{kSZ}$ & 1.60 & >0 \\
$\log(10^{10} A_\mathrm{s})$ & 3.044 & & $a_p$ &  6.90 & \\
$\Omega_\mathrm{b}h^2$ & 0.02237 &  &  $\beta_p$ & 2.20 &\\
$\Omega_\mathrm{c}h^2$ & 0.1200 &  & $a_c$ &  4.90 &\\
$n_\mathrm{s}$ & 0.9649 &  & $\beta_c$ & 2.20 & \\
$\tau_\mathrm{reio}$ & 0.0544 & $\mathcal{N}(0.0544, 0.0073)$ &  $a_\mathrm{dust}^\mathrm{TT}$ & 2.80 & >0 \\
\rule{0pt}{11pt}
 & & &  $a_\mathrm{dust}^\mathrm{TE}$ & 0.10 & >0  \\
 \rule{0pt}{11pt}
\bf{\lcdm}+$\mathbf{N_{\rm eff}}$ & & & $a_\mathrm{dust}^\mathrm{EE}$ & 0.10 & >0 \\
$N_{\rm eff}$  & 3.044  &  & $a_s$ & 3.10 & \\
 &  &  & $a_\mathrm{ps}^\mathrm{EE}$ & 0 & >0  \\
 \rule{0pt}{11pt}
 & &  & $a_\mathrm{ps}^\mathrm{TE}$ & 0 & \\
 & & &  $\xi$ & 0.10  & $\mathcal{U}(0,0.2)$ \\
\hline
\end{tabular} 
 \caption{Fiducial values and priors adopted for the cosmological and astrophysical foreground parameters used to generate the simulations. $\mathcal{N}(a,b)$ represents a Gaussian prior with mean $a$ and $\sigma = b$, while $\mathcal{U}(a,b)$ represents a uniform prior with extrema $a,b$. The cosmology is taken from the \emph{Planck} TTTEEE+lowE+lensing fits (second column of Table 1 of Ref.~\cite{planck2016-l06}). Astrophysical foregrounds are chosen to be consistent with results in Ref.~\cite{Dunkley:2013vu, ACT:2020frw} and they are representative of an ACT-like survey.} 	\label{tab:fid_params}
\end{table}

We generate 100 simulations of the LAT sky, drawing Gaussian realizations of CMB and foregrounds, using the SO framework {\ttfamily PSpipe}\footnote{\href{https://github.com/adrien-laposta/PSpipe/tree/maps2params_to_data_analysis}{\texttt{https://github.com/adrien-laposta/PSpipe/tree/maps2params\_to\_data\_analysis}}}, a pipeline for computing SO power spectra and covariance matrices, which is heavily based on the {\ttfamily pspy}\footnote{\href{https://github.com/adrien-laposta/pspy/tree/dev-alp}{\texttt{https://github.com/adrien-laposta/pspy/tree/dev-alp}}} code \cite{Louis:2020cqu}. The pipeline takes in input three main components: a set of fiducial CMB power spectra (computed with the fiducial cosmology listed in Table \ref{tab:fid_params}), passband-integrated foreground spectra (collecting all the terms shown in Figures \ref{fig:fg_comp_tt}, \ref{fig:fg_comp_ee} and \ref{fig:fg_comp_te} and computed with the fiducial astrophysics listed in Table \ref{tab:fid_params}), and noise spectra (shown in Figs. \ref{fig:fg_comp_tt}, \ref{fig:fg_comp_ee} and obtained with the SO noise calculator\footnote{\url{https://github.com/simonsobs/PSpipe/blob/master/project/data_analysis/python/so/so_noise_calculator_public_20180822.py}}). 
The CMB and foregrounds spectra are generated up to $\ell = 9000$, using respectively the \texttt{CAMB}\footnote{\url{https://github.com/cmbant/CAMB/tree/1.4.0}} Boltzmann solver~\cite{Lewis:1999bs,Howlett:2012mh} with the high accuracy settings presented in \cite{Bolliet:2023sst, McCarthy22},  and the SO library {\ttfamily fgspectra}\footnote{\url{https://github.com/simonsobs/fgspectra/tree/v1.1.0}}. All simulations are built with all systematic parameters fixed to the ideal value ($\Delta^{\nu} = 0$, $\alpha^{\nu} = 0$ and calibrations fixed to 1)\footnote{To make sure that the choice of the setup in the simulations does not impact the results, we run the opposite scenario compared to what reported in the main text as a consistency test. We generate one simulation with non-ideal bandpass shifts ($\Delta^{93} = 0.8$, $\Delta^{93} = -1$, $\Delta^{93} = 1.5$) and one with non-ideal polarization angles ($\alpha^{93} = 0.3$, $\alpha^{93} = 0.2$, $\alpha^{93} = 0.25$), using one of the 100 CMB and foreground parameters realizations. The reason behind the choice of these values for the systematic parameters is explained in Sections \ref{sec:bsh_fix} and \ref{sec:alpha_fix}.
The results of the runs performed on systematics-affected and systematics-free simulations agree. In some cases, parameters are shifted in the opposite direction due to the asymmetry of having a non-ideal value of systematics parameters in the data or in the theory.}. We empirically reconstruct the covariance matrix of the spectra from them and apply Monte Carlo corrections to the covariance matrix computed analytically, used in Eq. \ref{eq:logp}. 

Additionally to the set of 100 simulations, we also compute a smooth CMB + foreground spectrum. This is simply the prediction from \texttt{CAMB} and \texttt{fgspectra} of a spectrum using exactly the fiducial parameters in Table \ref{tab:fid_params} and ideal values for the systematics ones. This smooth spectrum can be analyzed in the same way as the 100 simulations and allows us to test that the average of the 100 realizations recovers unbiased results. In the following, we report all results from both type of spectra.

\subsection{Likelihood} \label{section:SO_likelihood}
Simulated spectra and sky and instrument models are fed to a multi-frequency likelihood for the SO LAT, implemented in the likelihood package {\ttfamily LAT\_MFLike}. 
  
As common practice in the analysis of small-scale CMB data \cite{planck2016-l05, Aiola_2020, SPT-3G:2022hvq}, the likelihood is approximated with a fiducial Gaussian \cite{Gerbino:2019okg}:
\begin{equation} \label{eq:logp}
	\log \mathcal{L}(\mathbf{C}^{\rm{data}}_{\ell}|\mathbf{C}^{\rm{th}}_{\ell}) \propto -\frac12 \left(\mathbf{C}^{\rm{data}}_{\ell} - \mathbf{C}^{\rm{th}}_{\ell}\right)_k^{\rm T} \mathbf{Cov}(C^{\rm{fid}}_{\ell})_{km}^{-1} \left(\mathbf{C}^{\rm{data}}_{\ell} - \mathbf{C}^{\rm{th}}_{\ell}\right)_m
\end{equation}
where $\mathbf{C}^{X}_{\ell} = [C^{\rm{X, TT},\nu_i \nu_j}_{\ell}, C^{\rm{X, TE,\nu_i \nu_j}}_{\ell}, C^{\rm{X, ET,\nu_i \nu_j}}_{\ell}, C^{\rm{X, EE,\nu_i \nu_j}}_{\ell}]$\footnote{When $\nu_i = \nu_j$, ET = TE. In this case, the ET spectrum is not included in the data vector to avoid double-counting.} for $\nu_i,\nu_j \in \{93,145,225\}$, $X=\rm{data/th}$, $\mathbf{Cov}(C^{\rm{fid}}_{\ell})$ is the covariance matrix of the set of simulated spectra, computed as explained in Sec.~\ref{section:simulated_data} and the $k,m$ are bin indices in multipole space.
We interface our likelihood  {\ttfamily LAT\_MFLike} to the MonteCarlo sampler {\ttfamily Cobaya} \cite{Torrado:2020dgo}.
To speed up the computation of theoretical CMB spectra on the selected scales ($30 \leq \ell \leq 9000$) we use the current public version of neural-network-emulated cosmological power spectra from \texttt{COSMOPOWER}\footnote{\href{https://github.com/alessiospuriomancini/cosmopower}{\texttt{https://github.com/alessiospuriomancini/cosmopower}, version 0.1.0}} \cite{SpurioMancini:2021ppk, Bolliet:2023sst}, trained with the \texttt{CLASS} Boltzmann solver \cite{class}\footnote{For this reason, the output for the ratio of the sound horizon to the angular diameter distance is not $\theta_{\rm MC}$ (used in \texttt{CAMB}) but $100 \theta_s$, see footnote 3 in Ref. \cite{Bolliet:2023sst}. We also note that the theoretical settings used for these networks are compatible with the ones used to generate our input CMB simulations done with \texttt{CAMB}. There are however still intrinsic differences between \texttt{CAMB} and \texttt{CLASS} which cause a small offset in our \texttt{COSMOPOWER} MCMC results. We recover the simulations inputs perfectly if we analyse either the full suite of 100 sims or the smooth spectra with \texttt{CAMB} or \texttt{COSMOPOWER} networks trained with \texttt{CAMB}~\cite{Jense:2024}. When using \texttt{COSMOPOWER} networks trained with \texttt{CLASS} - the only ones available when this work was started - we find an offset in recovering the inputs at the level of $\sim 0.2-0.5\sigma$ in some parameters. Since this happens coherently with or without instrumental systematics in the analysis, and since we only report shifts and not absolute mean values of parameters this offset is effectively subtracted out and is not propagated through in our results. This remaining discrepancy in Boltzmann codes has been fixed in \href{https://github.com/lesgourg/class_public/releases/tag/v3.2.2}{version 3.2.2} of \texttt{CLASS} and will be addressed elsewhere.}.

As mentioned above we consider two cosmological models, and therefore we fit the simulations assuming either \lcdm\, or \lcdm+$N_{\rm{eff}}$ for which we sample \{$\Omega_b h^2$, $\Omega_c h^2$,$H_0$, $A_s$, $n_s$, $\tau_{\rm reio}$\} + \{$N_\mathrm{eff}$\}. When assuming \lcdm\, we fix $N_\mathrm{eff}$ to the 
default standard model value of 
3.044 \cite{Bennett:2019ewm}. We always assume the presence of one massive neutrino with mass of $0.06\,\mathrm{eV}$ to mimic the minimal mass scenario allowed by flavour oscillation experiments. 

We do not include external data in the analysis as the scope of this work is to focus on SO LAT performance and requirements. We only make use of a Gaussian prior on $\tau_{\rm reio}$, $P(\tau_{\rm reio}) = \mathcal{N}(0.0544, 0.0073)$, informed by the \emph{Planck} satellite measurements of the large-scale polarization signal (\emph{Planck} TTTEEE+lowE+lensing with \texttt{Plik} \cite{planck2016-l06}), which is not accessible with the LAT.

In addition, we sample over the foreground parameters introduced in Sec.~\ref{sec:so_lat_sky} and selected systematic parameters detailed in Sec. \ref{sec:systematics}. As mentioned above, we do not explore the impact from uncertainties in the foreground model but assume that the foreground parametrization in the analysis is the same used to generate the simulations. This is done to isolate the effects of instrumental systematics from uncertainty in the astrophysical foreground modeling. 

On some parameters we impose priors to incorporate either physically motivated ranges or external information. When doing so, for both cosmological and foreground parameters, we report the prior probability distributions used during the MCMC sampling in Table \ref{tab:fid_params}, while those applied to the systematic parameters are reported in the following sections when introducing each case study.

\subsection{Grid of MCMC runs}\label{sec:runs}
\begin{table}
\centering
\begin{tabular} {l | l | l  }
\multicolumn{3}{c}{\makecell{\bf{List of runs for $\Lambda$CDM+$N_{\rm eff}$}}}\\
\hline
& \bf{case} &  \bf{systematic treatment} \\
\hline
\rule{0pt}{11pt}
\bf{benchmark} & fid &  \makecell[l]{all fixed to ideal values}  \\
\hline
\rule{0pt}{11pt}
 & $\Delta$sys &  $\Delta^{\nu}$ = \{0.8, -1, 1.5\} GHz \\
\bf{mismatched systematics} & $\alpha$sys & $\alpha$ =  \{$0.3^\circ, 0.2^\circ, 0.25^\circ$\}  \\
 & Csys & $\mathrm{Cal}^{\nu}$, $\mathrm{Cal}_{\rm E}^{\nu}  = \{1.01, 1.01, 1.01\}$ \\ 
\hline
\rule{0pt}{11pt}
&$\Delta \mathcal{N}$1 &  $\Delta^{\nu}  \in \mathcal{N}(0, 1)$ GHz \\
& $\Delta \mathcal{U}$ &  $\Delta^{\nu} \in \mathcal{U}(-20,20)$ GHz \\
\cline{2-3}
\rule{0pt}{11pt}
\bf{marginalized systematics} &$\alpha \mathcal{N}0.25$ &  $\alpha^{\nu} \in \mathcal{N}(0, 0.25)^\circ$ \\
&$\alpha \mathcal{U}$ &  $\alpha^{\nu}  \in \mathcal{U}(0,10)^\circ$ \\
\cline{2-3}
\rule{0pt}{11pt}
&C$\mathcal{N}$0.01 &  $\mathrm{Cal}^{\nu} \in \mathcal{N}(1, 0.01)$,  $\mathrm{Cal}_{\rm E}^{\nu} \in \mathcal{U}(0.9,1.1)$  \\ 
\hline
\end{tabular} 
\caption{MCMC runs for the $\Lambda$CDM+$N_{\rm eff}$ cosmology. Column one reports the analysis approach, the labels in the second column help to identify the runs throughout the text and in plots, while column three summarises how the specific systematics is treated (fixed or marginalized, and in the case of marginalization with which prior. When a systematic parameter is not explicitly mentioned, it means that it has been fixed to the ideal value).} \label{tab:runs_ideal}
\end{table}

We have devised a compilation of MCMC runs to explore different analysis options. We need to estimate parameters from each of the 100 simulations, plus the smooth spectra, with three different analysis approaches -- i.e., the benchmark case where systematics stay fixed to their ideal values, a case  where a mismatch is introduced in the value of the systematics parameters, and a case where the systematics are modelled and marginalized over. 

We summarise this grid of runs in Table \ref{tab:runs_ideal}. 
The first column of the table specifies the analysis approach,  e.g. fixing the systematic parameters to the fiducial value (``benchmark run''), or fixing them to the wrong value (``mismatched systematics'') or marginalizing over them (``marginalized systematics''). The priors adopted on the systematics parameters and the values at which they are fixed when not varying in each run are reported in the third column of the table. The labels used in plots and results assigned to each run are explicitly reported in the second column and incorporate compact description of the systematic treatment. For example, \texttt{$\Delta \mathcal{N} 1$} means that $\Delta^{\nu}$ has been marginalized with a Gaussian prior with $\sigma = 1$ GHz, while \texttt{$\Delta \mathcal{U}$} refers to a case adopting flat priors. Labels like \texttt{fid}/\texttt{$\Delta$sys} refer respectively to the benchmark run/run with $\Delta^{\nu}$ fixed to the wrong value, both using ideal simulations. 

We run MCMC analysis using the priors in this table on each of the 100 CMB+foreground realizations and on the realization-independent, smooth spectra. To distinguish the runs using the smooth spectra, we are attaching the tag \texttt{-smooth} at the end of each label in Table \ref{tab:runs_ideal}. When averaging over the runs on the 100 simulations, we will just use the labels without any additional tag. 

In the main body of the text we only show tables, figures and results for the \lcdm+$N_{\rm eff}$ model, results for \lcdm~are only obtained with the smooth spectra and discussed in Appendix~\ref{app:result_lcdm}.
To derive the empirical distribution for each parameter from the runs on 100 simulations,
we compute its mean and standard deviation for parameter $p$ as:
\begin{equation} \label{eq:avg}
    \bar{\mu}_p = \frac{\sum^{100}_{i = 1} \mu_{p,i}}{100}, \quad \bar{\sigma}_p =  \frac{\sum^{100}_{i = 1} \sigma_{p,i}}{100}
\end{equation}
where $\mu_{p,i}$ and $\sigma_{p,i}$ are the mean value and standard deviation of $p$ from the $i$th run.

\section{Results and discussion for the \lcdm+$N_{\rm eff}$ model} \label{sec:result}
In the following, we discuss in detail the results of our exploration. We divide the results in two subsections, each focused on a different way to assess the effects of systematics:
\begin{itemize}[noitemsep,topsep=0pt]
	\item Section \ref{sec:fix_sys_params} looks at the impact of having a mismatch between simulated spectra and the theory model. We expect biases in the estimation of cosmological and/or foreground parameters;
	\item Section \ref{sec:sampl_sys_params} explores marginalization over the systematic parameters either conservatively assuming a wide, uniform prior, or using a relatively narrow, and thus more informative, Gaussian prior, which might represent the information coming from, e.g. lab measurements of the systematics parameters. We expect a significant reduction of possible biases, at the price of possible degradation of the constraints on cosmological and/or foreground parameters.
\end{itemize}
The first case describes the situation we would face during the analysis of real data if systematics are ignored or unknown. With simulations we can quantify the effect of missing a systematics effect in the model. The second case is representative of the more realistic scenario in which the systematic effects are propagated in the pipeline with information from in-lab and in-field calibration and instrument characterization measurements (e.g., as done in previous analyses of ACT~\cite{Aiola_2020}).

The main results are shown in terms of:
\begin{itemize}[noitemsep,topsep=0pt]
\item The shift in the mean values of the one-dimensional posterior distributions of cosmological and foreground parameters, normalized to the $1\sigma$ width of the case under study: $(\bar{\mu}_{s} - \bar{\mu}_{b})/\bar{\sigma}_{s}$ where the subscript $b$ is indicating the values from the benchmark cases and $s$ the values from any other run which we are comparing with the benchmark (we eliminate the subscript $p$ from Eq. \ref{eq:avg});
\item The ratio of the $1\sigma$ widths of the one-dimensional posterior distributions $\bar{\sigma}_s/\bar{\sigma}_b$, to quantify the degradation in constraining power\footnote{We use the $\sigma$ of the posterior distribution reconstructed from the analysis of the 100 realizations, i.e. the width of the empirical posterior distribution. Note that this is different from the standard error on the estimate of the mean from the 100 realizations, i.e., $\sigma/\sqrt{N_{sims}}$.}.
\end{itemize}

We expect more prominent shifts in the cosmological and foreground parameters when introducing a mismatch between the simulation setup and the model in the analysis, and when marginalizing over the systematics parameters with a flat prior. The latter could also be responsible for the worst degradation in constraining power. The information conveyed by $\bar{\sigma}_s/\bar{\sigma}_b$ allows us to understand whether a negligible bias in a given parameter is to be ascribed to poor sensitivity to that parameter (higher $\bar{\sigma}_s$) rather than to a reduced impact of instrumental systematics.  

In the following we only include figures summarizing the key results
from the 100 CMB+foreground realizations. Tables with the numerical results of the average of all these runs are in Appendix~\ref{app:SO_tables}, where we include also tables for the corresponding runs using the smooth spectra. 

\subsection{Systematic effects from the incorrect determination of instrumental properties} \label{sec:fix_sys_params}

In this section, we evaluate the effect of a mismatch between how instrumental systematics are introduced in the simulations and how they are then modelled in the theory vector of the likelihood. We anticipate that the most relevant effect is induced by an unaccounted-for bandpass shift. This is explained by the frequency dependence of the foreground model: when evaluated at the incorrect (i.e., shifted) frequency range, the foreground (and, due to correlation effects, even the cosmological) parameters manifest a large bias.

\subsubsection{Bandpass shifts} \label{sec:bsh_fix}
\noindent\rule[0.5ex]{\linewidth}{1pt}
\textbf{Summary:} A mismatch of $\gtrsim 1$ GHz in the bandpass shift parameters induces shifts $> 10 \sigma$ in foreground parameters like $a_{\rm kSZ}, a_c, \beta_c, a_s, \xi$ but only $\sim -0.7 \sigma$ in $N_{\rm eff}$ and $H_0$ and -0.5$\sigma$ in $n_s$. Since these large shifts can severely hinder astrophysical constraints on SZ and CIB, as well as precise limits on cosmological parameters like $N_\mathrm{eff}$, it is essential to marginalize over bandpass shifts (as we do in Section \ref{sec:bsh}) instead of fixing them to a possibly wrong value.

\noindent\rule[0.5ex]{\linewidth}{1pt}

We introduce a mismatch between simulated and modelled systematics by fixing $\Delta^{93} = 0.8,\, \Delta^{145}=-1, \,\Delta^{225}= 1.5$ GHz in the theory model when fitting spectra simulated with $\Delta^{\nu} = 0$ GHz. The values chosen for the bandpass shifts are of the order of $1$~GHz, corresponding to a Fourier Transform Spectrometer calibration lower than 1\% per channel; these are better or similar to the ACT uncertainty~\cite{Thornton_2016, Madhavacheril:2019nfz, ACT:2020frw}, close to the uncertainty of the bandpass center measured by POLARBEAR~\cite{Matsuda:2019zgd} and close to the requirements for SO \cite{Bryan:2018mva, Ward:2018fjf}. In one of the channels, we introduce a negative shift to randomize the bandpass uncertainty considered here. The sign of the shift will not impact the effect studied here, in fact moving the band centres in opposite direction will allow us to explore a worse case scenario in terms of overall frequency uncertainty. 
We label the case with the wrong values of $\Delta^{\nu}$ as $\Delta$\texttt{sys}, while the reference case with ideal systematic parameters is labeled as \texttt{fid} (see Table \ref{tab:runs_ideal}).

\begin{figure}[t!]	
\centering
	{\includegraphics[width=1.\textwidth]{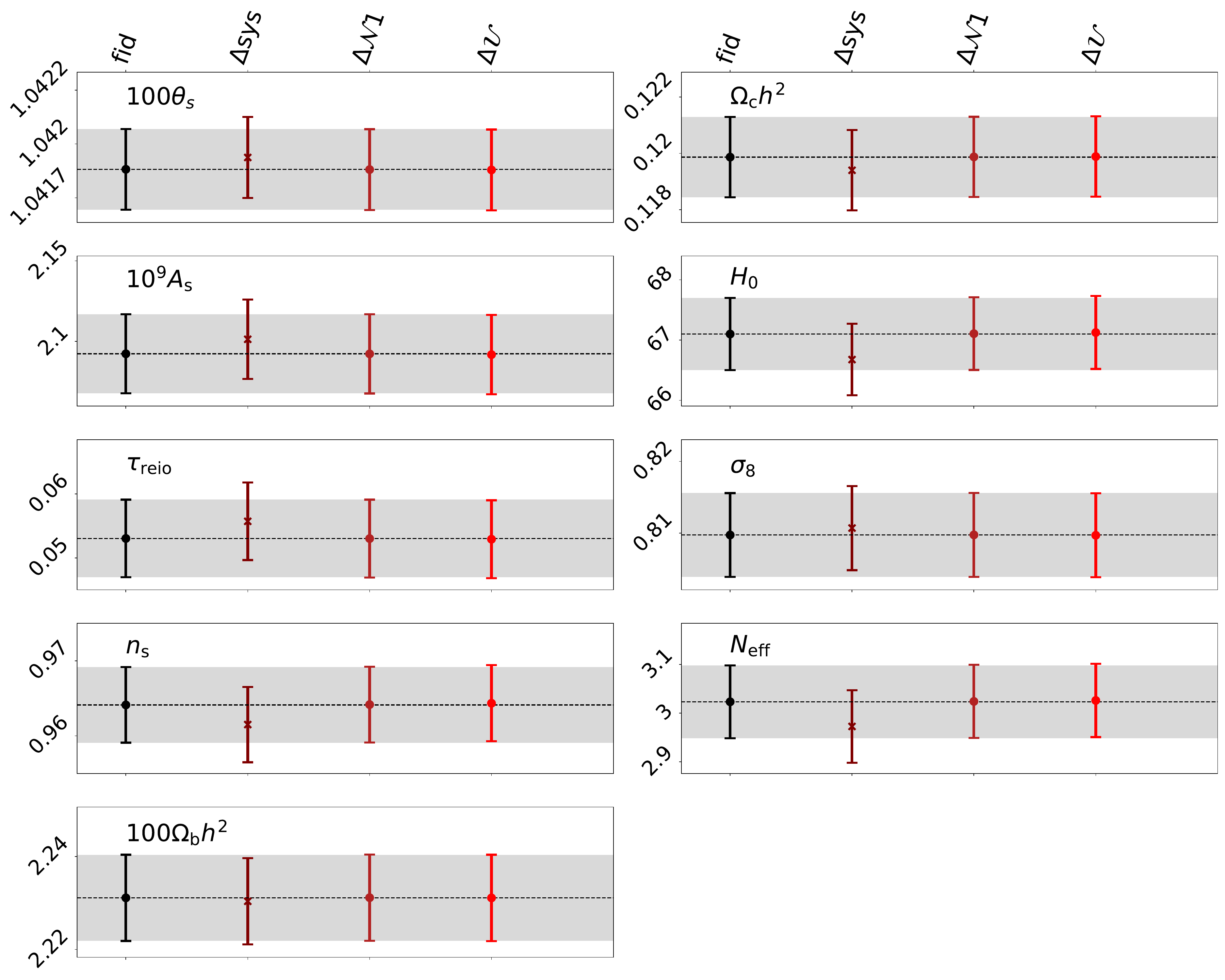}}
	{\caption{Average over 100 simulations of the mean values and 1$\sigma$ limits of the cosmological parameters recovered in all bandpass shifts runs (labels as listed in Table \ref{tab:runs_ideal}). The black dashed line indicates the mean value of the \texttt{fid} case, i.e. $\Lambda$CDM+$N_{\rm eff}$ with $\Delta^{\nu}$ and the other systematics fixed to the ideal fiducial values, used as benchmark case. The gray band corresponds to its  1$\sigma$ limit. The case with mismatches in $\Delta^{\nu}$ are indicated with a cross marker instead of a dot for the mean value. This case (second column) shows relevant biases in cosmological parameters, e.g., -0.7$\sigma$ in $N_{\rm eff}$ and $H_0$ and $-0.5\sigma$ in $n_s$, mostly due to movement correlated to the foreground biases (see Sec. \ref{sec:bsh_fix}). When marginalizing over $\Delta^{\nu}$ (third and fourth columns), we have negligible biases in cosmological parameters (as described later in Sec. \ref{sec:bsh}).} \label{fig:sim_bsh_cosmo}}
\end{figure}

\begin{figure}[hpt!]
	\centering
	{\includegraphics[width=1.\textwidth]{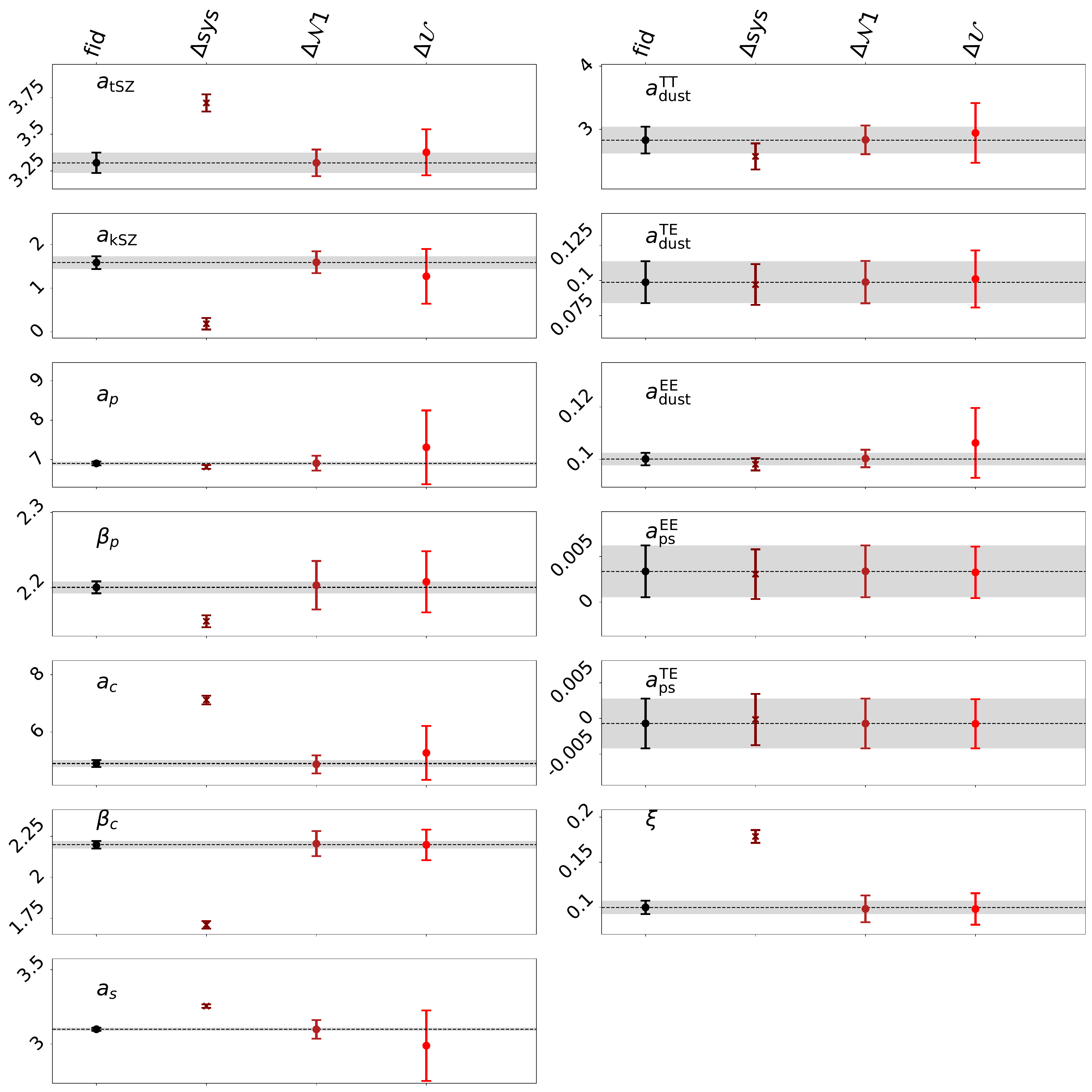}}
	{\caption{Same as in Figure \ref{fig:sim_bsh_cosmo}, but for foreground parameters. The case with mismatches in $\Delta^{\nu}$ (second column) shows $>10 \sigma$ biases for some foreground parameters, due to the fact that astrophysical components are frequency dependent. When marginalizing over $\Delta^{\nu}$ (third and fourth columns), the biases reduce but constraints get degraded, especially for wide priors on $\Delta^{\nu}$ (see Sec. \ref{sec:bsh}).} \label{fig:sim_bsh_fg_neff}}
\end{figure}

When setting a mismatch in the bandpass shifts there are deviations with module $> 0.5 \sigma$ in $N_{\rm eff}$, $n_s$, and $H_0$ with respect to the reference case \texttt{fid} (see Figure \ref{fig:sim_bsh_cosmo} and Tables \ref{tab:shift/sigma_bsh_neff}, \ref{tab:shift/sigma_bsh_smoothsim_neff}). As expected, the effect of the mismatch is more relevant for the foreground parameters. This can be appreciated in Figure \ref{fig:sim_bsh_fg_neff}, where we see that the recovered 1$\sigma$ constraints can be several $\sigma$s away from the reference value.
The $\sim 0.7 \sigma$ decrease in $N_{\rm eff}$ in the case with a mismatch in $\Delta^{\nu}$ could be driven by the $\sim 14 \sigma$ increase in $a_s$, due to a slight anticorrelation between the two parameters; this pushes $n_s$ and $H_0$ to lower values given their strong correlation with $N_{\rm eff}$ \citep{pdg_nu} (also visible in Figure \ref{fig:tr_cal_Neff}).

\FloatBarrier

\FloatBarrier

\subsubsection{Polarization angles} \label{sec:alpha_fix}

\noindent\rule[0.5ex]{\linewidth}{1pt}
\textbf{Summary:} With a mismatch at the level of $\sim 0.25^\circ$ in polarization angles, all parameters move by negligible amounts, shifting by $< 0.03 \sigma$. This level of polarization angle uncertainty is therefore sufficient for SO LAT science.

\noindent\rule[0.5ex]{\linewidth}{1pt}

We introduce a mismatch in the polarization angles $\alpha^{\nu}$ between the simulations and theory model, assuming $\alpha^{93} =  0.3^\circ,\, \alpha^{145} = 0.2^\circ,\, \alpha^{225} = 0.25^\circ$ in the theory model when fitting spectra simulated with $\alpha^{\nu} = 0^\circ$.
Values $\sim 0.25^\circ$ for polarization angles match the accuracy reached by ACT \cite{ACT:2020frw}. The values of $\alpha^{\nu}$ are chosen to average to $0.25^\circ$ between the different channels\footnote{We have assumed no frequency dependence of the polarization angles uncertainty, which could have made the values of $\alpha^{\nu}$ between the frequency channels much more different. This can happen, for example, in the presence of sinuous antennas \cite{sin-antenna} and half-wave plates \cite{pol-hwp}, which are not present in the design of LAT middle and high frequency channels \cite{Zhu:2021beu}.}. The labels for the case with $\alpha^{\nu}$ fixed to the wrong/fiducial values are $\alpha$\texttt{sys}/\texttt{fid} (see Table \ref{tab:runs_ideal}). 

\begin{figure}[h!]
	\centering
	{\includegraphics[width=1.\textwidth]{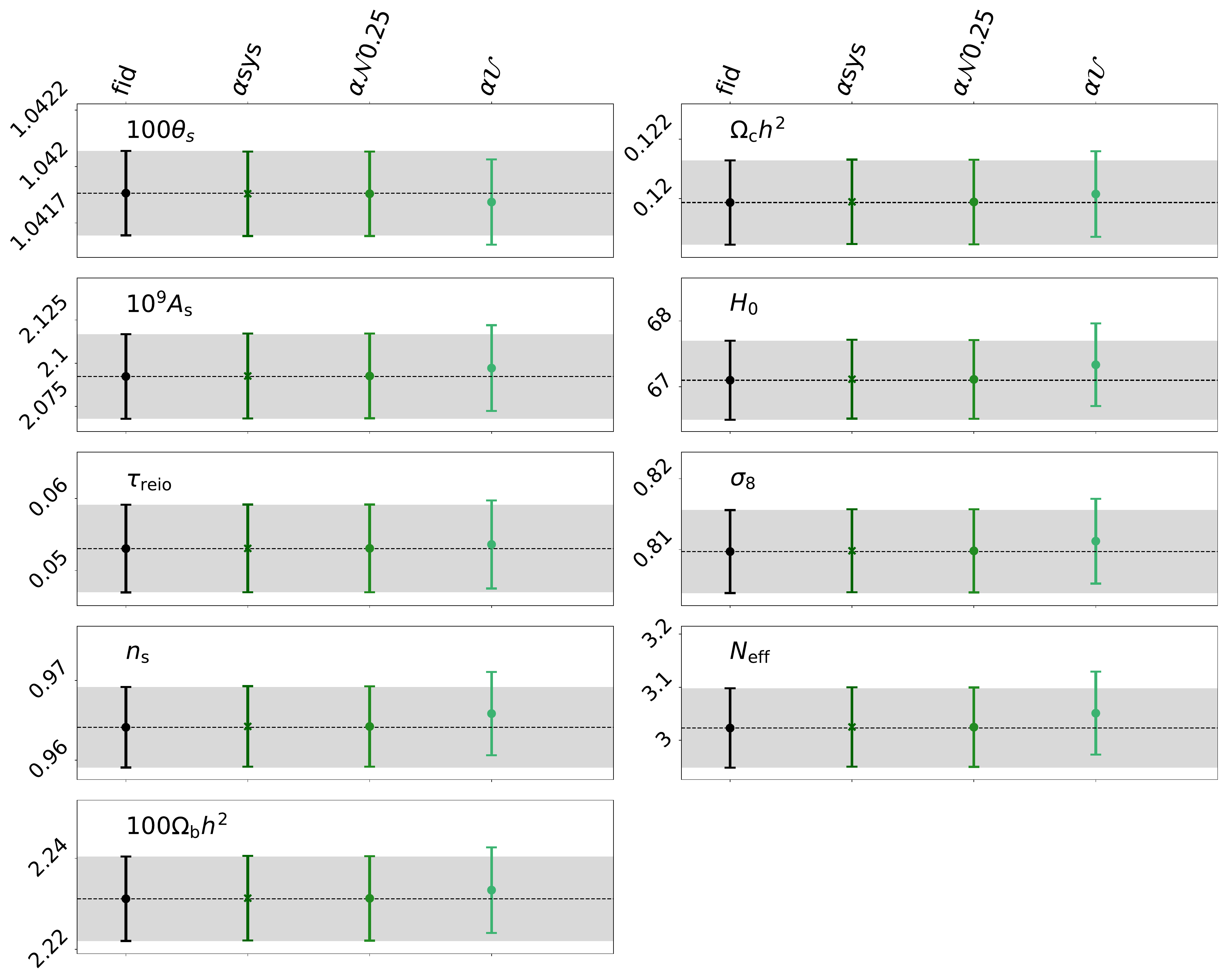}}
	{\caption{Average over 100 simulations of the mean values and 1$\sigma$ limits of the cosmological parameters recovered in all polarization angles runs (labels as listed in Table \ref{tab:runs_ideal}). The black dashed line indicates the mean value of the benchmark \texttt{fid} case, i.e. $\Lambda$CDM+$N_{\rm eff}$ with $\alpha^{\nu}$ and the other systematics fixed to the fiducial values. The gray band corresponds to its  1$\sigma$ limit. The cases with mismatches in $\alpha$ are indicated with a cross marker instead of a dot for the mean value. 
  The cases in which $\alpha^{\nu}$ has an offset of $\sim 0.25^\circ$ (second column) or is marginalized over with $\sigma = 0.25^\circ$ (third column) do not cause relevant biases, meaning that a level of $\sim 0.25^\circ$ uncertainty on $\alpha^{\nu}$ is acceptable for the SO LAT.
 Noticeable biases instead appear in cosmological parameters, especially a 0.36$\sigma$ shift in $N_{\rm eff}$, when we marginalize over $\alpha^{\nu}$ with a flat prior (last column; described in Sec. \ref{sec:alfa}). This is due to the fact that the posteriors of $\alpha^{\nu}$ peak around 1$^\circ$ (see Fig. \ref{fig:sim_a_syst}) causing much larger corrections in the model.} \label{fig:sim_a_cosmo}}
\end{figure}

\begin{figure}[h!]
	\centering
	{\includegraphics[width=1.\textwidth]{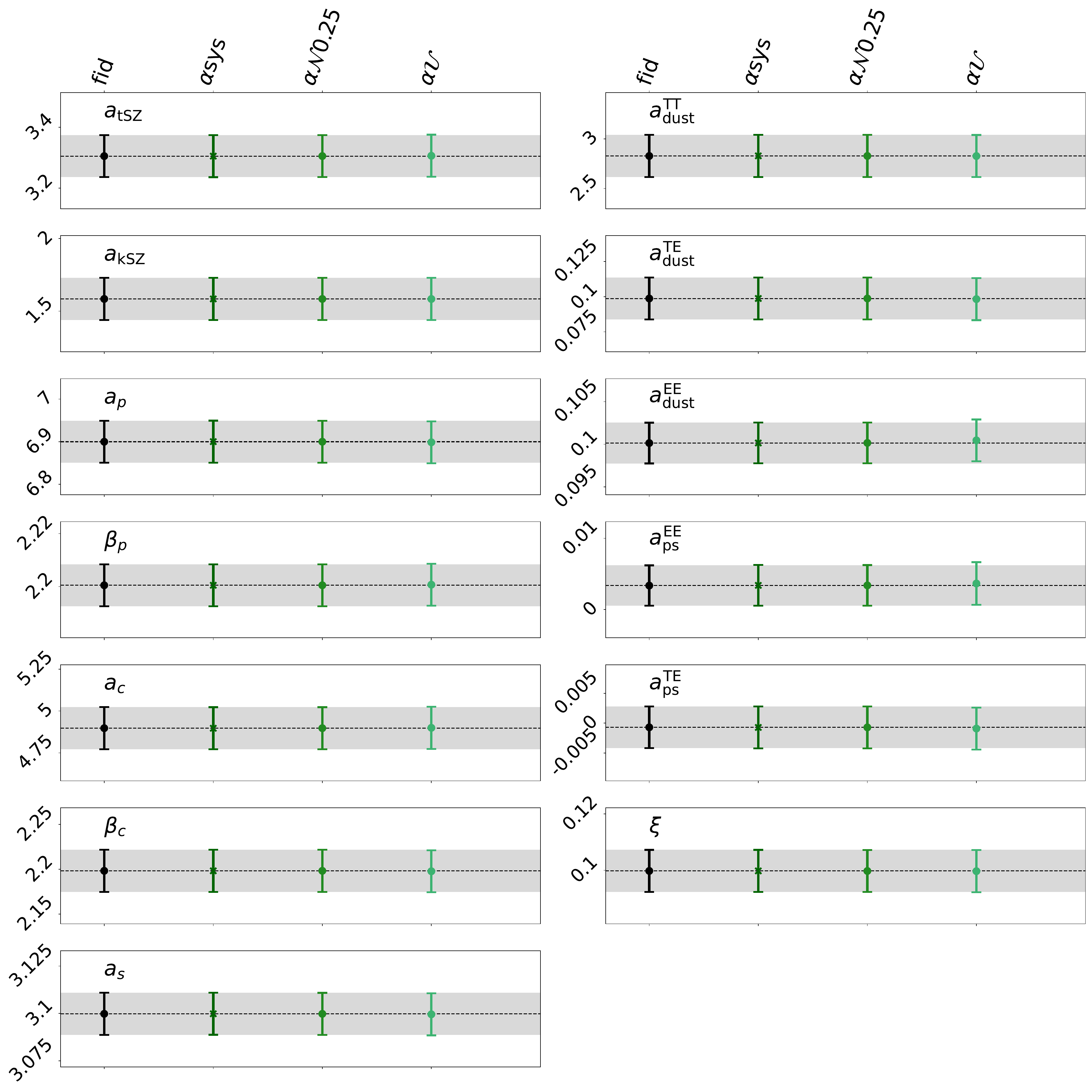}}
	{\caption{Same as Figure \ref{fig:sim_a_cosmo}, but for the foreground parameters. We see no relevant biases.} \label{fig:sim_a_fg}}
\end{figure}

As mentioned in Sec.~\ref{sec:systematics}, in the absence of BB power spectra, a non-zero value of the polarization angle leads to a rescaling of the polarization spectra. If the $\alpha^{\nu}$ values are close to each other (such as in our case), the model correction is roughly the same at all frequencies and therefore we expect similar effects on  cosmological and foreground parameters. Would the $\alpha^{\nu}$ values be very different from each other, foreground parameters would likely be affected more. Assuming a mismatch at the level of what has been considered in this analysis (the current ACT accuracy on polarization angles), the bias in the recovered cosmological and foreground parameters is negligible (see Figures \ref{fig:sim_a_cosmo}, \ref{fig:sim_a_fg} and Tables \ref{tab:shift/sigma_alfa_neff}, \ref{tab:shift/sigma_alfa_neff_smooth}). The most affected parameters (by only $\sim 0.025 \sigma$) are $N_{\rm eff}$ and $H_0$. A plausible explanation is the fact that $\alpha \lesssim 0.5^\circ$ corresponds to a calibration factor in polarization $\lesssim \cos(2^\circ)^2 \lesssim 0.9997$, i.e., only a mild miscalibration of $\lesssim 3 \times 10^{-4}$.

\FloatBarrier

\FloatBarrier

\subsubsection{Calibrations} \label{subsec:cal_fix}

\noindent\rule[0.5ex]{\linewidth}{1pt}
\textbf{Summary:} When we introduce a 1\% mismatch in calibration, fixing $\mathrm{Cal}^{\nu} = \mathrm{Cal}^{\nu}_{\rm E} =  1.01 \rightarrow 1/\mathrm{Cal}^{\nu} = 1/\mathrm{Cal}^{\nu}_{\rm E} = 0.99$, the  parameters measuring amplitudes of cosmological and foreground signals are biased towards higher values ($3.7 \sigma$ in $A_s$, $5.5 \sigma$ in $N_{\rm eff}$ and $5.2 \sigma$ in $a_s$). Due to their correlation with $N_{\rm eff}$, also $n_s$ and $H_0$ are biased by $5.3 \sigma$ and $6 \sigma$, respectively.
The mismatch in $\mathrm{Cal}_{\rm{E}}^{\nu}$ causes most of the shifts in $N_{\rm{eff}}$, $n_s$ and $H_0$, because of their stronger correlations with polarization efficiencies. 
Though this represents a pessimistic scenario, these behaviours can impact the science reach of the SO LAT, so we advise to marginalize over calibration parameters (see Section~\ref{subsec:cal_margin}).

\noindent\rule[0.5ex]{\linewidth}{1pt}

We introduce a $1\%$ mismatch in the overall per-frequency calibration factors $\mathrm{Cal}^{\nu}$ and polarization efficiencies $\mathrm{Cal}^{\nu}_{\rm E}$ fixing them to 1.01 in the theory model, when analysing simulations built with the reference value of 1. 
The level of mismatch of $1\%$ has been chosen based on uncertainties on calibrations from ACT~\cite{ACT:2020frw}, and a mismatch in the same direction for each channel provides the most conservative scenario. Fixing all calibrations and polarization efficiencies to a 1\% mismatch represents a pessimistic scenario, since for ACT~\cite{ACT:2020frw} the uncertainty on polarization efficiencies is $\sim 0.005$ and for SPT~\cite{SPT-3G:2021eoc} the uncertainty on calibrations is $\sim 0.005$ and the one on polarization efficiencies is $\sim 0.008 - 0.01$.

The labels for the case with calibrations fixed to the wrong/fiducial values are \texttt{Csys}/\texttt{fid}
(see Table \ref{tab:runs_ideal}). 

\begin{figure}[h!]
	\centering
	{\includegraphics[width=1.\textwidth]{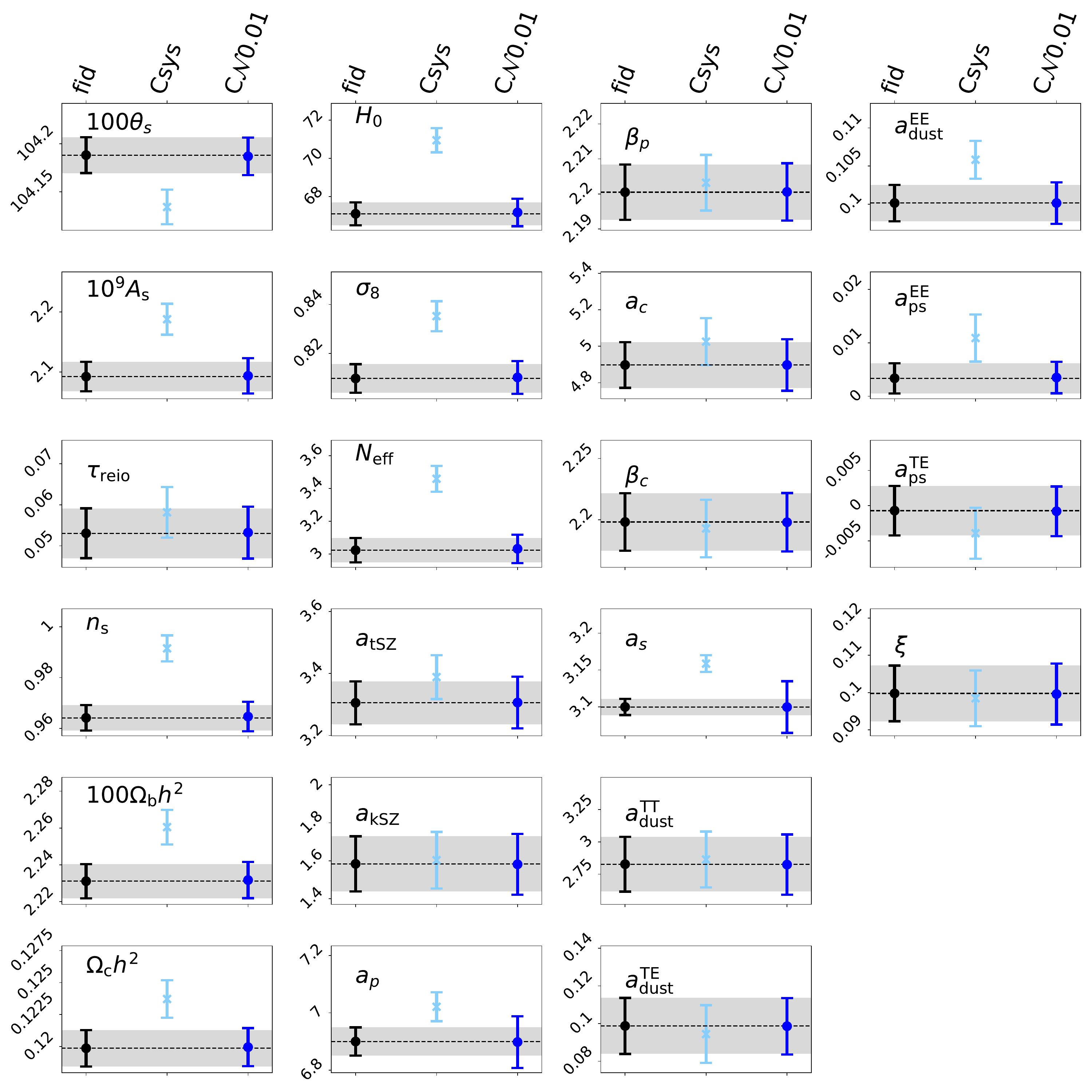}}
	\caption{Average over 100 simulations of the mean values and 1$\sigma$ limits of the cosmological and foreground parameters recovered in all calibrations runs (labels as listed in Table \ref{tab:runs_ideal}). The black dashed line indicates the mean value of the benchmark \texttt{fid} case, i.e. $\Lambda$CDM+$N_{\rm eff}$ with calibrations and the other systematics fixed to the fiducial values. The gray band corresponds to its  1$\sigma$ limit. The case with mismatches in $\mathrm{Cal}^{\nu}$ and Cal$^{\nu}_{\mathrm{E}}$ are indicated with a cross marker instead of a dot for the mean value.  When setting Cal$^{\nu}$ = Cal$^{\nu}_{\mathrm{E}}$ = 1.01 (cyan case) we have biases in amplitudes parameters, such as $3.7 \sigma$ in $A_s$ and $5 \sigma$ in $a_s$, and in parameters correlated with Cal$^{\nu}_{\mathrm{E}}$, like $5-6 \sigma$ in $N_{\rm eff}, n_s, H_0$. They get reduced when marginalizing over Cal$^{\nu}$ and Cal$^{\nu}_{\mathrm{E}}$ (dark blue case -- described later in Secs.~\ref{subsec:cal_fix} and \ref{subsec:cal_margin}). Marginalization over Cal$^{\nu}_{\mathrm{E}}$ slightly widens the posterior of $N_{\rm eff}$.} \label{fig:sim_nosyst_neff}
	\end{figure}

The main results are shown in Fig.~\ref{fig:sim_nosyst_neff} and in Tables \ref{tab:shift/sigma_Neff_cals_nosys_neff}, \ref{tab:shift/sigma_Neff_cals_smoothsim_neff}.
As expected from the way calibrations act on the model, when fixing $\mathrm{Cal}^{\nu} = \mathrm{Cal}^{\nu}_{\rm E} =  1.01 \rightarrow 1/\mathrm{Cal}^{\nu} = 1/\mathrm{Cal}^{\nu}_{\rm E} = 0.99$, the values of the parameters measuring amplitudes of cosmological and foreground signals show a bias towards higher values ($3.7 \sigma$ in $A_s$, $5.5 \sigma$ in $N_{\rm eff}$ and $5.2 \sigma$ in $a_s$). 
The shift in $N_{\rm eff}$ is mostly driven by a correlation with $\mathrm{Cal}^{\nu}_{\rm E}$, biasing also $n_s$ and $H_0$ by $5.3 \sigma$ and $6 \sigma$, respectively. As we will see in Section \ref{subsec:cal_margin}, polarization efficiencies are constrained with an uncertainty of $10^{-3}$, so a mismatch of $10^{-2}$ results in relevant biases of cosmological parameters. 
Marginalizing over the calibration parameters is thus preferable compared to fixing them.

\FloatBarrier

\subsection{Folding in the uncertainty in the determination of instrumental properties} \label{sec:sampl_sys_params}
	We now present the results of the Monte Carlo runs realized marginalizing over the three sets of systematics, one at the time. As detailed in the corresponding sections, we impose different priors on the systematics parameters, including uniform and Gaussian prior distributions. In particular, we note that in all cases the uniform prior range is much wider than the ranges probed with the Gaussian priors. As a result, the uniform prior here is representative of a scenario where we have less prior information on the relevant parameters.
 
 As mentioned above, we use 100 realizations of CMB, foreground and noise  without systematics (for more details, see Section \ref{section:simulated_data}). We also use the smooth spectra for all the cases. 
	We show triangle plots\footnote{For simplicity, we generate the triangle plots using results from the smooth spectra. These results are anyway consistent with the average over 100 simulations.} with 1- and 2-dimensional results for the most affected parameters when their correlation is relevant, and only 1$\sigma$ limits for the other parameters, for each case analyzed.
	
	\subsubsection{Bandpass shifts} \label{sec:bsh}
\noindent\rule[0.5ex]{\linewidth}{1pt}
\textbf{Summary:} When marginalizing over the bandpass shift parameters we note both biases and widening of the constraints in the foreground parameters. When applying Gaussian priors to $\Delta^\nu$ with $\sigma = 1$ GHz, the ratio $\bar{\sigma}_s/\bar{\sigma}_b$ for foregrounds is between 1 and 5.6. In the case of flat priors for $\Delta^\nu$ much wider than the Gaussian priors, the ratio can be as high as 19-21. The degradation of the constraints allows the biases on the mean values of the parameters to stay below $0.5 \sigma$ in all cases. In particular, marginalizing $\Delta^\nu$ with a 1 GHz uncertainty would keep biases on SZ parameters at a level lower than $0.1\sigma$ (with $\sigma$ larger by a factor 1.3/1.7 for $a_\mathrm{tSZ}$/$a_\mathrm{kSZ}$).  The posteriors of cosmological parameters are not affected, with insignificant widening of the constraints and shifts $< 0.05 \sigma$ in all cases.

\noindent\rule[0.5ex]{\linewidth}{1pt}
 
	We discuss here the effect of allowing the bandpass shifts $\Delta^{\nu}$ to vary, i.e., sampling them during the MCMC run using simulated ideal spectra, and we analyze the impact of various prior knowledges on $\Delta^{\nu}$. We impose the following priors: Gaussian priors $\mathcal{N}(0, 1)$ GHz (label: \texttt{$\Delta \mathcal{N}$1}) and flat priors $\mathcal{U}(-20,20)$ GHz (label: \texttt{$\Delta \mathcal{U}$}). As a test  performed just on the smooth spectra, we run a case with Gaussian priors centered on a wrong value: $\mathcal{N}(0.3, 1)$, $\mathcal{N}(0.5, 1)$, $\mathcal{N}(0.8, 1)$ for $\Delta^{93}$, $\Delta^{145}$ and $\Delta^{225}$ (label: \texttt{$\Delta \mathcal{N}$1sys-smooth}). 
    The 1$\sigma$ limits on all parameters are presented in Figures \ref{fig:sim_bsh_cosmo}, \ref{fig:sim_bsh_fg_neff} -- there to aid the comparison with the previous analyses with fixed systematics. In the plot we do not include the test case \texttt{$\Delta \mathcal{N}$1sys-smooth}, but the corresponding shifts and degradation of the constrainig power for cosmological and foreground parameters are presented in the third column of Table \ref{tab:shift/sigma_bsh_smoothsim_neff}.

\begin{figure}[!tp]  
\begin{center}
	{\includegraphics[width=0.8\textwidth]{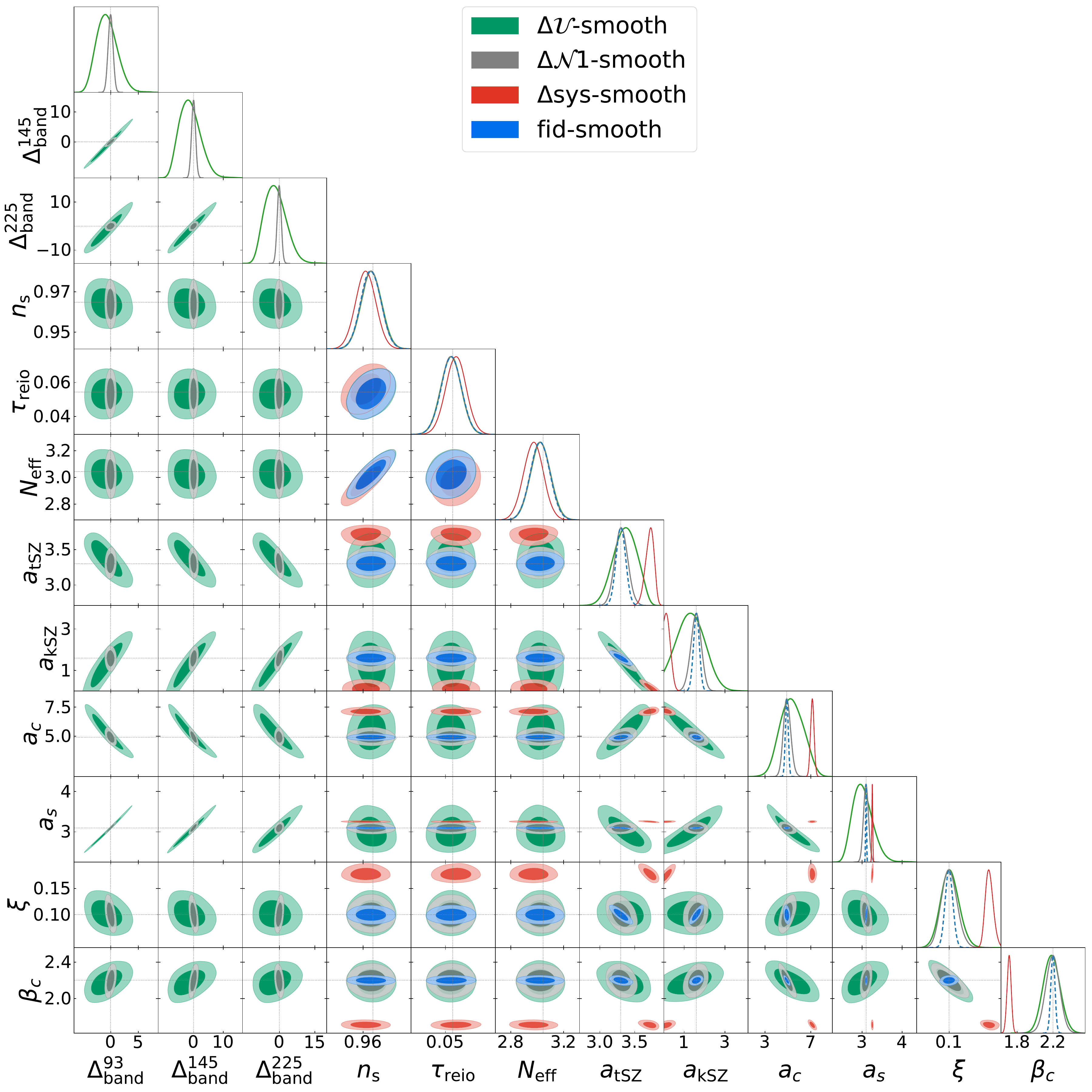}}
	\caption{1-dimensional posteriors and 2-dimensional contour plots at 68\% and 95\% C.L. for bandpass shift parameters $\Delta^{\nu}$ and the cosmological/foreground parameters most affected by the mismatch or marginalization over $\Delta^{\nu}$. We compare results for runs using smooth spectra with no bandpass shift analysed with: $\Delta^{\nu} = 0$ (\texttt{fid-smooth}, in dashed blue), $\Delta^{\nu} \neq 0$  ($\Delta$\texttt{sys-smooth}, in red) and $\Delta^{\nu}$ varied with different priors ($\Delta \mathcal{N}1$\texttt{-smooth}: Gaussian prior with $\sigma = 1$ GHz, in gray; $\Delta \mathcal{U}$\texttt{-smooth}: flat priors, in green). 
 The vertical dotted lines represent the input, fiducial value of the parameters in the simulation. We can notice the very strong correlations between $\Delta^{\nu}$ and foreground parameters. These cause degradation of the constraining power for foreground parameters, especially in the green case (flat priors). In the red case (mismatch), since all $\Delta^{\nu}$ parameters are fixed we see stronger biased but no degradation of the posteriors. } \label{fig:tr_bsh_simwbsh}
\end{center}
\end{figure}

The marginalization over $\Delta^{\nu}$ has a large impact on the foreground parameters which, as expected, depend strongly on our prior knowledge of the passbands. In particular, the constraints on the foreground parameters degrade visibly with broader priors on $\Delta^{\nu}$, i.e. in the case of flat priors (last column of Fig.~\ref{fig:sim_bsh_fg_neff} and Tables~\ref{tab:shift/sigma_bsh_neff}, \ref{tab:shift/sigma_bsh_smoothsim_neff}). This degradation consists of large shifts in the mean value of the foreground parameters and large broadening of their marginalized distribution, e.g., up to $\sim 19-21 \sigma$ larger than the benchmark case for $a_s, a_p$. We note that the marginalization over $\Delta^{\nu}$ with broad flat priors is the worst-case scenario, since we expect some external lab-based information on the passbands which would reduce the prior range. The impact on foreground parameters is due to the strong degeneracy with $\Delta^{\nu}$, as shown in Fig.~\ref{fig:tr_bsh_simwbsh}. In contrast, we have no correlation with the (frequency-independent) CMB signal, thus the constraints on cosmological parameters are less affected. In fact, there is no noticeable shift or degradation in the posteriors of cosmological parameters ($< 0.05 \sigma$ shifts for all parameters). In the case with Gaussian priors centered on the wrong value, the cosmological parameters do not experience any significant shift. Compared to the case \texttt{$\Delta \mathcal{N}$1}, the foreground parameters get shifted more, up to $0.7\sigma$ in $a_p$, with a similar degradation of the constraining power due to the uncertainty of 1 GHz on $\Delta^{\nu}$.

\FloatBarrier

\subsubsection{Polarization angles} \label{sec:alfa}

\noindent\rule[0.5ex]{\linewidth}{1pt}
\textbf{Summary:} When we marginalize over $\alpha^{\nu}$ using a Gaussian prior with $\sigma = 0.25^\circ$, the bias on cosmological and foreground parameters is irrelevant, $\lesssim 0.05 \sigma$. The marginalization over $\alpha^{\nu}$ with flat priors causes non-negligible shifts, at the level of $\sim 0.2-0.4\sigma$, on relevant cosmological parameters. \\
\noindent\rule[0.5ex]{\linewidth}{1pt}

\begin{figure}[tp!]
	\centering
	{\includegraphics[width=.8\textwidth]{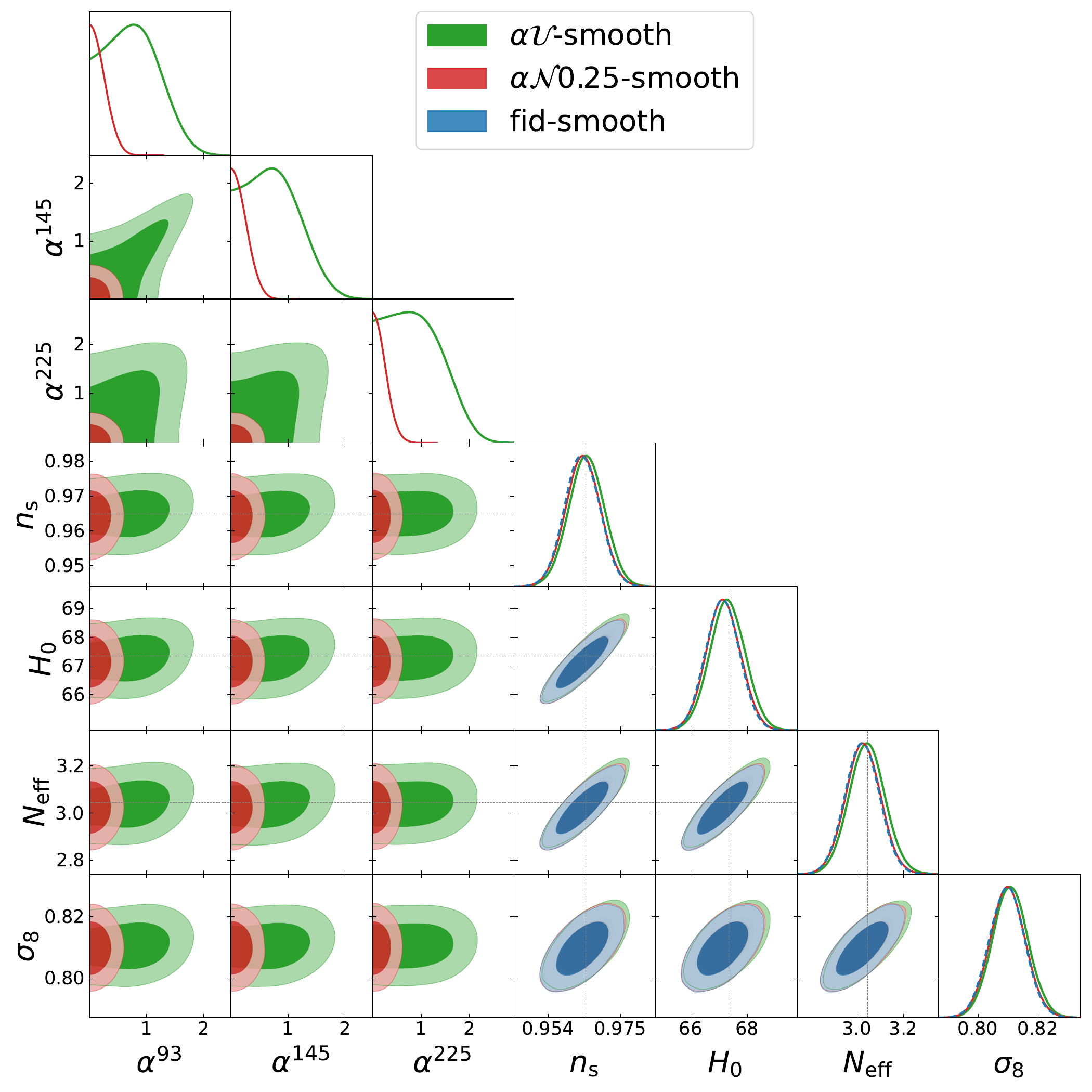}}
	{\caption{Same as Fig. \ref{fig:tr_bsh_simwbsh} but for the case in which we explore variations in polarization angles with Gaussian priors $\mathcal{N}(0,0.25)^\circ$ (in red) and flat priors $[0,10]^\circ$ (in green), while the dashed blue shows the benchmark \lcdm$+N_{\rm eff}$ case. The vertical dotted lines represent the fiducial parameters used to generate the simulations. We notice that the posteriors of $\alpha^{\nu}$ peak close to $1^\circ$ when using a flat positive prior, which causes biases in cosmological parameters  (see Fig.~\ref{fig:sim_a_cosmo}), especially in the ones shown here.} \label{fig:sim_a_syst}}
\end{figure}

Like for bandpass shifts, we study the effect of marginalizing over the polarization angles $\alpha^{\nu}$ on simulations with no systematics. We explore two types of priors: a Gaussian prior $\mathcal{N}(0,0.25)^\circ$ (label: $\alpha \mathcal{N}0.25$), and a flat positive prior $\mathcal{U}(0,10)^\circ$ (labeled as $\alpha \mathcal{U}$). Results are shown in Figures~\ref{fig:sim_a_cosmo}, \ref{fig:sim_a_fg} (last two columns) and \ref{fig:sim_a_syst}, and in Tables~\ref{tab:shift/sigma_alfa_neff}, \ref{tab:shift/sigma_alfa_neff_smooth}. 
Note that we only consider positive $\alpha^{\nu}$ when we use a flat prior. Indeed, in the absence of parity violating spectra (i.e., TB/BT and EB/BE), as it is our case, there is no sensitivity to the sign of $\alpha^{\nu}$ and considering a flat, positive prior allows to avoid bimodal distributions in the $\alpha^{\nu}$ posteriors. 

When we marginalize over $\alpha^{\nu}$ using the Gaussian prior, the bias on the cosmological and foreground parameters is irrelevant, $\lesssim 0.05 \sigma$ and the fiducial value of $\alpha^\nu$ is correctly recovered in this case (see Fig.~\ref{fig:sim_a_syst}). When we marginalize over $\alpha^{\nu}$ with flat priors we see $\sim 0.2-0.4\sigma$ shifts on cosmological parameters such as $N_{\rm{eff}}$ and $H_0$ (see last column of Fig.~\ref{fig:sim_a_cosmo}). 
As discussed at the end of Sec. \ref{sec:systematics}, a higher $N_{\rm{eff}}$ can compensate the lower polarization efficiency due to the posteriors of $\alpha^{\nu}$ peaking around $\sim 1^\circ$. An increase in $N_{\rm{eff}}$ causes shifts in the other cosmological parameters correlated with it (such as $n_s$, $\Omega_{b/c} h^2$, $H_0$, $\sigma_8$, $A_s e^{-2\tau}$). 
 
 It is worth noticing that the shifts in cosmological and foreground parameters are higher in the case we marginalize over $\alpha^{\nu}$ with flat priors with respect to the case we set a mismatch $\alpha^{\nu} \sim 0.25^\circ$ (Section \ref{sec:alpha_fix}). This is likely due to the fact that $\alpha^{\nu}$ is driven to explore very large values, up to 1$^\circ$ (see Fig.~\ref{fig:sim_a_syst}). 
Finally, we emphasize that the impact of values of $\alpha \sim 0.25^\circ$ is low only for the EE and TE/ET spectra included in this analysis. Indeed, the BB spectra can be affected significantly by the E-to-B mixing caused by non-ideal polarization angles \cite{Mirmelstein:2020pfk}.

\FloatBarrier

\subsubsection{Calibrations} \label{subsec:cal_margin}

\noindent\rule[0.5ex]{\linewidth}{1pt}
\textbf{Summary:} As seen above, calibration factors are correlated with those cosmological and foreground parameters that enhance or damp the overall amplitude of the spectra. Marginalization over calibrations thus broadens the posteriors of parameters like $A_s$, $a_p$, $a_s$, $a_{\rm{tSZ}}$. $N_{\rm eff}$, $H_0$ and $n_s$ posteriors are widened by $\sim 1.2$. When sampling $\mathrm{Cal}^{\nu}$ with Gaussian priors $\mathcal{N}(1, 0.01)$ and $\mathrm{Cal}_{\rm{E}}^{\nu}$ with flat priors, we obtain biases of $0.1\sigma$ in $N_{\rm{eff}}$, $n_s$ and $H_0$.

\begin{figure}[!htbp]  
		\centering 
		{\includegraphics[width=1\textwidth]{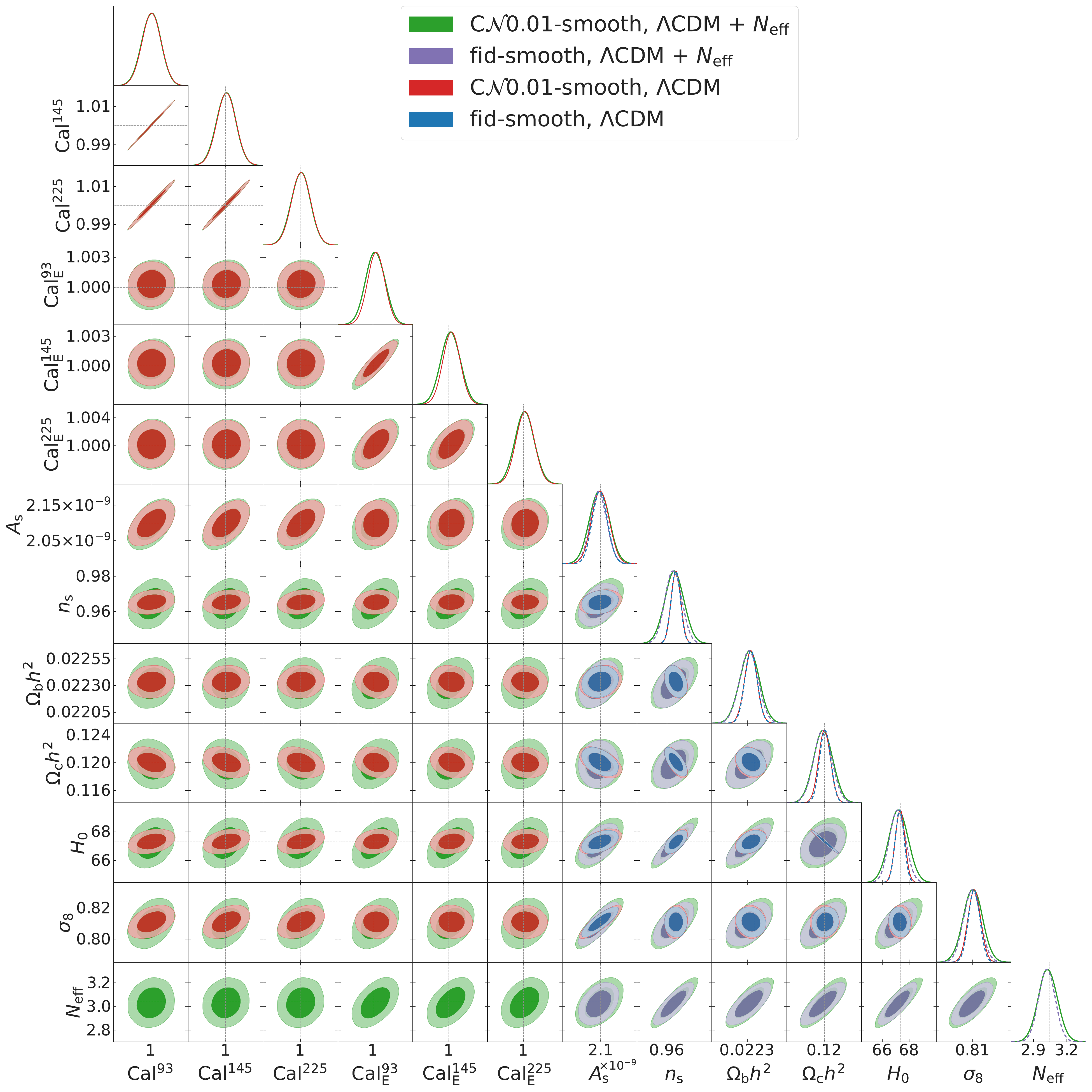}} 
			\caption{Same as Fig.~\ref{fig:tr_bsh_simwbsh} but for the case in which we explore variations in calibrations with Gaussian priors on $\mathrm{Cal}^{\nu}$ and flat priors on $\mathrm{Cal}_{\rm{E}}^{\nu}$ (in green), or with calibrations fixed to the fiducial, ideal value (in dashed purple) -- red and blue show the equivalent for a \lcdm\ cosmology. The vertical dotted lines represent the fiducial parameters used to generate the simulations, note that in this case the input is without systematics, i.e. centered on perfect calibration $\equiv 1$. We can notice that the $\mathrm{Cal}^{\nu}$ parameters are more strongly correlated with $A_s$ and $\sigma_8$ among the cosmological parameters, while the $\mathrm{Cal}_{\rm{E}}^{\nu}$ parameters are more correlated with $N_{\rm{eff}}$, $n_s$ and $H_0$. Due to this correlation, the $N_{\rm{eff}}$ posterior widens when marginalizing over $\mathrm{Cal}_{\rm{E}}^{\nu}$.} \label{fig:tr_cal_Neff}
		\end{figure}

\noindent\rule[0.5ex]{\linewidth}{1pt}
Here we allow both  $\mathrm{Cal}^{\nu}$ and $\mathrm{Cal}_{\rm E}^{\nu}$ to vary in the MCMC, analysing a simulation without any injected systematics.
We impose a Gaussian prior $\mathcal{N}(1, 0.01)$ on $\mathrm{Cal}^{\nu}$ and a flat prior $\mathcal{U}(0.9,1.1)$ on $\mathrm{Cal}_{\rm E}^{\nu}$ (label: C$\mathcal{N}0.01$) based on Ref.~\cite{ACT:2020frw}. 
The constraints are shown in the summary calibration figure, Fig.~\ref{fig:sim_nosyst_neff}. The bias and degradation in the constraints are quoted in Tables~\ref{tab:shift/sigma_Neff_cals_nosys_neff}, \ref{tab:shift/sigma_Neff_cals_smoothsim_neff}. We report the correlations between the most relevant parameters in Fig.~\ref{fig:tr_cal_Neff}. We also perform one exploration of correlated systematics, jointly marginalizing over calibrations and bandpass shifts, to investigate the possible interplay between an incorrect foreground model - induced by uncertainties in $\Delta^{\nu}$ - and the calibrations (see Appendix \ref{app:cal_marg_lcdm}). We find no correlations between the two classes of parameters. We expect the same result for the interplay between polarization angles and bandpass shifts, since in our framework the polarization angles act as calibrations per channel.
	
The marginalization over the calibration factors for each channel and the polarization efficiencies induces correlations between them and the cosmological and foreground amplitudes. The strongest correlations are between individual $\mathrm{Cal}^{\nu}$ parameters and between $\mathrm{Cal}^{\nu}$ and $A_s$, $a_s$ and $a_p$. The common calibration factors act on both temperature and polarization, so they can be constrained by the combination of all the auto and cross spectra. They are strongly correlated with the parameters which mostly impact the temperature spectra at small scales, such as the foreground amplitudes in temperature of radio and CIB Poisson and the primordial amplitude $A_s$ for the CMB component.
The polarization efficiencies are mildly correlated with $N_{\rm eff}$, $H_0$ and $a^{\rm{EE}}_{\rm{dust}}$.
These parameters impact the amplitude of the polarization spectra, $a^{\rm{EE}}_{\rm{dust}}$ being the amplitude of dust in EE and $N_{\rm eff}$ by changing the epoch of matter-radiation equality thus modifying the amplitude of oscillations at intermediate scales. These correlations induce a widening of $N_{\rm eff}$, $H_0$ and $n_s$ posteriors by $\sim 1.2$ and $a^{\rm{EE}}_{\rm{dust}}$ one by 1.1. The correlations between $\mathrm{Cal}^{\nu}$ and $\mathrm{Cal}_{\rm E}^{\nu}$ are mild since the combination of temperature and polarization spectra helps disentangle the two classes of calibration factors. 
 
We note that the simulated data are able to constrain $\mathrm{Cal}^{\nu}$ with a  1$\sigma$ sensitivity of $\sim 5 \times 10^{-3}$, tighter than the width of the Gaussian prior employed in the analysis. The constraints of the polarization efficiencies are even more stringent, with 1$\sigma \sim 1 \times 10^{-3}$.

\section{Discussion and conclusions} \label{sec:conclusion}
In this study, we have explored three classes of instrumental effects which could be a source of systematics for the SO LAT. We have presented the impact of bandpass shifts, $\Delta^{\nu}$, polarization angles, $\alpha^{\nu}$, and calibration factors, $\mathrm{Cal}^{\nu}$ and $\mathrm{Cal}_{\rm{E}}^{\nu}$. We have explored two analysis scenarios: when introducing a mismatch with respect to a benchmark scenario in these classes of parameters, and when marginalizing over them with different priors.
We assess their importance by looking at the posterior distributions of cosmological and foreground parameters, i.e., if and how the mean value of the posteriors moves and how their width broadens. 

The SO target specification of 0.5-1\% uncertainty on the bandpass central frequency \cite{Bryan:2018mva, Ward:2018fjf} corresponds to an uncertainty $\sim 1$ GHz for the frequency channels we have used. When marginalizing over bandpass shifts with a Gaussian prior $\mathcal{N}(0, 1)$ GHz, we find no significant bias on the cosmological parameters ($< 0.02 \sigma$), both in the \lcdm+$N_{\rm eff}$ model and the \lcdm~one. We do find that marginalization over bandpass shifts $\Delta^{\nu}$ is strictly necessary for accurate parameter recovery: if we set a mismatch in $\Delta^{\nu}$ between 0.8 and 1.5 GHz, we observe a bias of $0.7 \sigma$ in $N_{\rm eff}$, or biases up to $0.7 \sigma$ for $A_s$, $0.6 \sigma$ for $\Omega_b h^2$ and $1 \sigma$ for $\sigma_8$ in the \lcdm~model. The biases from improperly fixing the bandpass shifts are much larger for the foreground terms ($>10 \sigma$ in some cases), due to their frequency dependence (see Fig. \ref{fig:sim_bsh_fg_neff}). 

The most significant effect of $\Delta^{\nu}$ is on foreground parameters, whose constraints get degraded when marginalizing over $\Delta^{\nu}$. For the SZ parameters, when running with an uncertainty of 1 GHz on $\Delta^{\nu}$, we get (Table \ref{tab:shift/sigma_bsh_neff}) a bias of $0.02 \sigma/0.01 \sigma$ for $a_{\rm{tSZ}}/a_{\rm{kSZ}}$ and $\sigma$ ratio $\sim 1.3/1.7$. When considering a flat prior on $\Delta^{\nu}$, the bias gets $0.39 \sigma/-0.43 \sigma$ for $a_{\rm{tSZ}}/a_{\rm{kSZ}}$ and $\sigma$ ratio $\sim 2.3/4.3$. If we aim to keep the bias on the SZ parameters $< 0.5 \sigma$ we can also accept bandpass shifts higher than $\sim 1$ GHz, though worsening the SZ constraints. Cosmological parameters constraints are not degraded at all when marginalizing over bandpass shifts.

When considering polarization angles miscalibrated at a $\sim 0.25^\circ$ level (derived from calibrations performed by previous experiments \citep{ACT:2020frw}), cosmological and foreground parameters are not significantly affected. This is true both when introducing a mismatch $\sim 0.25^\circ$  and when marginalizing over $\alpha^{\nu}$ with a gaussian prior with $\sigma = 0.25^\circ$. In both cases, the biases on all parameters are below $0.05\sigma$ and there is no relevant degradation in the constraining power\footnote{This level of uncertainty is suitable for the analysis we have performed but more stringent requirements may be required to constrain isotropic birefringence.}.

We also considered a worse case scenario of marginalizing over $\alpha^{\nu}$ with flat priors. When considering $\Lambda$CDM+$N_{\rm{eff}}$, this causes a 0.36$\sigma$ shift on $N_{\rm{eff}}$.
This is due to the fact that $\alpha^{\nu} \neq 0^\circ$ scales down the polarization spectra, favoring higher $N_{\rm{eff}}$ to counteract the reduction of power via the delay of matter-radiation equality (see Section \ref{sec:alfa}). An increase in $N_{\rm{eff}}$ causes similar shifts in other cosmological parameters correlated with it (e.g., $n_s$, $\Omega_{b/c} h^2$, $H_0$, $\sigma_8$, $A_s$). 
The shifts are in general higher in the case of marginalization over $\alpha^{\nu}$ with flat priors since the posteriors for $\alpha^{\nu}$ peak towards $1^\circ$, which is higher than the value we set in the case with mismatch.  
In the \lcdm~case,  marginalizing over $\alpha^{\nu}$ with a flat prior biases cosmological and foreground parameters by $\lesssim 0.1 \sigma$.

We conclude that it is better to marginalize over polarization angles with a gaussian prior with $\sigma \ll 1^\circ$, which is achievable by the current precision of polarization angles measurements.

A plausible strategy for treating calibrations at the likelihood level in SO is to follow a procedure similar to the last ACT analysis \cite{ACT:2020frw}. We reproduced that by marginalizing over $\mathrm{Cal}^{\nu}$ with Gaussian priors $\mathcal{N}(1, 0.01)$ and over $\mathrm{Cal}_{\rm{E}}^{\nu}$ with flat priors [0.9,1.1].
The calibrations are correlated with the cosmological and foreground parameters that enhance or damp the acoustic peaks or the overall spectra. The marginalization over the calibrations thus broadens the posteriors of the most correlated parameters, such as $A_s$, $a_p$, $a_s$ and $a_{\rm{tSZ}}$  (see Figs. \ref{fig:tr_cal_bsh_small}, \ref{fig:sim_nosyst_neff}). Polarization efficiencies are mostly correlated with $H_0$, $n_s$ and $N_{\rm eff}$. In the \lcdm+$N_{\rm{eff}}$ case, $N_{\rm{eff}}$ and $H_0$ are shifted by $0.1\sigma$, $n_s$ by $0.09 \sigma$.
In \lcdm, we get biases of $0.09\sigma$ in $H_0$ and $-0.09$ in $\Omega_c h^2$ and higher biases in $a_s$ and in $a_p$ ($0.09/0.08 \sigma$) compared to \lcdm+$N_{\rm{eff}}$.

It is important to marginalize over calibrations and polarization efficiencies. Indeed, when considering mismatches of $0.01$ in $\mathrm{Cal}^{\nu}$ and $\mathrm{Cal}^{\nu}_{\rm E}$, we get noticeable biases on the aforementioned parameters, especially $5\sigma$ in $a_s$ and $2.4\sigma$ in $a_p$. In \lcdm+$N_{\rm eff}$, $N_{\rm{eff}}$ is biased by $5.5\sigma$ due to the correlation with $\mathrm{Cal}^{\nu}_{\rm E}$. In the \lcdm~model $H_0$ is biased by $2.7 \sigma$, driving a bias of $-3.1 \sigma$ in $\Omega_c h^2$ (see Figure \ref{fig:tr_cal_Neff}), always because of correlations with $\mathrm{Cal}_{\rm E}^{\nu}$. Though a 1\% mismatch on all calibrations represents a worst-case scenario, this example shows the danger in improperly fixing the calibration parameters. 

When marginalizing jointly over calibrations and bandpass shifts, the posterior distributions of the foreground parameters broaden due to the uncertainty on $\Delta^{\nu}$. For those parameters more correlated with calibrations, such as $a_s$ or $a_{\rm{tSZ}}$, the degradation of constraining power is stronger than in the case of marginalization over $\Delta^{\nu}$ only (see Tables \ref{tab:shift/sigma_cal_lcdm} and \ref{tab:shift/sigma_bsh_lcdm}).

We note that a $\chi^2$ analysis can also help identify untracked systematic effects which may prevent the model from describing the data accurately. As an example, we checked that a $\Delta \chi^2 \sim 80$ with respect to the benchmark case is obtained when the smooth data are fit with the systematic parameters $\Delta^{\nu}$ fixed to incorrect values. Obviously, not only does the marginalization over the systematic parameters reduce the bias levels on cosmological and foreground parameters, but also dramatically reduces the $\Delta \chi^2$ with respect to the benchmark case. This happens also when imposing gaussian priors centered on the incorrect values of the systematic parameters, further stressing the importance of marginalizing over the most relevant instrumental systematic effects like bandpass shifts.

This paper also serves as validation for the LAT power spectrum - likelihood pipeline.

In this work, we have not treated other relevant systematic effects like beam systematics, which are left to future studies. A more complex treatment of bandpass systematics, e.g. the effects of more realistic bandpass shape, can eventually be connected to the shift of the bandpass \textit{effective frequency}. We have also neglected the frequency dependence of polarization angles, which has been shown not to be problematic for the SO SAT \cite{Abitbol:2020fvn}.

Future experiments like CMB-S4 \cite{Abazajian:2019eic} will have to match more stringent requirements on systematic parameters: for example, a precision $< 0.5$\% on bandpass shifts and $\lesssim 0.1^\circ$ for polarization angles \cite{Abazajian:2019eic}. To be adapted to next-generation experiments, this kind of analysis will have to be repeated considering their more ambitious scientific goals.

\acknowledgments
We thank Xavier Garrido and Thibaut Louis for several useful discussions and invaluable help during the analysis. We acknowledge the use of \texttt{numpy} \citep{harris2020array}, \texttt{matplotlib} \citep{Hunter:2007} and \texttt{getdist} \citep{Lewis:2019xzd} software packages. We acknowledge the CINECA award under the ISCRA initiative, for the availability of high performance computing resources and the Hawk high-performance computing cluster at the Advanced Research Computing at Cardiff (ARCCA).
SG, HTJ, IH, BBe and EC acknowledge support from the Horizon 2020 ERC Starting Grant (Grant agreement No 849169); SG and EC also acknowledge support from STFC and  UKRI (grant numbers ST/W002892/1 and ST/X006360/1); MG is funded by the European Union (ERC, RELiCS, project number 101116027). Views and opinions expressed are however those of the author(s) only and do not necessarily reflect those of the European Union or the European Research Council Executive Agency. Neither the European Union nor the granting authority can be held responsible for them. MG and GP acknowledge support from the PRIN (Progetti di ricerca di Rilevante Interesse Nazionale) number 2022WJ9J33. MG, LP, ML and GP~acknowledge the financial support from the COSMOS network (www.cosmosnet.it) through the ASI (Italian Space Agency) Grants 2016-24-H.0 and 2016-24-H.1-2018. DA acknowledges support from the Beecroft Trust. GF acknowledges the support of the European Research Council under the Marie Sk\l{}odowska Curie actions through the Individual Global Fellowship No.~892401 PiCOGAMBAS. CLR acknowledges support from the Australian Research Council’s Discovery Projects scheme (DP200101068). BBo acknowledges  support from the European Research Council (ERC) under the European Union's Horizon 2020 research and innovation programme (Grant agreement No. 851274). This manuscript has been authored by Fermi Research Alliance, LLC under Contract No. DE-AC02-07CH11359 with the U.S. Department of Energy, Office of Science, Office of High Energy Physics. This work was supported in part by a grant from the Simons Foundation (Award No 457687, B.K.).

\bibliographystyle{JHEP}
\bibliography{draft,Planck_bib}

\appendix
\section{Foreground models} \label{app:fg_models}
\subsection{Kinematic Sunyaev-Zel’dovich} \label{sec:ksz}
The kSZ measures the Doppler shift of CMB photons scattering off
electrons with bulk velocity, so it has contributions from the motion of galaxy clusters at later times, from fluctuations in the electron density \citep{Ostriker:1986fua}, and in the ionization fraction (e.g., \citep{Gruzinov:1998un, McQuinn:2005ce, Iliev:2006sw}). The power is modeled as a template rescaled by an amplitude:
\begin{equation} \label{eq:ksz}
\mathcal{B}_{\ell, \mathrm{kSZ}}^{\mathrm{TT}, \nu_{1}\nu_{2}} = a_\mathrm{kSZ} \, \mathcal{B}^{\mathrm{kSZ}}_{0,\ell}
\end{equation}
where $\mathcal{B}^{\mathrm{kSZ}}_{0,\ell}$ is a template spectrum for the predicted blackbody kSZ emission for a model with $\sigma_8 = 0.8$, normalized to 1 $\mu K^2$ at $\ell_0 = 3000$ (further described in \citep{Dunkley:2013vu}), and $a_\mathrm{kSZ}$ is the normalization amplitude.

\subsection{Cosmic infrared background} \label{sec:cib}
Cosmic infrared background is caused by the redshifted thermal dust emission from high redshift star-forming galaxies. It is modeled as the sum of a Poisson part:
\begin{equation} \label{eq:cibp}
\mathcal{B}^{\mathrm{TT},\nu_1 \nu_2}_{\ell, \mathrm{CIBP}} = a_p \frac{\ell(\ell+1)}{\ell_0(\ell_0+1)} 
\left( \frac{\mu(\nu_1, \beta_p) \, \mu(\nu_2, \beta_p)}{\mu^2(\nu_0, \beta_p)}
 \right) \mu K^2 ,
\end{equation}
and a clustered one:
\begin{equation} \label{eq:cibc}
	\mathcal{B}^{\mathrm{TT},\nu_1 \nu_2}_{\ell, \mathrm{CIBC}} = a_c \mathcal{B}^{\mathrm{CIBC}}_{0,\ell} \ell^{\alpha_{CIB}}
	\left( \frac{\mu(\nu_1, \beta_c) \, \mu(\nu_2, \beta_c)}{\mu^2(\nu_0, \beta_c)}
	\right) \mu K^2 .
\end{equation} 
They are both modeled as a modified black body:
\begin{equation} \label{eq:mod_bb}
	\mu(\nu, \beta) = \nu^{\beta} B_\nu(T_d) \, g(\nu)
\end{equation} 
where $\beta_p$ and $\beta_c$ are emissivity indices for the Poisson and clustered dust terms respectively, $T_d = 9.60 $K is the effective dust temperature\footnote{At the frequencies considered here we are in Reyleigh-Jeans regime, thus we are not sensible to the modified black body temperature, which is degenerate with the spectral index.} \cite{ACT:2020frw} and the function $g(\nu) = \left. \left( \frac{\partial B_\nu(T)}{\partial T} \right)^{-1} \right\vert_{T_{\mathrm{CMB}}}$ converts from flux to thermodynamic unit. The
CIBC term is a hybrid of Planck and \citep{Addison:2011se}: it follows the Planck model $\mathcal{B}^{\mathrm{CIBC}}_{0,\ell}$  below $\ell$ = 3000 and scales as $\alpha_{CIB} = 0.8$ for $\ell >$  3000 \citep{ACT:2020frw}. We take $\nu_0 = 150$ GHz as the reference frequency and $\ell_0$ = 3000 as the normalization multipole for the template. We model Eq. \eqref{eq:cibp} with $\ell(\ell+1)$ instead of $\ell^2$ since we are dealing with binned multipoles. 

\subsection{Radio point sources} \label{sec:radio}
The radio emission is due to power from unresolved radio sources which are not bright enough to be masked. The power spectrum in temperature and polarization is:
\begin{equation} \label{eq:radio}
\mathcal{B}^{\mathrm{XY},\nu_1 \nu_2}_{\ell, \mathrm{radio}} = a_{\mathrm{radio}}^{\mathrm{XY}} \frac{\ell(\ell+1)}{\ell_0(\ell_0+1)} \left( \frac{\nu_1 \nu_2}{\nu_0^2}\right)^{-0.5}
\left( \frac{g(\nu_1) \, g(\nu_2)}{g^2(\nu_0)}
\right) \, \mu K^2 ,	
\end{equation}
where $g(\nu)$ is the same conversion factor already described. We have the same parametrization both in temperature and polarization, while amplitudes are $a_{\mathrm{radio}}^{\mathrm{TT}} = a_\mathrm{s}$ in temperature and  $a_{\mathrm{radio}}^{\mathrm{TE/EE}} = a_\mathrm{ps}^\mathrm{TE/EE}$ in polarization.

\subsection{Thermal Sunyaev-Zel’dovich} \label{sec:tSZ}
The thermal Sunyaev-Zel’dovich effect is caused by the CMB photons interacting with high energy electrons by inverse Compton scattering. This causes a spectral distortion of the CMB, which is most apparent when observing galactic clusters.
Its power spectrum in temperature is given by:
\begin{equation} \label{eq:tsz}
\mathcal{B}_{\ell, \mathrm{tSZ}}^{\mathrm{TT}, \nu_{1}\nu_{2}} = a_\mathrm{tSZ} \, \mathcal{B}^{\mathrm{tSZ}}_{0,\ell}  \, \frac{f(\nu_1) f(\nu_2)}{f^2(\nu_0)}
\end{equation}
where $\mathcal{B}^{\mathrm{tSZ}}_{0,\ell}$ is the tSZ template, normalized to unity at $\ell_0$ = 3000, $ a_\mathrm{tSZ}$ is the amplitude and $f(\nu_1) = x \, \coth(x/2)-4$, with $x = \frac{h \nu}{k_B T_{\mathrm{CMB}}}$ and $T_{\mathrm{CMB}}$ = 2.725 K.

\subsection{Dust} \label{sec:dust}
The diffuse polarized Galactic dust emission is modeled as a modified black body in frequency times a power law in $\ell$, with a different spectral index for temperature and polarization:
\begin{equation} \label{eq:dust_tt}
\mathcal{B}^{\mathrm{TT},\nu_1 \nu_2}_{\ell, \mathrm{dust}} = a^{\mathrm{TT}}_{\mathrm{dust}} \left( \frac{\ell}{500} \right)^{-0.6}
\left( \frac{\mu(\nu_1, \beta_g) \, \mu(\nu_2, \beta_g)}{\mu^2(\nu_0, \beta_g)}
\right) \mu K^2 ,
\end{equation}
\begin{equation} \label{eq:dust_teee}
	\mathcal{B}^{\mathrm{TE/EE},\nu_1 \nu_2}_{\ell, \mathrm{dust}} = a^{\mathrm{TE/EE}}_{\mathrm{dust}} \left( \frac{\ell}{500} \right)^{-0.4}
	\left( \frac{\mu(\nu_1, \beta_g) \, \mu(\nu_2, \beta_g)}{\mu^2(\nu_0, \beta_g)}
	\right) \mu K^2 .
\end{equation}
Here the normalization is $\ell_0 = 500$, the spectral index of the modified black body is $\beta_g = 1.5$ and the effective dust temperature is $T_d = 19.6$ K~\cite{ACT:2020frw}. 

\subsection{tSZ-CIB cross-correlation} 	\label{sec:tsz-cib}
\begin{figure}[ht!]
 	\centering
 \includegraphics[width=0.8\textwidth]{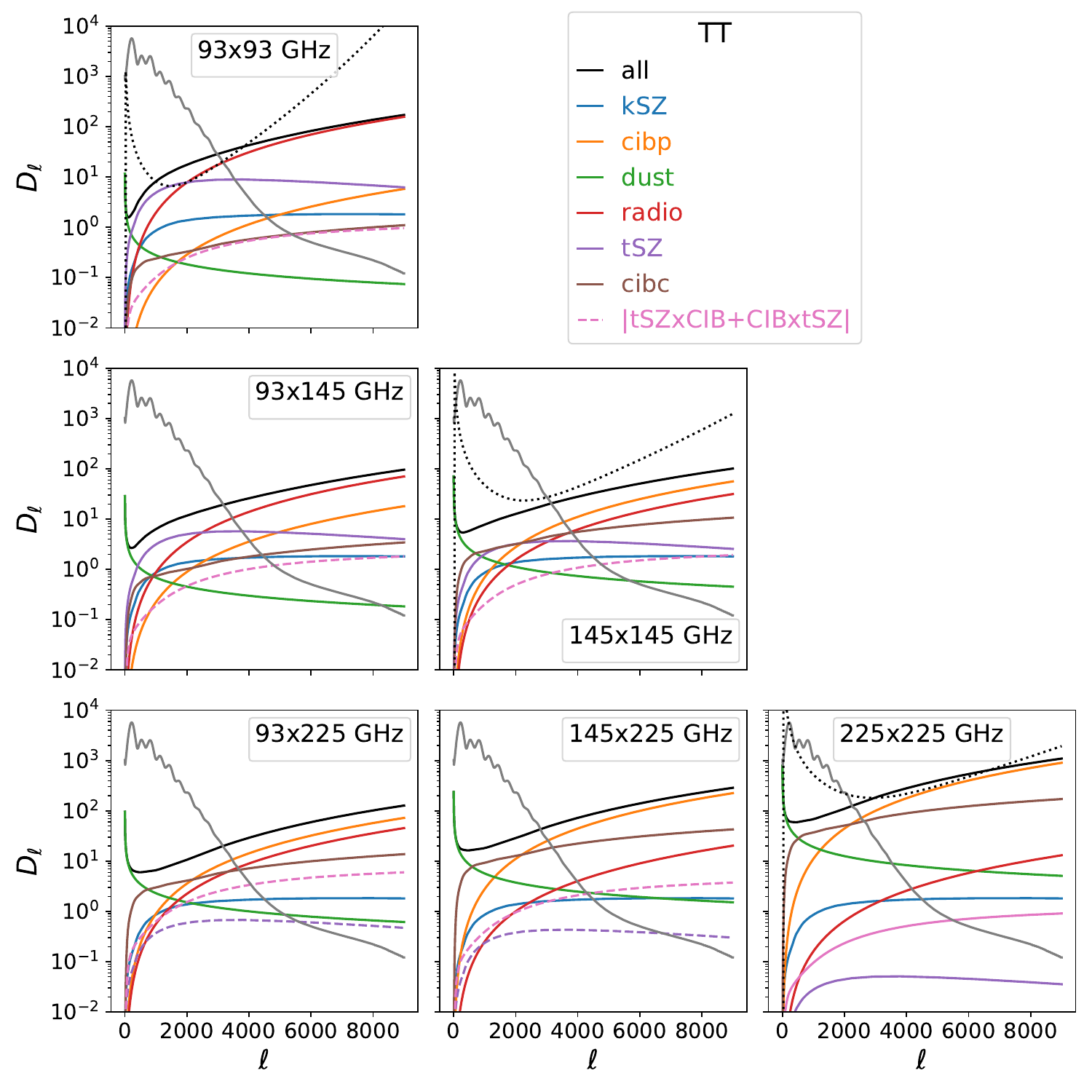}
  {\caption{Foreground components TT power spectra for the fiducial values listed in Table \ref{tab:fid_params}. The dashed lines are the absolute value of the spectra while the gray solid line is the CMB. The sum of all the foreground components is the black solid line. The black dotted line is the noise curve for the autospectra.} \label{fig:fg_comp_tt}}
\end{figure}

Since both tSZ and CIB trace galaxy clusters, there is a correlation between the two effects. The correlation is negative at 150 GHz (as you can see in Fig. \ref{fig:fg_comp_tt}):
\begin{equation} 	\label{eq:tsz-cib}
\mathcal{B}^{\mathrm{TT},\nu_1 \nu_2}_{\ell, \mathrm{tSZ-CIB}} = - \xi \sqrt{a_\mathrm{tSZ}\, a_c} \frac{2 f'(\nu_{1,2})}{f(\nu_0)} \mathcal{B}^{\mathrm{tSZ-CIB}}_{0,\ell}
\end{equation}
where $f'(\nu_{1,2}) = f(\nu_{1}) \, \mu(\nu_2,\beta_c) + f(\nu_{2}) \, \mu(\nu_1,\beta_c)$. The template $\mathcal{B}^{\mathrm{tSZ-CIB}}_{0,\ell}$ comes from \cite{Addison-tSZcib}. Also in this case the template is normalized at $\ell_0 = 3000$.

\begin{figure}
\label{fig:fg_comp_pol}
\centering
 	\subfloat[]{\includegraphics[width=0.8\textwidth]{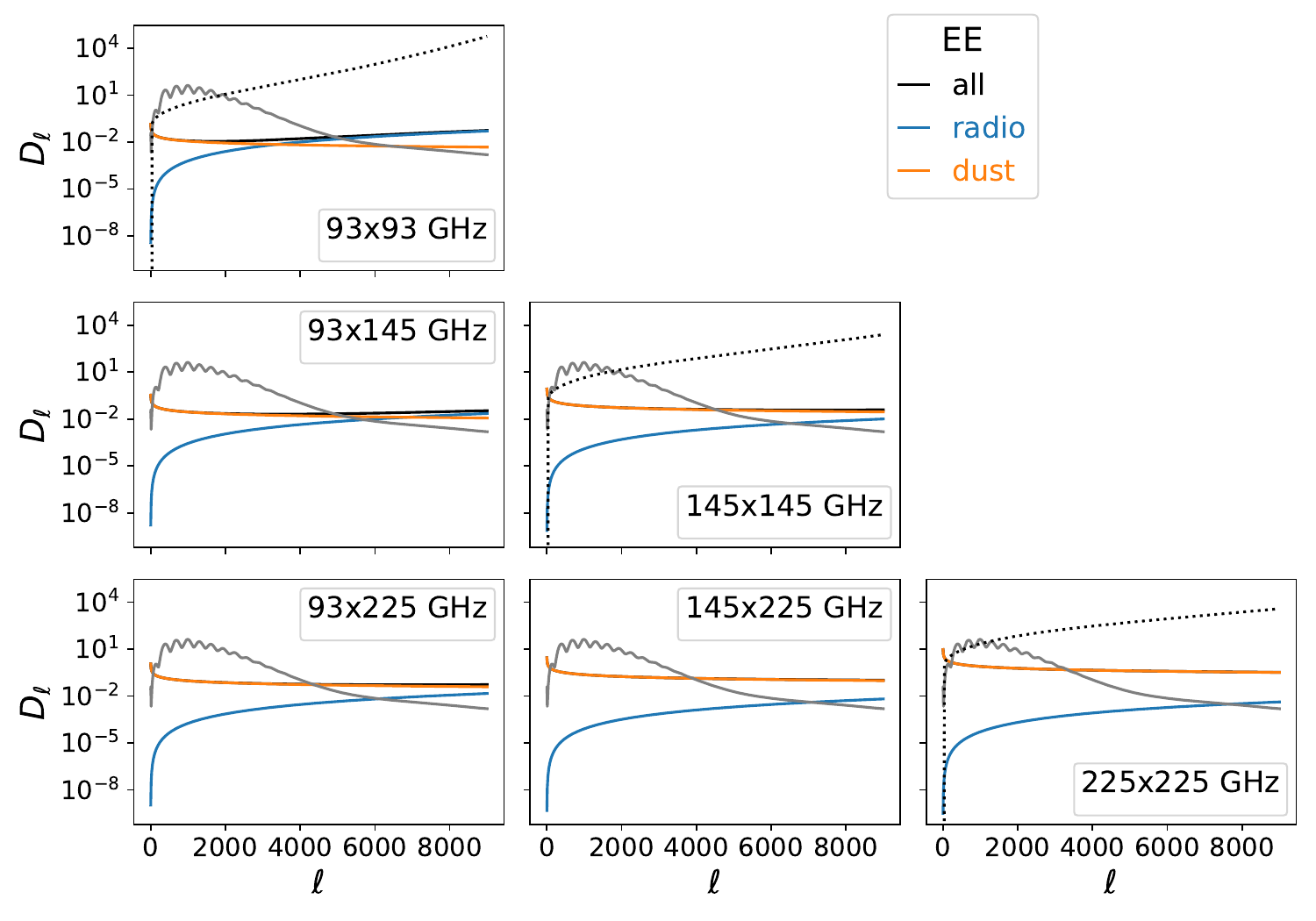}\label{fig:fg_comp_ee}}\quad
 	\subfloat[]{\includegraphics[width=0.8\textwidth]{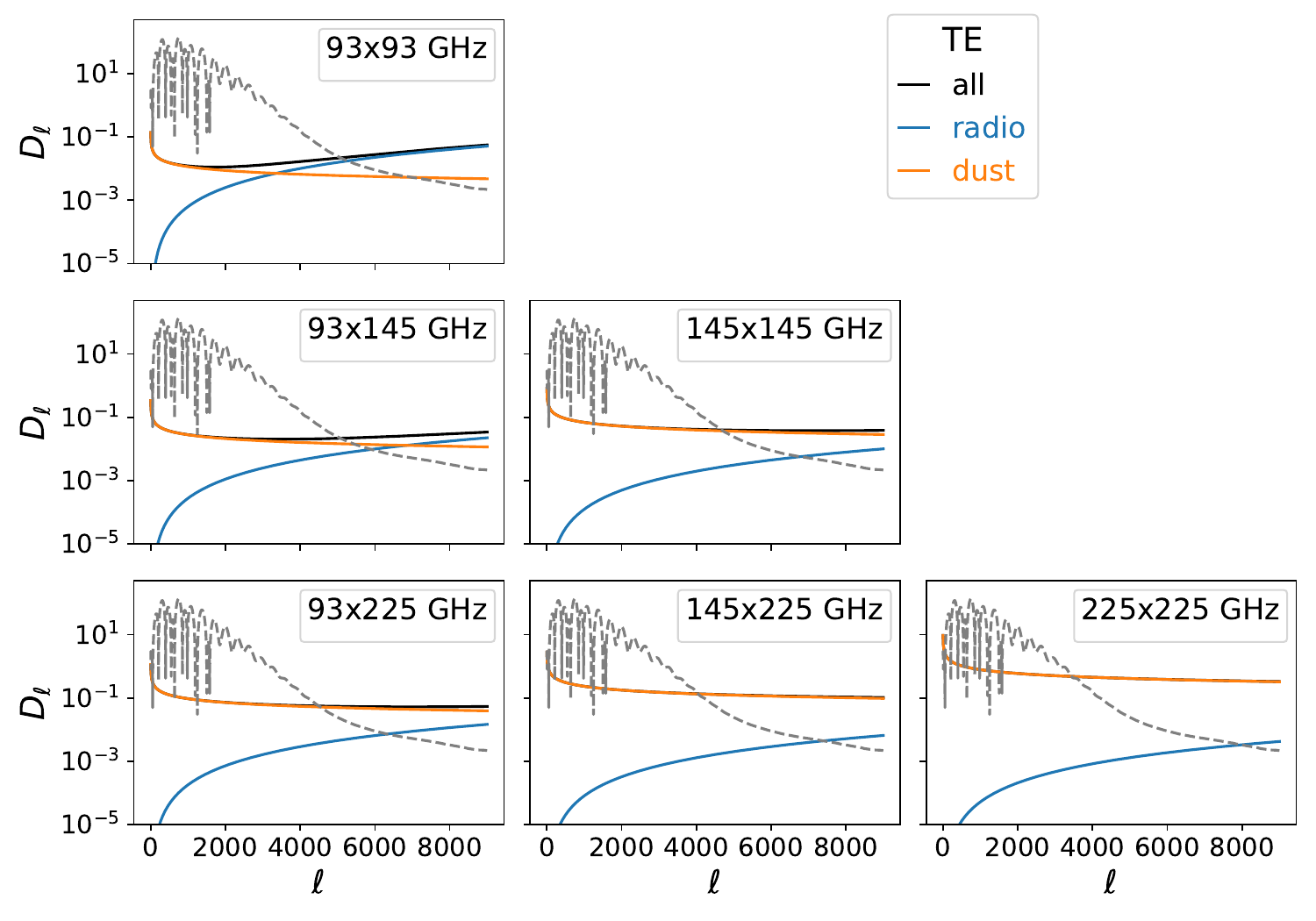}\label{fig:fg_comp_te}}\quad
 	{\caption{Foreground components EE and TE power spectra for the fiducial values listed in Table \ref{tab:fid_params} (except for $a_\mathrm{ps}^\mathrm{TE/EE}$, which have been set to 0.001 to show the radio component). The dashed lines are the absolute value of the spectra while the gray solid line is the CMB. The sum of all the foreground components is the black solid line. The black dotted line is the noise curve for the autospectra.} \label{fig:fg_comp_pol}}
 \end{figure}

\section{Results and discussion for the \lcdm~model} \label{app:result_lcdm}

The presentation of the results for the cases assuming \lcdm~follows the same structure as in Section \ref{sec:result}. All the \lcdm~runs are performed only on the smooth spectra.

\begin{table}
\centering
\begin{tabular} {l | l | l  }
\multicolumn{3}{c}{\makecell{\bf{List of runs for $\Lambda$CDM}}}\\
\hline
& \bf{case} &  \bf{systematic treatment} \\
\hline
\rule{0pt}{11pt}
\bf{benchmark} & fid &  \makecell[l]{all fixed to ideal values}  \\
\hline
\rule{0pt}{11pt}
 & $\Delta$sys &  $\Delta^{\nu}$ = \{0.8, -1, 1.5\} GHz \\
\bf{mismatched systematics} & $\alpha$sys & $\alpha$ =  \{$0.3^\circ, 0.2^\circ, 0.25^\circ$\}  \\
 & Csys & $\mathrm{Cal}^{\nu}$, $\mathrm{Cal}_{\rm E}^{\nu}  = \{1.01, 1.01, 1.01\}$ \\ 
\hline
\rule{0pt}{11pt}
\multirow{7}{*}{\bf{marginalized systematics}}&$\Delta \mathcal{N}$1 &  $\Delta^{\nu}  \in \mathcal{N}(0, 1)$ GHz \\
& $\Delta \mathcal{U}$ &  $\Delta^{\nu} \in \mathcal{U}(-10,10)$ GHz \\
\cline{2-3}
\rule{0pt}{11pt}
&$\alpha \mathcal{N}0.25$ &  $\alpha^{\nu} \in \mathcal{N}(0, 0.25)^\circ$ \\
&$\alpha \mathcal{U}$ &  $\alpha^{\nu}  \in \mathcal{U}(0,10)^\circ$ \\
\cline{2-3}
\rule{0pt}{11pt}
&C$\mathcal{N}$0.01 &  $\mathrm{Cal}^{\nu} \in \mathcal{N}(1, 0.01)$,  $\mathrm{Cal}_{\rm E}^{\nu} \in \mathcal{U}(0.9,1.1)$ \\ 
\cline{2-3}
\rule{0pt}{21pt}
\rule{0pt}{21pt}
&C$\mathcal{N}$0.01$\Delta \mathcal{N}$1 & {\makecell[l]{$\mathrm{Cal}^{\nu} \in \mathcal{N}(1, 0.01)$, $\mathrm{Cal}_{\rm E}^{\nu} \in \mathcal{U}(0.9,1.1)$, \\ $\Delta^{\nu} \in \mathcal{N}(0, 1)$ GHz }}  \\[9pt]
\hline
\end{tabular}
\caption{Summary table of all the runs using ideal spectra. The theoretical model assumed is $\Lambda$CDM. The labels used throughout the text are in the second column. Since for \lcdm~we are using just smooth spectra, throughout the text and in Tables and figures these labels are followed by the tag \texttt{-smooth}. In the third column we present how a specific systematic is treated (fixed or marginalized). When a systematic parameter is not mentioned, it means that it has been fixed to the ideal value.} \label{tab:runs_ideal_lcdm}
\end{table}

\subsection{Systematic effects from the incorrect determination of instrumental properties} \label{app:fix_sys_params_lcdm}

\subsubsection{Bandpass shifts}
We introduce a mismatch between input spectra and theory model by fixing $\Delta^{93} = 0.8,\, \Delta^{145}=-1, \,\Delta^{225}= 1.5$ GHz in the theory model when fitting spectra simulated with $\Delta^{\nu} = 0$ GHz.  We label this case as \texttt{$\Delta$sys-smooth}, while the reference case in which the systematic parameters are fixed to the fiducial values is labeled as \texttt{fid-smooth} (see Table \ref{tab:runs_ideal_lcdm}). 

\begin{figure}[h!]	
\centering
	{\includegraphics[width=1.\textwidth]{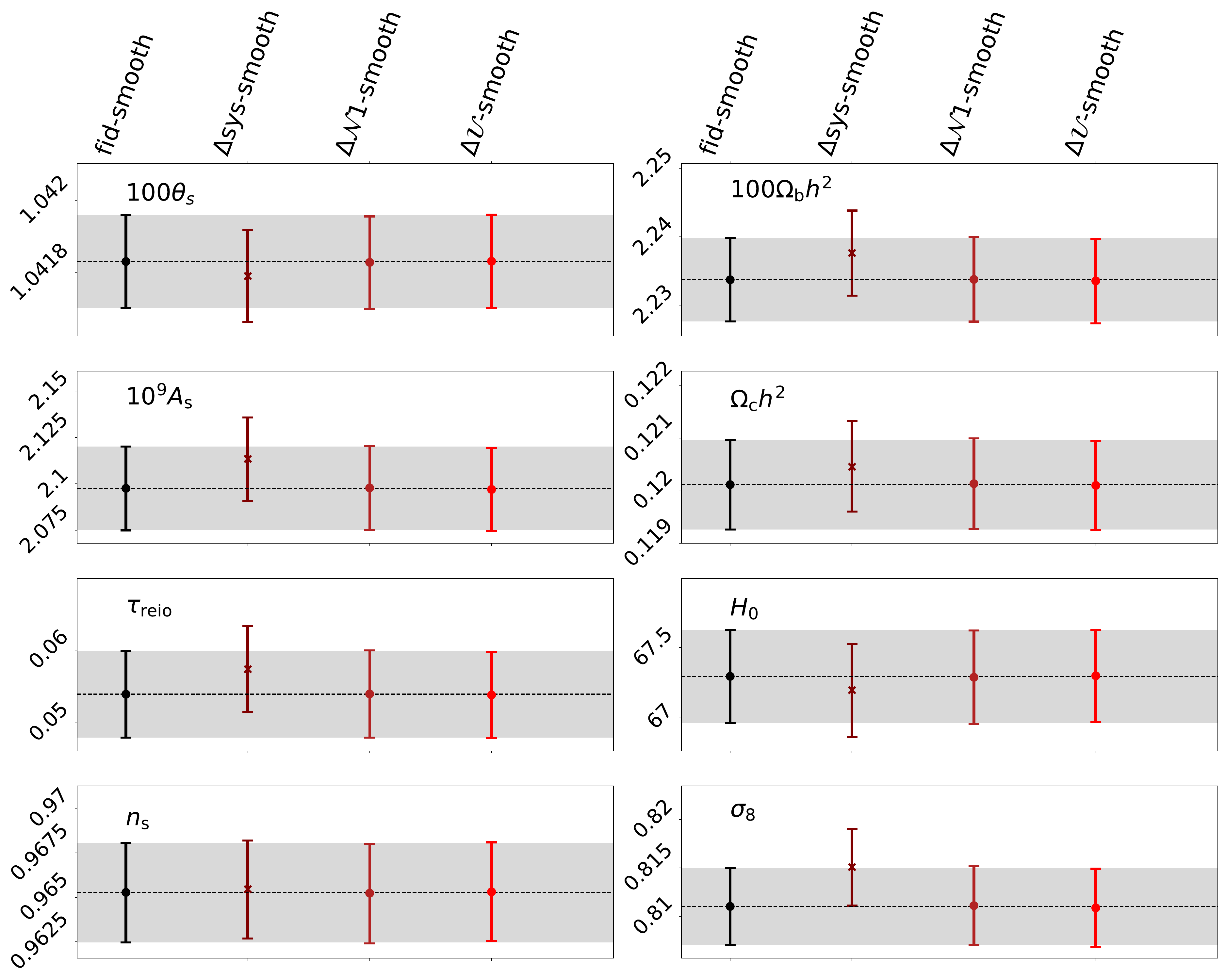}}
	{\caption{Mean values and 1$\sigma$ limits for the cosmological parameters, for all the bandpass shift cases using smooth ideal spectra, indicated by the corresponding label (see Table \ref{tab:runs_ideal_lcdm} for a list of the  $\Lambda$CDM cases analyzed and their labels). The black dashed line indicates the mean value of the \texttt{fid-smooth} case, i.e. $\Lambda$CDM with $\Delta^{\nu}$ and the other systematics fixed to the ideal fiducial values, used as benchmark for $\Lambda$CDM. The crosses indicate the values at which a parameter has been fixed, if not sampled. The cases with mismatches in $\Delta^{\nu}$ are indicated with a cross marker instead of a dot for the mean value. This case (second column) shows relevant biases in cosmological parameters, e.g. 1$\sigma$ bias in $\sigma_8$, due to larger biases in foreground parameters (see Fig. \ref{fig:sim_bsh_fg_lcdm}). When marginalizing over $\Delta^{\nu}$ (third and fourth columns), we have negligible biases in cosmological parameters.} \label{fig:sim_bsh_cosmo_lcdm}}
\end{figure}

\begin{figure}[h!]
	\centering
	{\includegraphics[width=1.\textwidth]{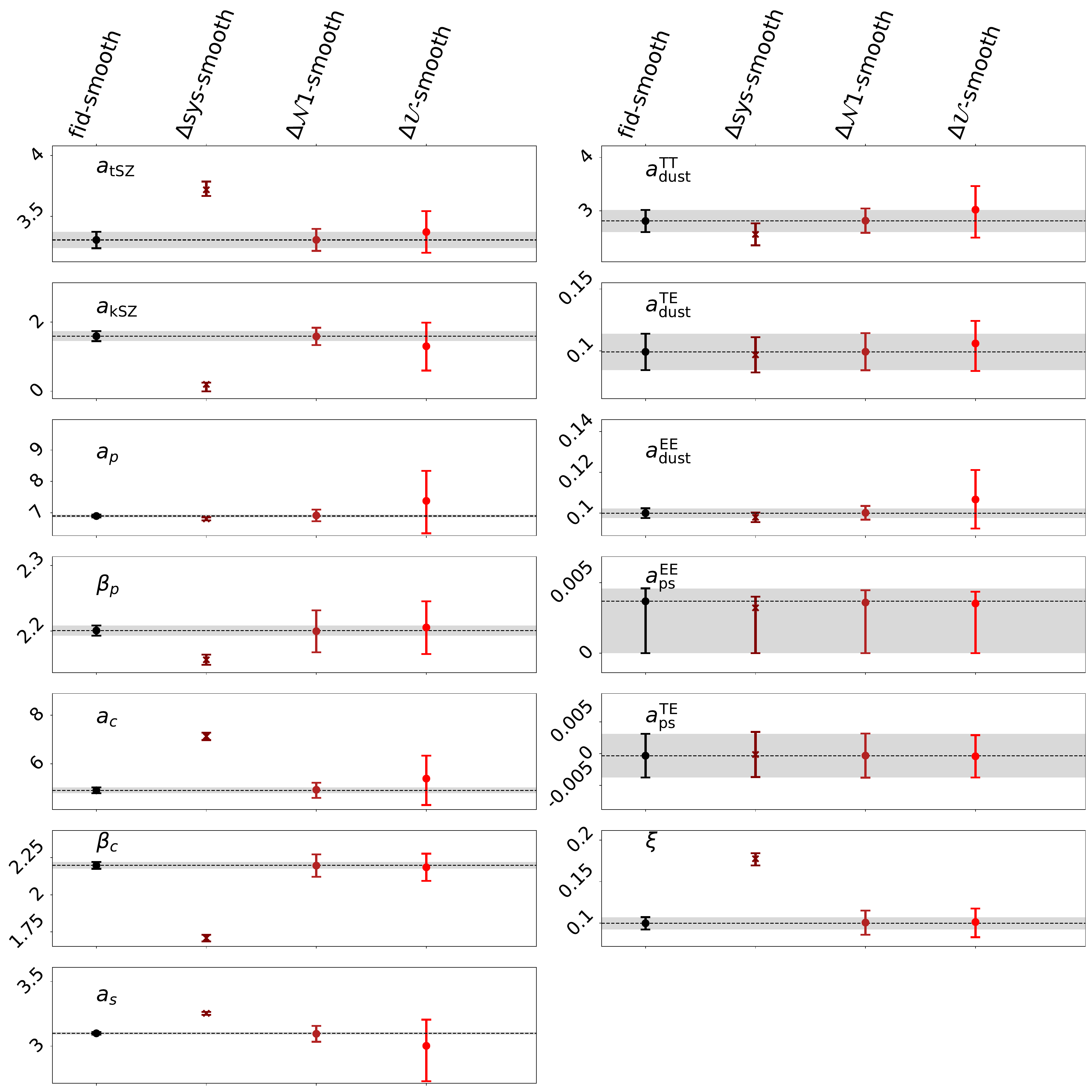}}
	{\caption{Same as in Figure \ref{fig:sim_bsh_cosmo_lcdm}, but for foreground parameters. See Table \ref{tab:runs_ideal_lcdm} for a list of the  $\Lambda$CDM cases analyzed and their labels. The case with mismatches in $\Delta^{\nu}$ (second column) shows $>10 \sigma$ biases for some foreground parameters, due to their frequency dependence. This causes biases in cosmological parameters, too (see Fig. \ref{fig:sim_bsh_cosmo_lcdm}). When marginalizing over $\Delta^{\nu}$ (third and fourth columns), the biases reduce but constraints get degraded, especially for wider priors on $\Delta^{\nu}$.} \label{fig:sim_bsh_fg_lcdm}}
\end{figure}

\begin{figure}[h!]
	\centering
	{\includegraphics[width=1.\textwidth]{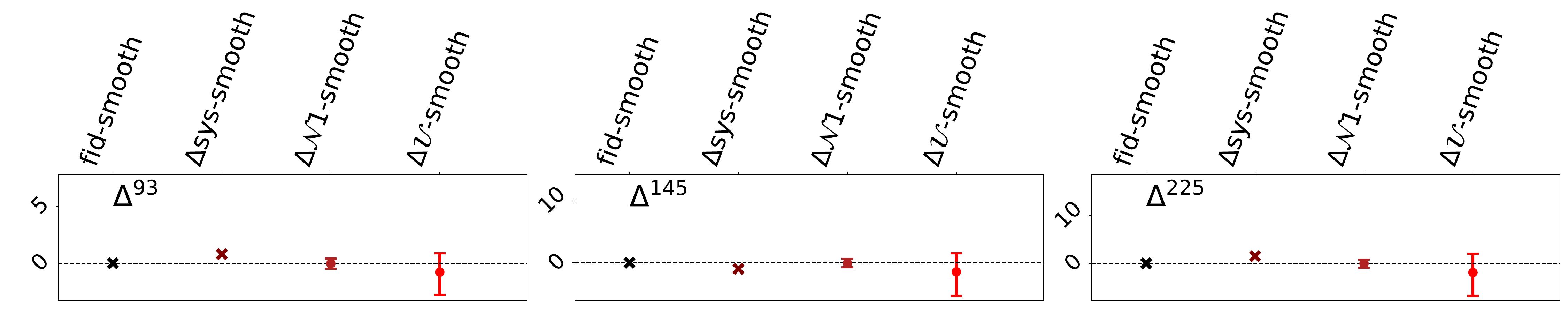}}
	{\caption{Same as in Figure \ref{fig:sim_bsh_cosmo_lcdm}, but for the bandpass shifts parameters. All the other systematic parameters are not shown, since they are fixed to their fiducial, ideal value. The crosses indicate the values at which each parameter has been fixed, in the case it has not been sampled. The black dashed line indicates the reference fiducial value for each $\Delta^{\nu}$, employed in the spectra used in this analysis. See Table \ref{tab:runs_ideal_lcdm} for a list of the  $\Lambda$CDM cases analyzed and their labels. The constraints on bandpass shifts become several GHz wide when using a flat prior.} \label{fig:sim_bsh_syst_lcdm}}
\end{figure}

The same conclusions as the \lcdm+$N_{\rm eff}$ case apply: the effect of the mismatch is more relevant for the foreground parameters. This can be appreciated in Figure \ref{fig:sim_bsh_fg_lcdm}, where the recovered 1$\sigma$ constraints can lay much more than 1$\sigma$ away from the reference value. The cosmological parameters are biased as well, even though by $\lesssim 1 \sigma$ (see Figure \ref{fig:sim_bsh_cosmo_lcdm}). The values at which the bandpass shifts have been fixed for the cases under consideration are shown in Figure \ref{fig:sim_bsh_syst_lcdm}. 

In Table \ref{tab:shift/sigma_bsh_lcdm} (first column) we report
the shifts in the mean of the marginalized parameters with respect to the case with no mismatch, normalized to the 1$\sigma$ error on the parameters. The effect on the foreground parameters can be quantitatively appreciated.

\FloatBarrier

\FloatBarrier

\FloatBarrier

\subsubsection{Polarization angles}

We introduce a mismatch in the polarization angles $\alpha^{\nu}$ between input spectra and theory vector, assuming $\alpha^{93} =  0.3^\circ,\, \alpha^{145} = 0.2^\circ,\, \alpha^{225} = 0.25^\circ$ for all channels in the theory model when fitting spectra with $\alpha^{\nu} = 0^\circ$.  
The labels for the case with $\alpha^{\nu}$ fixed to the wrong/fiducial values are \texttt{$\alpha$sys-smooth}/\texttt{fid-smooth} (see Table \ref{tab:runs_ideal_lcdm}).  

The same conclusions as in the \lcdm+$N_{\rm eff}$ case apply here, i.e. a negligible level of bias ($< 0.05 \sigma$) on all parameters due to the small mismatch in $\alpha^{\nu}$. 

In Table \ref{tab:shift/sigma_alfa_lcdm} (first column) we report  
the shifts in the mean of the marginalized parameters with respect to the case with no mismatch, normalized to the $\sigma$ of the parameters.

\begin{figure}[h!]
	\centering
	{\includegraphics[width=1.\textwidth]{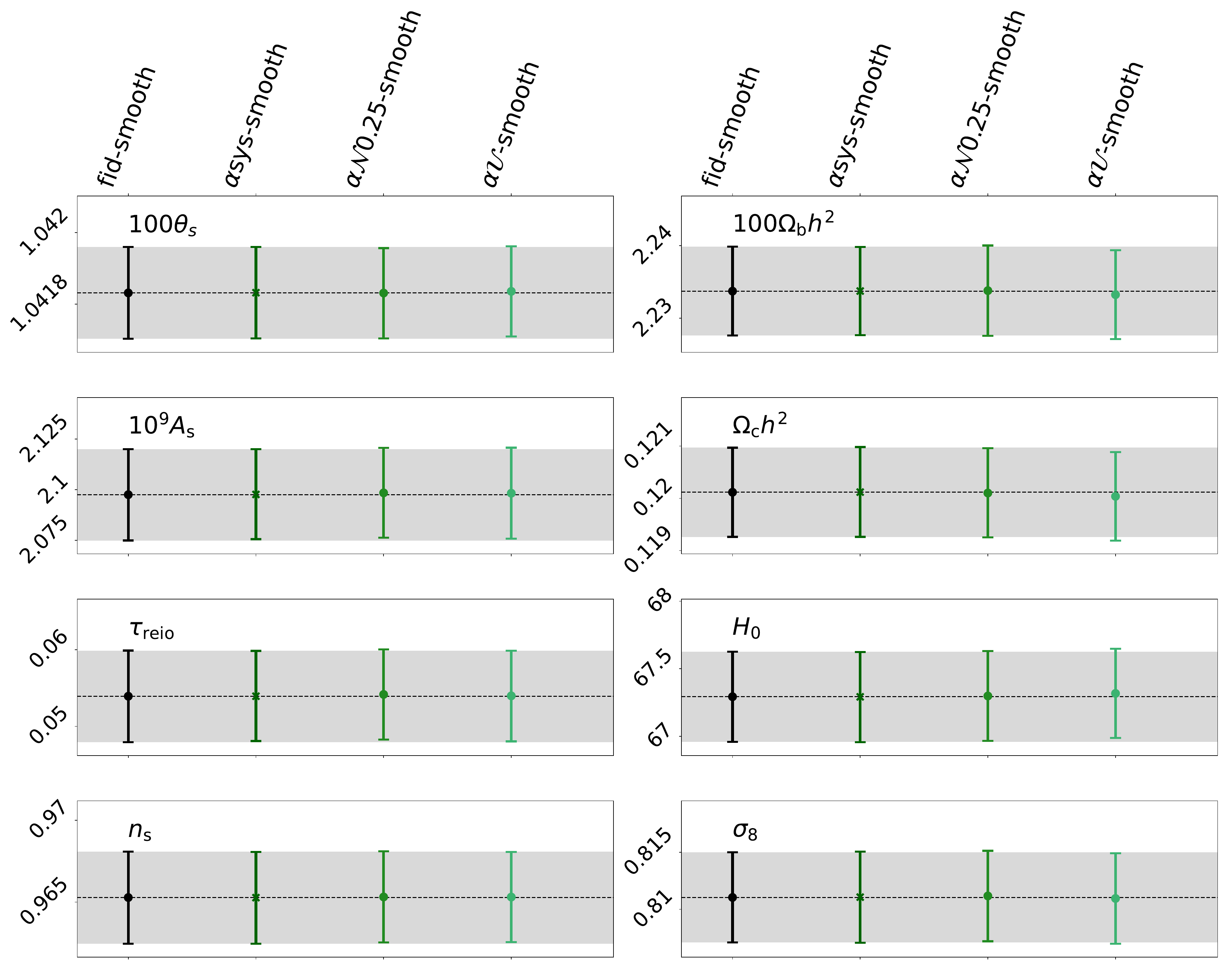}}
	{\caption{Mean values and 1$\sigma$ limits for the cosmological parameters, for all the polarization angles cases using smooth ideal spectra, indicated by the corresponding label (see Table \ref{tab:runs_ideal_lcdm} for a list of the $\Lambda$CDM cases analyzed and their labels). The black dashed line indicates the mean value of the \texttt{fid-smooth} case, i.e. $\Lambda$CDM with $\alpha^{\nu}$ and the other systematics fixed to the fiducial, ideal values, used as benchmark for $\Lambda$CDM. The cases with mismatches in $\alpha$ are indicated with a cross marker instead of a dot for the mean value. We don't have relevant biases in cosmological parameters, even in the case with marginalization over $\alpha^{\nu}$ with flat priors.} \label{fig:sim_a_cosmo_lcdm}}
\end{figure}
\FloatBarrier

\begin{figure}[h!]
	\centering
	{\includegraphics[width=1.\textwidth]{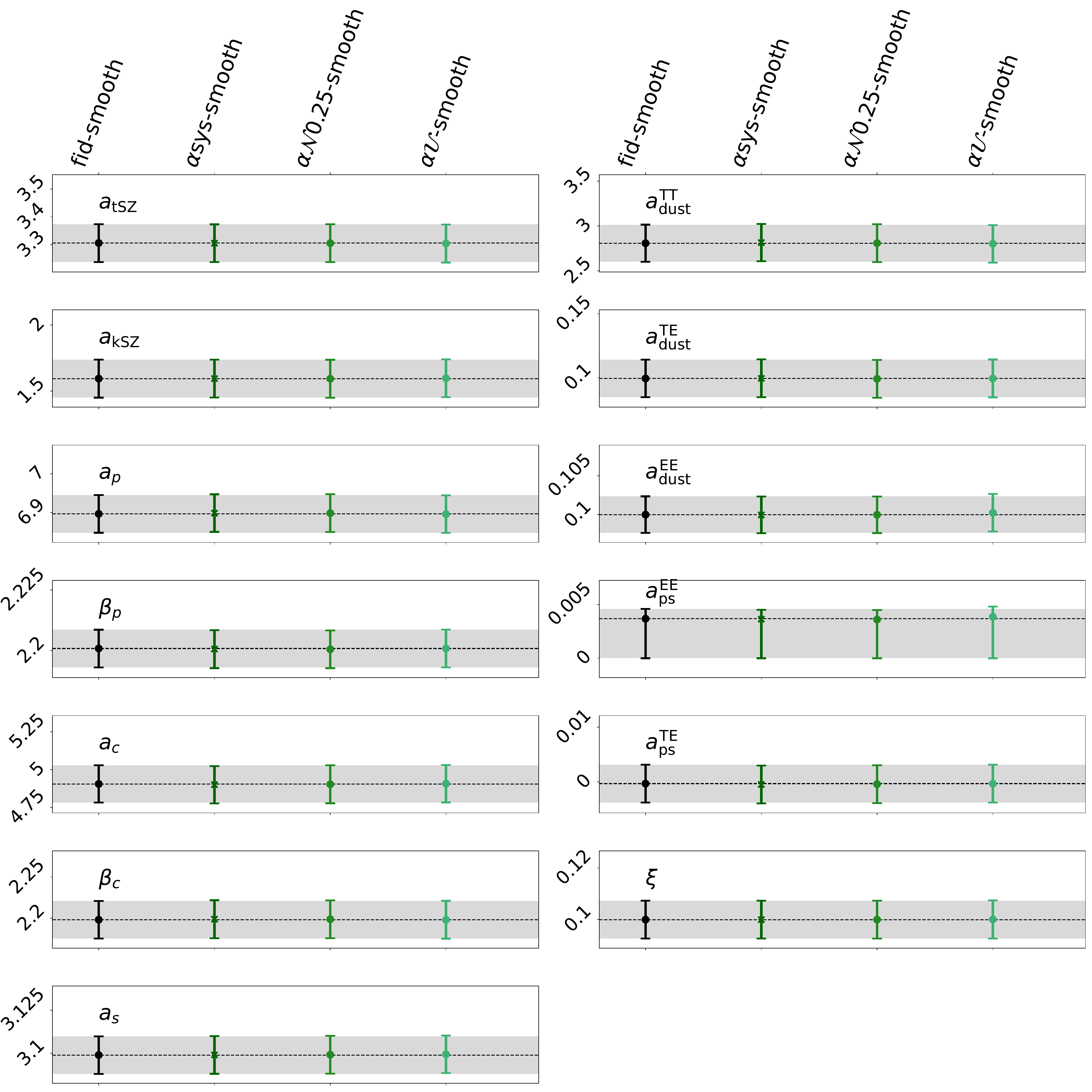}}
	{\caption{Same as Figure \ref{fig:sim_a_cosmo_lcdm}, but for the foreground parameters. See Table \ref{tab:runs_ideal_lcdm} for a list of the  $\Lambda$CDM cases analyzed and their labels.  We do not have noticeable biases in foreground parameters, including in the case with marginalization over $\alpha^{\nu}$ with flat priors.} \label{fig:sim_a_fg_lcdm}}
\end{figure}
\FloatBarrier

\begin{figure}[h!]
	\centering
	{\includegraphics[width=1.\textwidth]{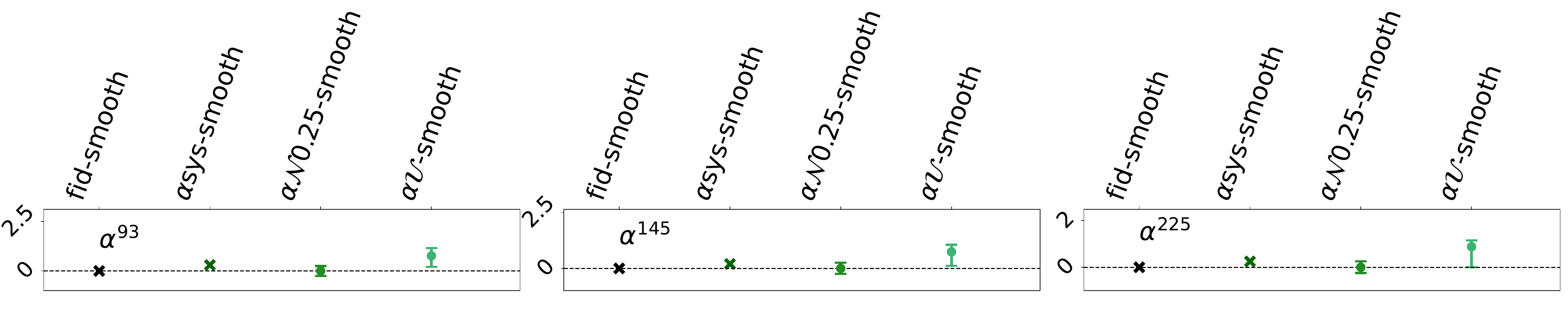}}
	{\caption{Same as in Figure \ref{fig:sim_a_cosmo_lcdm}, but for the polarization angles parameters. All the other systematic parameters are not shown, since they are fixed to their fiducial, ideal value. The crosses indicate the values each parameter has been fixed to, in the case it has not been sampled. The black dashed line indicates the reference fiducial value for each $\alpha^{\nu}$, employed in the spectra used in this analysis. See Table \ref{tab:runs_ideal_lcdm} for a list of the  $\Lambda$CDM cases analyzed and their labels. The posterior peaks towards 1$^\circ$ in the case using flat priors.} \label{fig:sim_a_syst_lcdm}}
\end{figure}
\FloatBarrier

\subsubsection{Calibrations}
We introduce a mismatch in the calibration factors of each channel $\mathrm{Cal}^{\nu}$ and polarization efficiencies $\mathrm{Cal}_{\rm{E}}^{\nu}$ fixing them to 1.01 in the theory model compared to the reference value 1 in the input spectra. The labels for the case with calibrations fixed to the wrong/fiducial values are \texttt{Csys-smooth}/\texttt{fid-smooth} (see Table \ref{tab:runs_ideal_lcdm}). 

\begin{figure}[h!]
	\centering
	{\includegraphics[width=1.\textwidth]{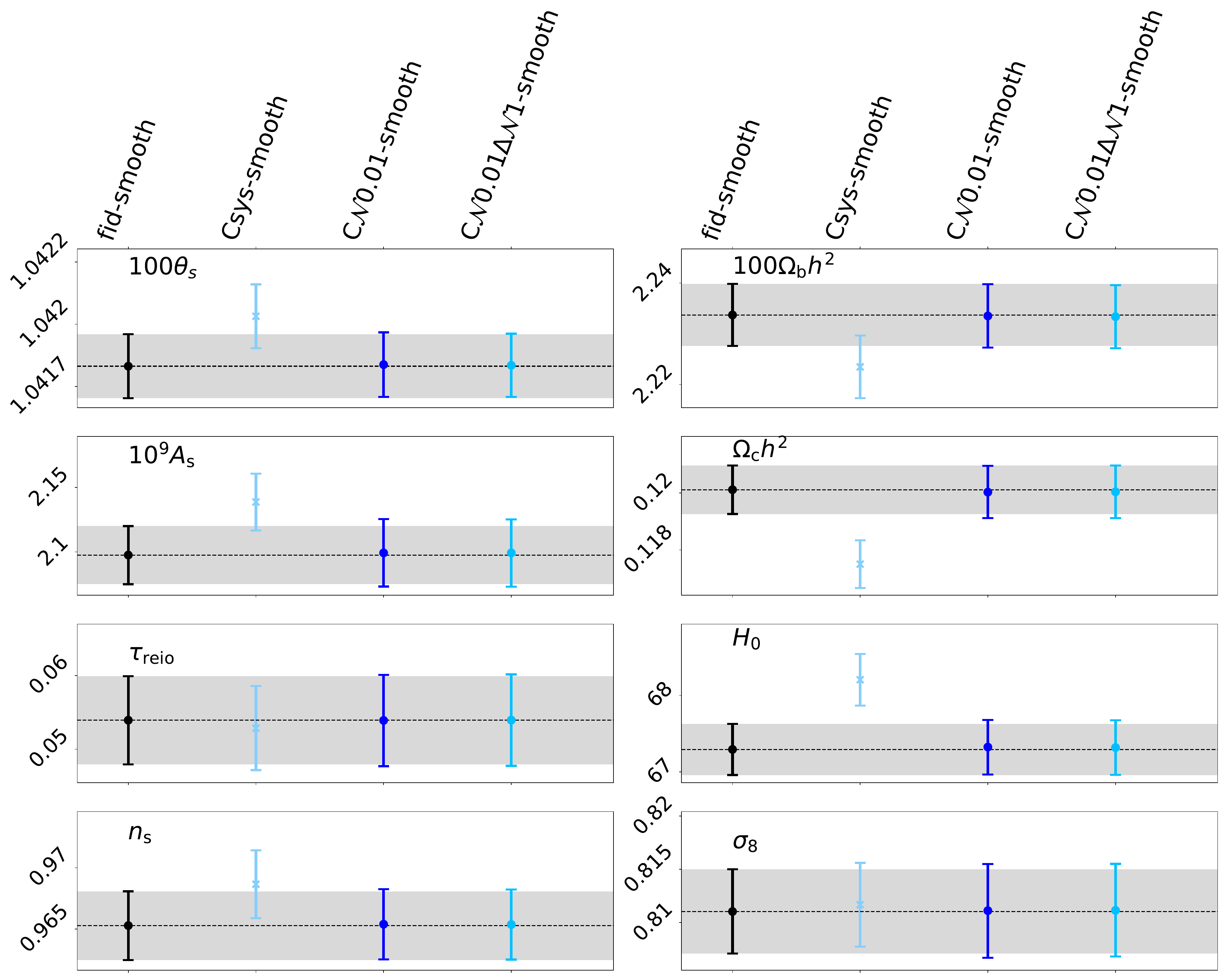}}
	\caption{Mean values and 1$\sigma$ limits for the cosmological parameters, for all the calibration cases using smooth ideal spectra (see Table \ref{tab:runs_ideal_lcdm} for a list of the $\Lambda$CDM cases analyzed and their labels). The black dashed line indicates the mean value of the \texttt{fid-smooth} case, i.e. $\Lambda$CDM with all systematics fixed to the ideal, fiducial values, used as benchmark for $\Lambda$CDM. The cases with mismatches in the calibrations are indicated with a cross marker instead of a dot for the mean value.  When setting Cal$^{\nu}$ = Cal$^{\nu}_{\rm E}$ = 1.01 (second column) we have biases in amplitudes parameters, such as $1.8\sigma$ in $A_s$, and parameters determining the amplitude of polarization spectra, such as $2.7 \sigma$ in $H_0$ and $-3.1 \sigma $ in $\Omega_c h^2$. They get reduced when marginalizing over the calibrations (third and fourth columns). The last column corresponds to marginalization over both calibrations and $\Delta^{\nu}$ (with Gaussian prior with $\sigma = 1$ GHz). No relevant bias is shown. See Sec. \ref{app:cal_marg_lcdm} for more details.} \label{fig:sim_nosyst_cosmo_lcdm}
	\end{figure}

\begin{figure}[h!]
		\centering
		{\includegraphics[width=1.\textwidth]{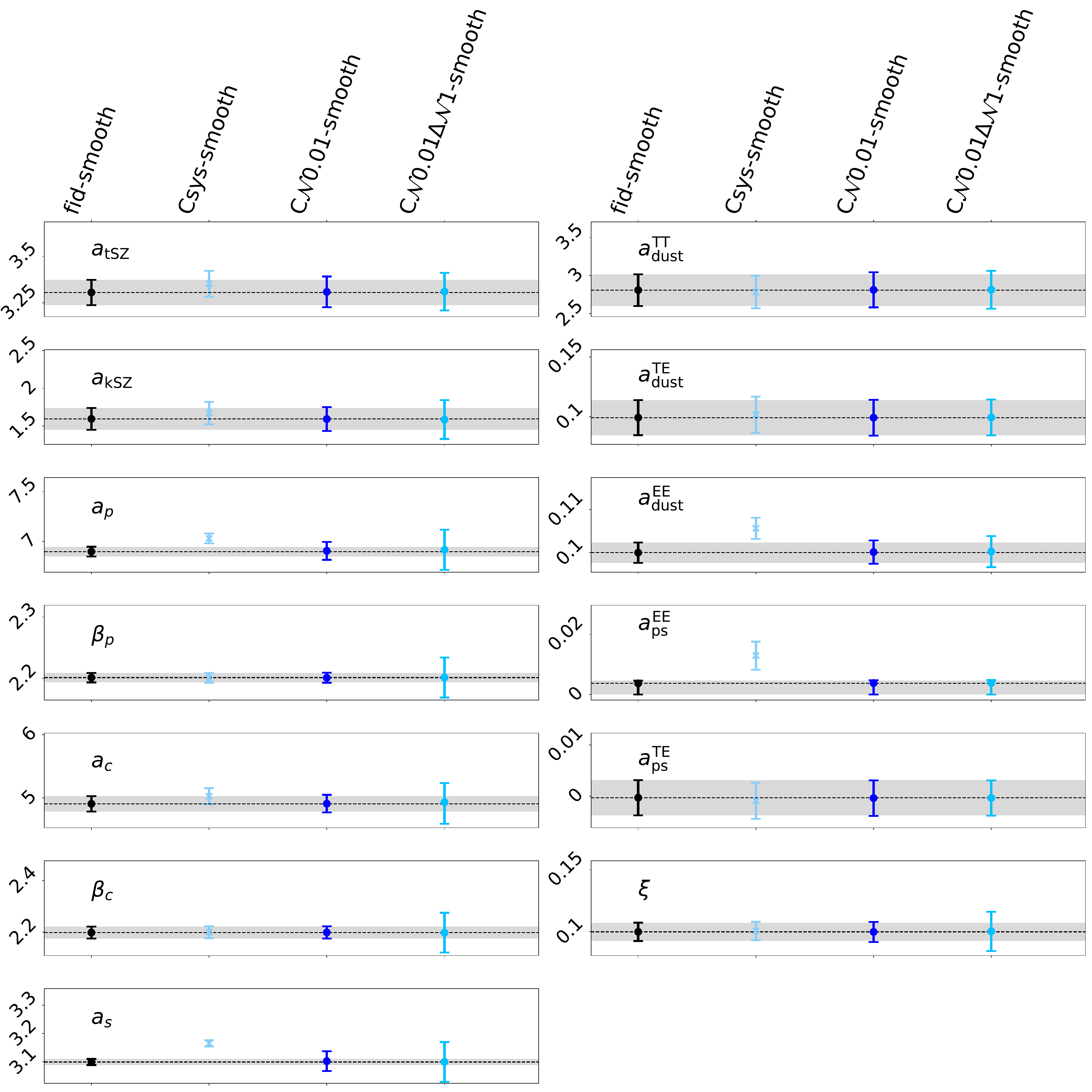}}
		{\caption{Same as Figure \ref{fig:sim_nosyst_cosmo_lcdm}, but for the foreground parameters. See Table \ref{tab:runs_ideal_lcdm} for a list of the  $\Lambda$CDM cases analyzed and their labels. When setting Cal$^{\nu}$ = Cal$^{\nu}_{\rm E}$ = 1.01 (second column) we have biases in amplitudes parameters, such as $5.9 \sigma$ in $a_s$ and $2.6 \sigma$ in $a_p$. They get reduced when marginalizing over the calibrations (third and fourth column), while their posterior get wider. The last column corresponds to marginalization over both calibrations and $\Delta^{\nu}$ (with Gaussian prior with $\sigma = 1$ GHz). In this case, the foreground parameters are dominated by the marginalization over $\Delta^{\nu}$ (compare with the third column of Fig. \ref{fig:sim_bsh_fg_lcdm}). See Sec. \ref{app:cal_marg_lcdm} for more details.} \label{fig:sim_nosyst_fg_lcdm}}
	\end{figure}

\begin{figure}[h!]
		\centering
		{\includegraphics[width=1.\textwidth]{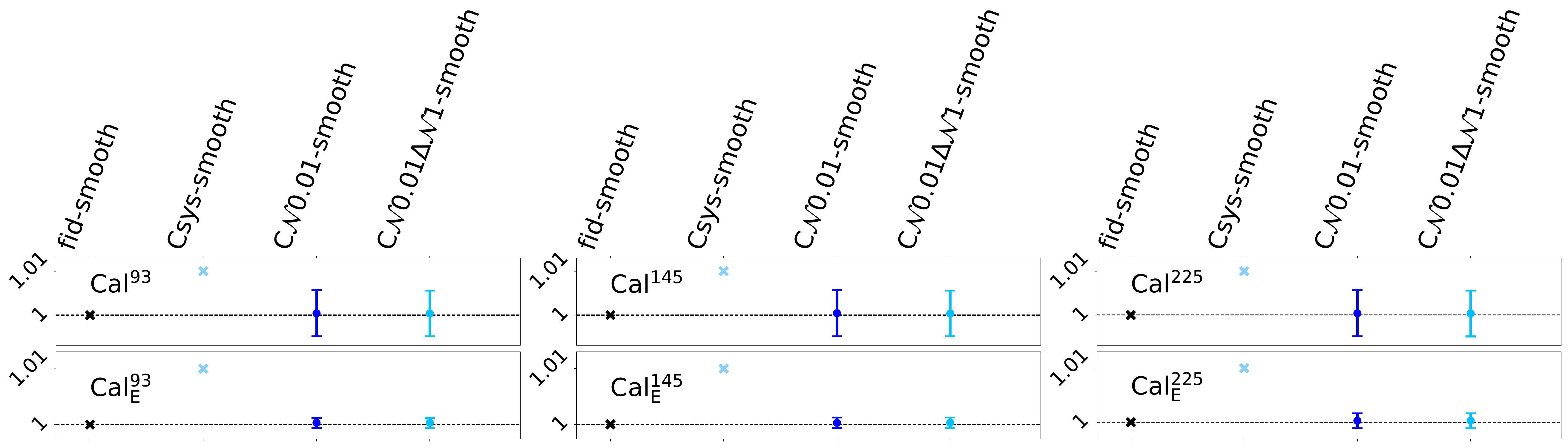}}
		{\caption{Same as in Figure \ref{fig:sim_nosyst_cosmo_lcdm}, but for the calibration parameters. The crosses indicate the values each parameter has been fixed to, in the case it has not been sampled. The black dashed line indicates the reference fiducial value for each systematic parameter, employed in the spectra used in this runs. See Table \ref{tab:runs_ideal_lcdm} for a list of the  $\Lambda$CDM cases analyzed and their labels. There is no difference in the calibrations constraints with or without marginalization over $\Delta^{\nu}$} \label{fig:sim_nosyst_syst_lcdm}}
	\end{figure}

Very similar conclusions as in the \lcdm+$N_{\rm eff}$ case apply here, i.e. the most shifted parameters are those that modify the amplitude of the spectra in both temperature and polarization.  We get noticeable biases on the parameters most correlated with $\mathrm{Cal}^{\nu}$, especially $5.9\sigma$ in $a_s$ and $2.6\sigma$ in $a_p$. Due to the correlation with $\mathrm{Cal}^{\nu}_{\rm E}$, $H_0$ is biased by $2.7 \sigma$, driving a bias of $-3.1 \sigma$ in $\Omega_c h^2$ (see Figure \ref{fig:tr_cal_Neff}). See Figures \ref{fig:sim_nosyst_cosmo_lcdm}, \ref{fig:sim_nosyst_fg_lcdm} for the limits on the cosmological and foreground parameters and Figure \ref{fig:sim_nosyst_syst_lcdm} to visualize the values at which the calibrations have been fixed.

The bias levels and the amount of degradation in the posteriors due to the mismatches are reported in the first column of Table \ref{tab:shift/sigma_cal_lcdm}.

\FloatBarrier

\FloatBarrier

\FloatBarrier

\subsection{Folding in the uncertainty in the determination of instrumental properties} \label{app:sampl_sys_params_lcdm}

\subsubsection{Bandpass shifts}
We discuss here the effect of sampling the bandpass shifts $\Delta^{\nu}$ using smooth ideal spectra. We analyze the impact of considering various levels of prior knowledge on $\Delta^{\nu}$. The 1$\sigma$ limits on all parameters are in Figures \ref{fig:sim_bsh_cosmo_lcdm}, \ref{fig:sim_bsh_fg_lcdm} and \ref{fig:sim_bsh_syst_lcdm}.  
	
The conclusions already described in Section \ref{sec:bsh} for \lcdm+$N_{\rm eff}$ apply also to the \lcdm~case. Indeed, bandpass shifts mostly affect foreground parameters. The assumption of a different cosmological model is almost irrelevant as far as the foreground parameters are concerned.
 
The reference Table is \ref{tab:shift/sigma_bsh_lcdm}. 
	
\subsubsection{Polarization angles}
We study the effect of marginalizing over the polarization angles $\alpha^{\nu}$ using ideal, smooth spectra. Details on the MCMC analysis are summarized in Table \ref{tab:runs_ideal_lcdm} and their results are presented in Figures \ref{fig:sim_a_cosmo_lcdm}, \ref{fig:sim_a_fg_lcdm}, \ref{fig:sim_a_syst_lcdm}. We use the same prior and label choice as the \lcdm+$N_{\rm eff}$ case, described in Section \ref{sec:alfa}.

There are no relevant parameter shifts or degradation in the constraining power when comparing the cases of varying $\alpha^{\nu}$ (with both Gaussian and flat positive priors) and those with polarization angles fixed to the fiducial values. The largest biases are of the order of $-0.09 \sigma$ for $\Omega_c h^2$, $0.07 \sigma$ for $H_0$ and $0.1 \sigma$ for $a_{\rm dust}^{\rm EE}$ in the case with flat priors on $\alpha$. 

Table \ref{tab:shift/sigma_alfa_lcdm} (second and third columns) does not show any degradations in constraining power even in the case of less informative priors on the polarization angles.

\subsubsection{Calibrations} \label{app:cal_marg_lcdm}

We discuss the \lcdm~analysis performed sampling over $\mathrm{Cal}^{\nu}$ and  $\mathrm{Cal}_{\rm E}^{\nu}$ with the following priors: Gaussian prior $\mathcal{N}(1, 0.01)$ on $\mathrm{Cal}^{\nu}$ and flat prior $\mathcal{U}(0.9,1.1)$ on $\mathrm{Cal}_{\rm E}^{\nu}$ (label: C$\mathcal{N}0.01$\texttt{-smooth}). This prior choice is motivated in Section \ref{section:SO_likelihood}. Additionally, we  explore one more option considering jointly calibration and bandpass shifts. When also $\Delta^{\nu}$ is sampled, we use a Gaussian prior $\mathcal{N}(0, 1)$ GHz on bandpass shifts (label: C$\mathcal{N}0.01 \Delta \mathcal{N} 1$\texttt{-smooth}). 
The constraints are shown in Figures \ref{fig:sim_nosyst_cosmo_lcdm}, \ref{fig:sim_nosyst_fg_lcdm} and \ref{fig:sim_nosyst_syst_lcdm}. 

In the \lcdm~case, the polarization efficiencies are mildly correlated with $\Omega_c h^2$, $a^{\rm{EE}}_{\rm{dust}}$ and $a^{\rm{EE}}_{\rm{ps}}$. These parameters impact the amplitude of the polarization spectra: $a^{\rm{EE}}_{\rm{dust}}$ and $a^{\rm{EE}}_{\rm{ps}}$ are the amplitudes of the two foreground component in EE (see Figure \ref{fig:fg_comp_ee}); $\Omega_c h^2$ modifies the amplitude of oscillations at intermediate scales by changing the epoch of matter-radiation equality. 
 
The 1$\sigma$ limits of cosmological and foreground parameters are shown in Figures \ref{fig:sim_nosyst_cosmo_lcdm} and \ref{fig:sim_nosyst_fg_lcdm}, with shifts such as $0.09 \sigma$ in $H_0$, $0.06 \sigma$ in $A_s$, $0.09 \sigma$ in $a_s$
and $0.08 \sigma$ in $a_p$, due to the complex correlations between calibrations and parameters. The bias and degradation of the constraints are listed in Table \ref{tab:shift/sigma_cal_lcdm}.
	
We now move to discuss the interplay between calibrations and bandpass shifts. The latter have a strong impact on the foreground parameters, whose amplitudes are also correlated with the calibrations. 
As usual when marginalizing over $\Delta^{\nu}$, we observe a degradation of the constraints on the foreground posteriors, proportional to the accuracy on $\Delta^{\nu}$, together with larger shifts with respect to the reference case. The bias levels and degradation of the foreground parameter constraints can be explained as due to the combination of the effects of calibrations and prior on $\Delta^{\nu}$, see e.g. the constraints of the $\Delta \mathcal{N}$1\texttt{-smooth} case in Figure \ref{fig:sim_bsh_fg_lcdm} (Gaussian prior on $\Delta^{\nu}$ with $\sigma$ = 1 GHz) vs. the C$\mathcal{N}0.01 \Delta \mathcal{N} 1$\texttt{-smooth} in Fig. \ref{fig:sim_nosyst_fg_lcdm} (calibrations + Gaussian prior on $\Delta^{\nu}$ with $\sigma$ = 1 GHz). The cosmological parameters are only mildly affected (see Figure \ref{fig:sim_nosyst_cosmo_lcdm}).  

 The calibration parameters are recovered without any difference with respect to the case without $\Delta^{\nu}$ marginalization, see Figure \ref{fig:sim_nosyst_syst_lcdm}.  
 
 The bias and degradation of the constraints for the two cases with bandpass shifts are presented in the last column of Table \ref{tab:shift/sigma_cal_lcdm}.
	
	\begin{figure}[!htbp]  
		\centering 
		\includegraphics[width=1\textwidth]{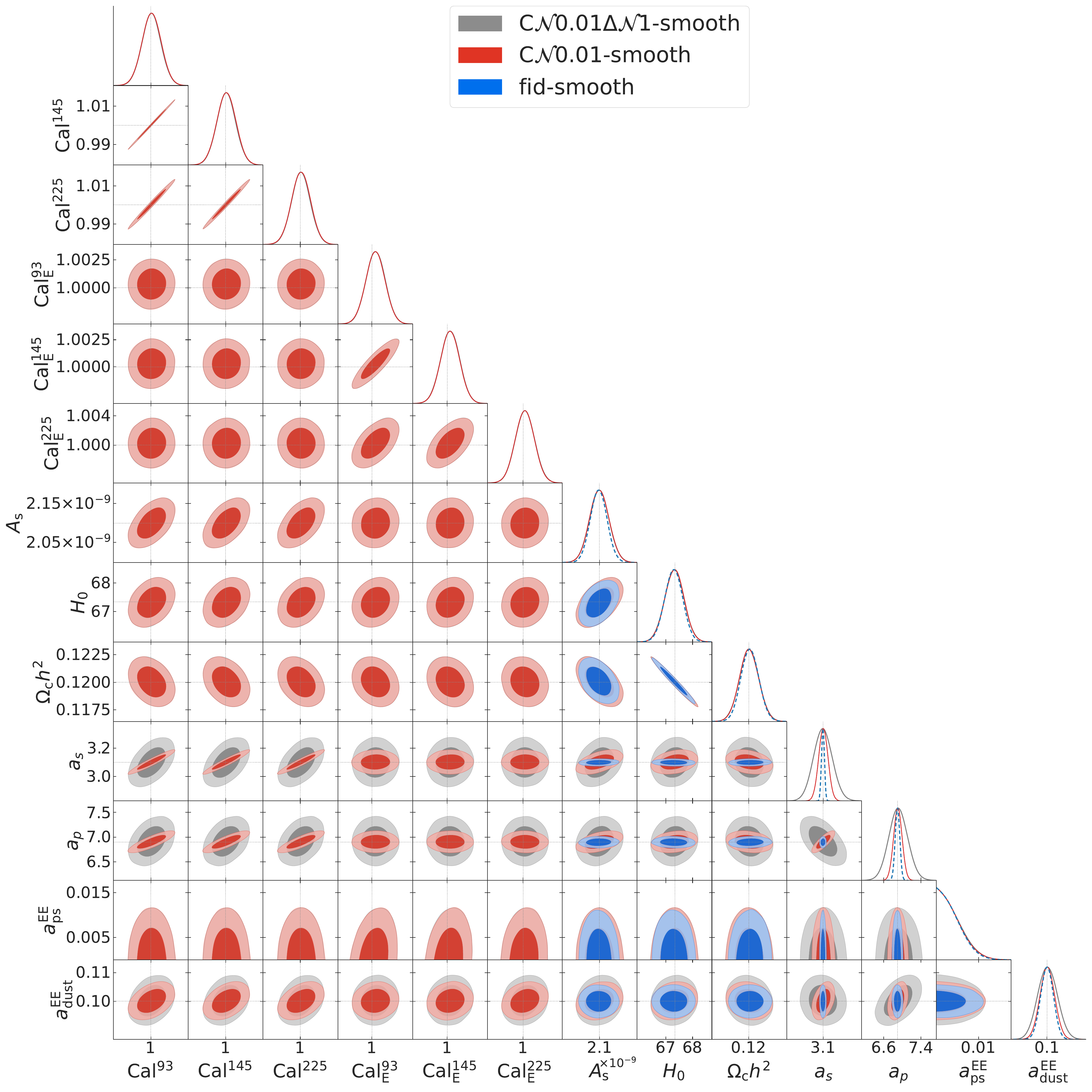}
		\caption{Posteriors for $\mathrm{Cal}^{\nu}$ and  $\mathrm{Cal}_{\rm E}^{\nu}$ and the cosmological and  foreground parameters most correlated with them. We compare results for $\Lambda$CDM with Gaussian priors on $\mathrm{Cal}^{\nu}$ and flat priors on $\mathrm{Cal}_{\rm{E}}^{\nu}$ (C$\mathcal{N}0.01$\texttt{-smooth}, in red), to the one with systematic parameters fixed to fiducial, ideal value (\texttt{fid-smooth}, in dashed blue). There are also the posteriors for the cases with calibrations and bandpass shifts (with Gaussian priors $\sigma = 1$ GHz, C$\mathcal{N}0.01 \Delta \mathcal{N} 1$\texttt{-smooth}, in gray). These results are obtained using smooth spectra without systematics, see Table \ref{tab:runs_ideal_lcdm}. The vertical dotted lines represent the fiducial parameters used to generate the simulation, summarized in Table \ref{tab:fid_params}. See the text for more discussion. The most relevant correlation in the case in which only calibrations are marginalized (red contours) is between Cal$^{\nu}$ and $a_s$, $a_p$, which are the amplitudes of the most relevant foreground components (see Fig. \ref{fig:fg_comp_tt}). When marginalizing over both calibrations and $\Delta^{\nu}$, this correlation is attenuated and the foreground posteriors widen due to degeneracies with bandpass shifts.} \label{fig:tr_cal_bsh_small}
	\end{figure}
\FloatBarrier

\section{Biases and degradation of constraints of cosmological and foreground parameters due to the effect of systematics} \label{app:SO_tables}

The effects of a mismatch in the systematic parameters or of the marginalization over them are reported in the following tables. The effects are quantified in terms of the shift in the mean of the marginalized distributions of the parameters with respect to a reference case (case $b$ in the tables), normalized to the posterior width $\sigma$ of the case under scrutiny (case $s$): $\frac{\bar{\mu}_s - \bar{\mu}_b}{\bar{\sigma}_s}$, and the ratio of their uncertainties: $\frac{\bar{\sigma}_s}{\bar{\sigma}_b}$. We recall that the bar indicates average over 100 realizations. When we quote results for the smooth spectra, we remove the $\bar{\phantom{x}}$. Information on the shift shows how much the parameters are biased due to the effect of the systematics. The ratio of uncertainties shows how much the constraints of each parameter are degraded. See Section \ref{sec:result} for more details.

\subsection{Tables for the \lcdm~+$N_{\rm eff}$ model}

\begin{table}[!htbp] 
 \renewcommand\thetable{C.1.1} 
\footnotesize
\centering 
\begin{tabular}{| P{1.3  cm} | P{1.55 cm}| P{1.55 cm} | P{1.5 cm}| P{1.5 cm} | P{1.5 cm}| P{1.5 cm} |} 
    \hline 
    Params & \multicolumn{2}{|c|}{\makecell{$s$: $\Delta$sys \\ $b$: fid }} &  \multicolumn{2}{|c|}{\makecell{$s$: $\Delta \mathcal{N}$1 \\ $b$: fid}} &  \multicolumn{2}{|c|}{\makecell{$s$: $\Delta \mathcal{U}$ \\ $b$: fid}} \\
    \hline
    & $\frac{\bar{\mu}_s - \bar{\mu}_b}{\bar{\sigma}_s}$ & $\frac{\bar{\sigma}_s}{\bar{\sigma}_b}$ & $\frac{\bar{\mu}_s - \bar{\mu}_b}{\bar{\sigma}_s}$ & $\frac{\bar{\sigma}_s}{\bar{\sigma}_b}$& $\frac{\bar{\mu}_s - \bar{\mu}_b}{\bar{\sigma}_s}$ & $\frac{\bar{\sigma}_s}{\bar{\sigma}_b}$\\
    \hline 
$A_\mathrm{s}$ & 0.366 & 1.004 & -0.002 & 1.002 & -0.022 & 1.003\\ 
$100 \theta_s$ & 0.291 & 1.007 & -0.005 & 1.001 & -0.014 & 1.001\\ 
$n_\mathrm{s}$ & -0.519 & 0.998 & 0.009 & 1.003 & 0.045 & 1.007\\ 
$\Omega_\mathrm{b}h^2$ & -0.081 & 0.999 & 0.004 & 1.001 & -0.002 & 1.002\\ 
$\Omega_\mathrm{c}h^2$ & -0.326 & 1.002 & 0.006 & 1.002 & 0.014 & 1.002\\ 
$\tau_\mathrm{reio}$ & 0.442 & 0.998 & -0.003 & 1.003 & -0.019 & 1.003\\ 
$\sigma_8$ & 0.165 & 1.005 & 0.001 & 1.001 & -0.009 & 1.001\\ 
$H_0$ & -0.714 & 0.991 & 0.010 & 1.006 & 0.042 & 1.010\\ 
$N_\mathrm{eff}$ & -0.680 & 0.996 & 0.010 & 1.005 & 0.037 & 1.009\\ 
\hline
$a_\mathrm{tSZ}$ & 6.927 & 0.849 & 0.017 & 1.311 & 0.392 & 2.273\\ 
$a_\mathrm{kSZ}$ & -10.794 & 0.900 & 0.006 & 1.730 & -0.431 & 4.305\\ 
$a_p$ & -1.783 & 1.007 & 0.033 & 3.815 & 0.445 & 19.019\\ 
$\beta_p$ & -5.726 & 0.999 & 0.090 & 4.075 & 0.193 & 5.124\\ 
$a_c$ & 14.475 & 1.219 & -0.041 & 2.507 & 0.412 & 7.481\\ 
$\beta_c$ & -21.220 & 0.983 & 0.102 & 3.259 & 0.024 & 3.977\\ 
$a_s$ & 13.909 & 1.006 & -0.010 & 5.597 & -0.385 & 21.270\\ 
$a_\mathrm{dust}^\mathrm{TT}$ & -1.251 & 0.975 & 0.053 & 1.075 & 0.401 & 2.229\\ 
$a_\mathrm{dust}^\mathrm{TE}$ & -0.113 & 0.975 & 0.015 & 1.016 & 0.259 & 1.361\\ 
$a_\mathrm{dust}^\mathrm{EE}$ & -0.864 & 0.980 & 0.071 & 1.395 & 0.461 & 5.571\\ 
$a_\mathrm{ps}^\mathrm{EE}$ & -0.130 & 0.955 & 0.001 & 1.003 & -0.032 & 0.991\\ 
$a_\mathrm{ps}^\mathrm{TE}$ & 0.155 & 1.028 & 0.002 & 1.001 & 0.001 & 0.984\\ 
$\xi$ & 11.180 & 0.947 & -0.080 & 1.985 & -0.061 & 2.330\\ 
\hline
\end{tabular} 
\caption{Average over 100 simulations of $\frac{\bar{\mu}_s - \bar{\mu}_b}{\bar{\sigma}_s}$, i.e. the shift in the mean of the marginalized distributions of the parameters with respect to a reference case, normalized to the $\sigma$ of the case to test, and of $\frac{\bar{\sigma}_s}{\bar{\sigma}_b}$, i.e. the ratio of the uncertainties. They are indicated with the labels presented in Table \ref{tab:runs_ideal}. The cases under scrutiny ($s$ for ``systematics'') are $\Lambda\text{CDM}$+$N_{\rm{eff}}$ with $\Delta^{\nu} \neq 0$ GHz, with Gaussian priors with $\sigma = 1$ GHz on $\Delta^{\nu}$, and with flat priors on $\Delta^{\nu}$, using ideal simulations.} \label{tab:shift/sigma_bsh_neff}
\end{table}
\FloatBarrier

\begin{table}[!htbp] 
 \renewcommand\thetable{C.1.2} 
\footnotesize
\centering 
\begin{tabular}{| P{1.1  cm} | P{1.4 cm}| P{1.4 cm} | P{1.4 cm}| P{1.4 cm} | P{1.4 cm}| P{1.4 cm} | P{1.4 cm}| P{1.4 cm} |} 
    \hline 
    Params & \multicolumn{2}{|c|}{\makecell{$s$: $\Delta$sys-smooth \\ $b$: fid-smooth }} &  \multicolumn{2}{|c|}{\makecell{$s$: $\Delta \mathcal{N}$1-smooth \\ $b$: fid-smooth}} &  \multicolumn{2}{|c|}{\makecell{$s$: $\Delta \mathcal{N}$1sys-smooth \\ $b$: fid-smooth}} &  \multicolumn{2}{|c|}{\makecell{$s$: $\Delta \mathcal{U}$-smooth \\ $b$: fid-smooth}} \\
    \hline
     & $\frac{{\mu}_s - {\mu}_b}{{\sigma}_s}$ & $\frac{{\sigma}_s}{{\sigma}_b}$ & $\frac{{\mu}_s - {\mu}_b}{{\sigma}_s}$ & $\frac{{\sigma}_s}{{\sigma}_b}$ & $\frac{{\mu}_s - {\mu}_b}{{\sigma}_s}$ & $\frac{{\sigma}_s}{{\sigma}_b}$ & $\frac{{\mu}_s - {\mu}_b}{{\sigma}_s}$ & $\frac{{\sigma}_s}{{\sigma}_b}$ \\
    \hline 
$A_\mathrm{s}$ & 0.392 & 1.003 & 0.007 & 0.993 & 0.020 & 0.997 & -0.015 & 0.992\\ 
$100 \theta_s$ & 0.274 & 1.018 & -0.024 & 1.002 & -0.016 & 1.009 & -0.030 & 1.005\\ 
$n_\mathrm{s}$ & -0.510 & 1.013 & 0.014 & 1.009 & -0.014 & 1.005 & 0.049 & 1.011\\ 
$\Omega_\mathrm{b}h^2$ & -0.075 & 1.017 & 0.007 & 0.999 & 0.004 & 1.013 & -0.002 & 1.012\\ 
$\Omega_\mathrm{c}h^2$ & -0.317 & 1.009 & 0.020 & 1.004 & 0.011 & 1.003 & 0.023 & 1.006\\ 
$\tau_\mathrm{reio}$ & 0.463 & 0.997 & 0.002 & 0.991 & 0.012 & 0.997 & -0.016 & 0.992\\ 
$\sigma_8$ & 0.191 & 1.008 & 0.017 & 1.000 & 0.021 & 1.000 & 0.003 & 0.997\\ 
$H_0$ & -0.703 & 1.008 & 0.011 & 1.011 & -0.009 & 1.014 & 0.041 & 1.015\\ 
$N_\mathrm{eff}$ & -0.666 & 1.016 & 0.020 & 1.016 & 0.001 & 1.015 & 0.043 & 1.018\\ 
\hline
$a_\mathrm{tSZ}$ & 6.724 & 0.890 & 0.071 & 1.338 & -0.221 & 1.322 & 0.362 & 2.399\\ 
$a_\mathrm{kSZ}$ & -10.032 & 0.965 & -0.075 & 1.778 & 0.372 & 1.741 & -0.390 & 4.603\\ 
$a_p$ & -1.822 & 1.011 & 0.073 & 3.841 & -0.717 & 3.732 & 0.433 & 19.995\\ 
$\beta_p$ & -5.636 & 0.997 & 0.002 & 4.078 & -0.103 & 4.092 & 0.084 & 5.041\\ 
$a_c$ & 14.469 & 1.228 & 0.096 & 2.596 & -0.386 & 2.491 & 0.443 & 7.985\\ 
$\beta_c$ & -21.276 & 0.982 & -0.039 & 3.319 & 0.028 & 3.282 & -0.150 & 3.999\\ 
$a_s$ & 13.735 & 1.008 & -0.037 & 5.611 & 0.560 & 5.623 & -0.330 & 22.387\\ 
$a_\mathrm{dust}^\mathrm{TT}$ & -1.234 & 0.989 & 0.020 & 1.086 & -0.267 & 1.065 & 0.376 & 2.325\\ 
$a_\mathrm{dust}^\mathrm{TE}$ & -0.101 & 0.986 & -0.004 & 1.029 & -0.135 & 1.008 & 0.289 & 1.398\\ 
$a_\mathrm{dust}^\mathrm{EE}$ & -0.837 & 0.988 & 0.058 & 1.404 & -0.612 & 1.363 & 0.423 & 5.826\\ 
$a_\mathrm{ps}^\mathrm{EE}$ & -0.152 & 0.924 & 0.000 & 0.992 & -0.002 & 0.994 & -0.039 & 0.978\\ 
$a_\mathrm{ps}^\mathrm{TE}$ & 0.155 & 1.027 & 0.016 & 0.997 & -0.000 & 1.013 & -0.010 & 0.968\\ 
$\xi$ & 10.385 & 1.008 & 0.042 & 2.031 & 0.053 & 2.026 & 0.106 & 2.367\\ 
\hline
\end{tabular} 
\caption{Shift in the mean of the marginalized distributions of the parameters with respect to a reference case, normalized to the $\sigma$ of the case to test $\frac{\mu_s - \mu_b}{\sigma_s}$, and ratio of the uncertainties $\frac{\sigma_s}{\sigma_b}$. They are indicated with the labels presented in Table \ref{tab:runs_ideal}, except for the case $\Delta \mathcal{N}$1sys-smooth in the third column, which is a test case done only with smooth spectra. The cases under scrutiny ($s$ for ``systematics'') are $\Lambda\text{CDM}$+$N_{\rm{eff}}$ with $\Delta^{\nu}  \neq 0$ GHz, with Gaussian priors with $\sigma = 1$ GHz on $\Delta^{\nu}$, with Gaussian priors centered on a wrong value: $\mathcal{N}(0.3, 1)$, $\mathcal{N}(0.5, 1)$, $\mathcal{N}(0.8, 1)$ for $\Delta^{93}$, $\Delta^{145}$ and $\Delta^{225}$, and with flat priors on $\Delta^{\nu}$, all using smooth spectra.} \label{tab:shift/sigma_bsh_smoothsim_neff}
\end{table}
\FloatBarrier

\begin{table}[!htbp] 
 \renewcommand\thetable{C.1.3}
\footnotesize
 \centering 
	\begin{tabular}{| P{1.3  cm} | P{1.55 cm}| P{1.55 cm} | P{1.55 cm}| P{1.55 cm} | P{1.5 cm}| P{1.5 cm} |} 
		\hline 
		Params & \multicolumn{2}{|c|}{\makecell{$s$: $\alpha$sys \\ $b$: fid}} &  
  \multicolumn{2}{|c|}{\makecell{$s$: $\alpha \mathcal{N}$0.25 \\ $b$: fid}} & \multicolumn{2}{|c|}{\makecell{$s$: $\alpha \mathcal{U}$ \\ $b$: fid }} \\
		\hline
    & $\frac{\bar{\mu}_s - \bar{\mu}_b}{\bar{\sigma}_s}$ & $\frac{\bar{\sigma}_s}{\bar{\sigma}_b}$ & $\frac{\bar{\mu}_s - \bar{\mu}_b}{\bar{\sigma}_s}$ & $\frac{\bar{\sigma}_s}{\bar{\sigma}_b}$& $\frac{\bar{\mu}_s - \bar{\mu}_b}{\bar{\sigma}_s}$ & $\frac{\bar{\sigma}_s}{\bar{\sigma}_b}$\\
		\hline 
$A_\mathrm{s}$ & 0.011 & 1.002 & 0.010 & 1.001 & 0.192 & 1.014\\ 
$100 \theta_s$ & -0.013 & 1.001 & -0.013 & 1.000 & -0.209 & 1.010\\ 
$n_\mathrm{s}$ & 0.022 & 1.000 & 0.020 & 1.000 & 0.333 & 1.035\\ 
$\Omega_\mathrm{b}h^2$ & 0.013 & 1.000 & 0.010 & 1.001 & 0.204 & 1.013\\ 
$\Omega_\mathrm{c}h^2$ & 0.014 & 1.001 & 0.013 & 1.001 & 0.199 & 1.016\\ 
$\tau_\mathrm{reio}$ & 0.004 & 1.001 & 0.004 & 1.001 & 0.092 & 1.004\\ 
$\sigma_8$ & 0.016 & 1.001 & 0.014 & 1.001 & 0.243 & 1.022\\ 
$H_0$ & 0.025 & 1.000 & 0.022 & 1.000 & 0.379 & 1.049\\ 
$N_\mathrm{eff}$ & 0.025 & 1.000 & 0.022 & 1.001 & 0.363 & 1.046\\ 
\hline
$a_\mathrm{tSZ}$ & 0.000 & 0.999 & 0.001 & 0.999 & 0.017 & 1.001\\ 
$a_\mathrm{kSZ}$ & -0.000 & 0.999 & -0.001 & 0.999 & -0.002 & 1.000\\ 
$a_p$ & -0.000 & 0.999 & -0.002 & 0.999 & -0.026 & 1.000\\ 
$\beta_p$ & 0.003 & 0.999 & 0.001 & 0.999 & 0.026 & 1.000\\ 
$a_c$ & -0.001 & 1.000 & 0.001 & 0.999 & 0.017 & 1.000\\ 
$\beta_c$ & -0.002 & 0.999 & -0.001 & 0.999 & -0.017 & 1.000\\ 
$a_s$ & -0.001 & 1.000 & -0.001 & 0.999 & -0.024 & 1.001\\ 
$a_\mathrm{dust}^\mathrm{TT}$ & 0.004 & 1.000 & -0.000 & 0.999 & -0.005 & 1.000\\ 
$a_\mathrm{dust}^\mathrm{TE}$ & 0.004 & 1.000 & -0.000 & 1.001 & -0.029 & 1.006\\ 
$a_\mathrm{dust}^\mathrm{EE}$ & 0.003 & 0.999 & 0.006 & 1.000 & 0.131 & 1.024\\ 
$a_\mathrm{ps}^\mathrm{EE}$ & 0.012 & 1.006 & 0.006 & 1.004 & 0.103 & 1.059\\ 
$a_\mathrm{ps}^\mathrm{TE}$ & -0.005 & 1.001 & -0.003 & 1.001 & -0.058 & 1.002\\ 
$\xi$ & -0.001 & 0.998 & -0.002 & 0.998 & -0.013 & 0.999\\ 
		\hline 
	\end{tabular} 
	\caption{The cases to compare are indicated with the labels presented in Table \ref{tab:runs_ideal}. Here the reference case ($b$ for ``benchmark'') is $\Lambda\text{CDM}$+$N_{\rm{eff}}$ with polarization angles $\alpha^{\nu}$ and all the other systematic parameters fixed to the fiducial, ideal values, like in the simulated spectra. The cases under scrutiny ($s$ for ``systematics'') are  $\Lambda\text{CDM}$+$N_{\rm{eff}}$ with $\alpha^{\nu} \neq 0^\circ$, with Gaussian priors with $\sigma = 0.25^\circ$ on $\alpha^{\nu}$ and with flat, positive priors on $\alpha^{\nu}$.} \label{tab:shift/sigma_alfa_neff}
\end{table}
\FloatBarrier

\begin{table}[!htbp] 
 \renewcommand\thetable{C.1.4}
\footnotesize
 \centering 
	\begin{tabular}{| P{1.3  cm} | P{1.55 cm}| P{1.55 cm} | P{1.55 cm}| P{1.55 cm} | P{1.5 cm}| P{1.5 cm} |} 
		\hline 
		Params & \multicolumn{2}{|c|}{\makecell{$s$: $\alpha$sys-smooth \\ $b$: fid-smooth}} &  
  \multicolumn{2}{|c|}{\makecell{$s$: $\alpha \mathcal{N}$0.25-smooth \\ $b$: fid-smooth}} & \multicolumn{2}{|c|}{\makecell{$s$: $\alpha \mathcal{U}$-smooth \\ $b$: fid-smooth }} \\
		\hline
		 & $\frac{{\mu}_s - {\mu}_b}{{\sigma}_s}$ & $\frac{{\sigma}_s}{{\sigma}_b}$ & $\frac{{\mu}_s - {\mu}_b}{{\sigma}_s}$ & $\frac{{\sigma}_s}{{\sigma}_b}$ & $\frac{{\mu}_s - {\mu}_b}{{\sigma}_s}$ & $\frac{{\sigma}_s}{{\sigma}_b}$ \\
		\hline 
$A_\mathrm{s}$ & 0.022 & 1.009 & 0.030 & 1.007 & 0.157 & 1.001\\ 
$100 \theta_s$ & -0.012 & 1.011 & -0.031 & 1.015 & -0.156 & 1.021\\ 
$n_\mathrm{s}$ & 0.016 & 1.001 & 0.047 & 1.007 & 0.242 & 1.028\\ 
$\Omega_\mathrm{b}h^2$ & 0.020 & 1.016 & 0.025 & 1.016 & 0.151 & 1.029\\ 
$\Omega_\mathrm{c}h^2$ & 0.005 & 1.009 & 0.036 & 1.004 & 0.147 & 1.015\\ 
$\tau_\mathrm{reio}$ & 0.014 & 1.003 & 0.019 & 1.006 & 0.081 & 0.992\\ 
$\sigma_8$ & 0.017 & 1.017 & 0.044 & 1.005 & 0.191 & 1.015\\ 
$H_0$ & 0.028 & 1.008 & 0.043 & 1.016 & 0.277 & 1.048\\ 
$N_\mathrm{eff}$ & 0.021 & 1.016 & 0.050 & 1.010 & 0.267 & 1.046\\ 
\hline
$a_\mathrm{tSZ}$ & 0.011 & 1.012 & 0.003 & 1.003 & 0.017 & 1.015\\ 
$a_\mathrm{kSZ}$ & -0.003 & 1.008 & -0.009 & 1.005 & -0.002 & 1.017\\ 
$a_p$ & 0.008 & 1.003 & 0.002 & 1.000 & 0.004 & 1.000\\ 
$\beta_p$ & 0.001 & 0.992 & 0.004 & 0.989 & 0.006 & 0.995\\ 
$a_c$ & -0.003 & 1.009 & 0.002 & 1.012 & 0.001 & 1.008\\ 
$\beta_c$ & -0.006 & 0.999 & -0.005 & 0.998 & -0.009 & 1.004\\ 
$a_s$ & -0.013 & 1.001 & 0.010 & 0.995 & -0.037 & 0.992\\ 
$a_\mathrm{dust}^\mathrm{TT}$ & 0.011 & 1.012 & -0.002 & 1.026 & 0.001 & 1.019\\ 
$a_\mathrm{dust}^\mathrm{TE}$ & 0.010 & 1.014 & -0.004 & 1.001 & -0.019 & 1.017\\ 
$a_\mathrm{dust}^\mathrm{EE}$ & 0.035 & 0.998 & 0.039 & 1.005 & 0.126 & 1.015\\ 
$a_\mathrm{ps}^\mathrm{EE}$ & 0.019 & 1.009 & -0.002 & 1.005 & 0.050 & 1.027\\ 
$a_\mathrm{ps}^\mathrm{TE}$ & 0.006 & 1.002 & 0.009 & 0.993 & -0.040 & 0.993\\ 
$\xi$ & 0.009 & 0.999 & 0.003 & 1.003 & 0.006 & 1.007\\ 
  \hline 
	\end{tabular} 
	\caption{The cases to compare are indicated with the labels presented in Table \ref{tab:runs_ideal}. Here the reference case ($b$ for ``benchmark'') is $\Lambda\text{CDM}$+$N_{\rm{eff}}$ with polarization angles $\alpha^{\nu}$ and all the other systematic parameters fixed to the fiducial, ideal values, like in the simulated spectra. We are using smooth spectra in this case. The cases under scrutiny ($s$ for ``systematics'') are  $\Lambda\text{CDM}$+$N_{\rm{eff}}$ with $\alpha^{\nu} \neq 0^\circ$, with Gaussian priors with $\sigma = 0.25^\circ$ on $\alpha^{\nu}$ and with flat, positive priors on $\alpha^{\nu}$.} \label{tab:shift/sigma_alfa_neff_smooth}
\end{table}
\FloatBarrier

\begin{table}[!htbp]
 \renewcommand\thetable{C.1.5}
\footnotesize
 \centering 
	\begin{tabular}{| P{1.3  cm} | P{1.5 cm}| P{1.5 cm} | P{1.5 cm}| P{1.5 cm} |} 
		\hline 
		Params & \multicolumn{2}{|c|}{\makecell{$s$: Csys  \\ $b$: fid}} &  \multicolumn{2}{|c|}{\makecell{$s$: C$\mathcal{N}$0.01 \\ $b$: fid}}  \\
		\hline	
    & $\frac{\bar{\mu}_s - \bar{\mu}_b}{\bar{\sigma}_s}$ & $\frac{\bar{\sigma}_s}{\bar{\sigma}_b}$& $\frac{\bar{\mu}_s - \bar{\mu}_b}{\bar{\sigma}_s}$ & $\frac{\bar{\sigma}_s}{\bar{\sigma}_b}$\\
		\hline 
  $A_\mathrm{s}$ & 3.709 & 1.049 & 0.048 & 1.202\\ 
$100 \theta_s$ & -3.015 & 0.958 & -0.063 & 1.046\\ 
$n_\mathrm{s}$ & 5.349 & 1.015 & 0.090 & 1.150\\ 
$\Omega_\mathrm{b}h^2$ & 3.133 & 1.009 & 0.061 & 1.057\\ 
$\Omega_\mathrm{c}h^2$ & 2.616 & 1.036 & 0.062 & 1.049\\ 
$\tau_\mathrm{reio}$ & 0.827 & 1.012 & 0.028 & 1.043\\ 
$\sigma_8$ & 4.162 & 1.047 & 0.065 & 1.142\\ 
$H_0$ & 6.066 & 1.057 & 0.101 & 1.204\\ 
$N_\mathrm{eff}$ & 5.523 & 1.056 & 0.102 & 1.157\\ 
\hline
$a_\mathrm{tSZ}$ & 1.175 & 1.022 & 0.017 & 1.200\\ 
$a_\mathrm{kSZ}$ & 0.128 & 1.024 & -0.017 & 1.100\\ 
$a_p$ & 2.402 & 1.021 & -0.010 & 1.826\\ 
$\beta_p$ & 0.342 & 1.003 & 0.012 & 1.032\\ 
$a_c$ & 0.999 & 1.024 & 0.010 & 1.125\\ 
$\beta_c$ & -0.228 & 1.000 & -0.008 & 1.018\\ 
$a_s$ & 5.184 & 1.021 & -0.004 & 3.162\\ 
$a_\mathrm{dust}^\mathrm{TT}$ & 0.166 & 1.021 & -0.009 & 1.100\\ 
$a_\mathrm{dust}^\mathrm{TE}$ & -0.282 & 1.031 & -0.010 & 1.005\\ 
$a_\mathrm{dust}^\mathrm{EE}$ & 2.283 & 1.040 & 0.006 & 1.142\\ 
$a_\mathrm{ps}^\mathrm{EE}$ & 1.642 & 1.632 & 0.044 & 1.046\\ 
$a_\mathrm{ps}^\mathrm{TE}$ & -0.889 & 1.031 & -0.023 & 1.006\\ 
$\xi$ & -0.178 & 0.996 & -0.021 & 1.092\\ 
		\hline 
	\end{tabular} 
	\caption{The cases to compare are indicated with the labels presented in Table \ref{tab:runs_ideal}. Here the reference case ($b$ for ``benchmark'') is $\Lambda\text{CDM} + N_{\mathrm{eff}}$ with all systematic parameters fixed to the ideal value, like in the simulated spectra. The cases under scrutiny ($s$ for ``systematics'') are $\Lambda\text{CDM}+ N_{\mathrm{eff}}$ with $\mathrm{Cal}^{\nu} = \mathrm{Cal}_{\rm E}^{\nu} = 1.01$ (introducing a mismatch) and $\Lambda\text{CDM}+ N_{\mathrm{eff}}$ with Gaussian priors on the calibrations per channel $\mathrm{Cal}^{\nu}$ and flat priors on the polarization efficiencies $\mathrm{Cal}_{\rm E}^{\nu}$, both using the 100 simulated spectra without any injected systematics.} \label{tab:shift/sigma_Neff_cals_nosys_neff}
\end{table}
\FloatBarrier

\begin{table}[!htbp]
 \renewcommand\thetable{C.1.6}
\footnotesize
 \centering 
	\begin{tabular}{| P{1.3  cm} | P{1.5 cm}| P{1.5 cm} | P{1.5 cm}| P{1.5 cm} |} 
		\hline 
		Params & \multicolumn{2}{|c|}{\makecell{$s$: Csys-smooth \\ $b$: fid-smooth}} &  \multicolumn{2}{|c|}{\makecell{$s$: C$\mathcal{N}$0.01-smooth  \\ $b$: fid-smooth }}  \\
		\hline
		 &  $\frac{{\mu}_s - {\mu}_b}{{\sigma}_s}$ & $\frac{{\sigma}_s}{{\sigma}_b}$ & $\frac{{\mu}_s - {\mu}_b}{{\sigma}_s}$ & $\frac{{\sigma}_s}{{\sigma}_b}$ \\
		\hline 
$A_\mathrm{s}$ & 3.722 & 1.043 & 0.085 & 1.194\\ 
$100 \theta_s$ & -2.983 & 0.969 & -0.063 & 1.055\\ 
$n_\mathrm{s}$ & 5.338 & 1.021 & 0.094 & 1.157\\ 
$\Omega_\mathrm{b}h^2$ & 3.109 & 1.022 & 0.067 & 1.062\\ 
$\Omega_\mathrm{c}h^2$ & 2.593 & 1.040 & 0.044 & 1.053\\ 
$\tau_\mathrm{reio}$ & 0.850 & 1.004 & 0.026 & 1.032\\ 
$\sigma_8$ & 4.133 & 1.054 & 0.086 & 1.137\\ 
$H_0$ & 6.054 & 1.067 & 0.115 & 1.210\\ 
$N_\mathrm{eff}$ & 5.501 & 1.068 & 0.102 & 1.167\\ 
\hline
$a_\mathrm{tSZ}$ & 1.191 & 1.029 & 0.042 & 1.203\\ 
$a_\mathrm{kSZ}$ & 0.124 & 1.043 & -0.006 & 1.106\\ 
$a_p$ & 2.392 & 1.026 & 0.039 & 1.826\\ 
$\beta_p$ & 0.345 & 0.999 & 0.001 & 1.017\\ 
$a_c$ & 0.996 & 1.037 & 0.031 & 1.127\\ 
$\beta_c$ & -0.236 & 1.002 & -0.003 & 1.006\\ 
$a_s$ & 5.187 & 1.008 & 0.044 & 3.163\\ 
$a_\mathrm{dust}^\mathrm{TT}$ & 0.164 & 1.036 & -0.005 & 1.119\\ 
$a_\mathrm{dust}^\mathrm{TE}$ & -0.276 & 1.045 & 0.004 & 1.001\\ 
$a_\mathrm{dust}^\mathrm{EE}$ & 2.259 & 1.046 & 0.047 & 1.138\\ 
$a_\mathrm{ps}^\mathrm{EE}$ & 1.633 & 1.640 & 0.040 & 1.036\\ 
$a_\mathrm{ps}^\mathrm{TE}$ & -0.874 & 1.031 & -0.013 & 1.007\\ 
$\xi$ & -0.174 & 1.007 & -0.005 & 1.092\\ 
  \hline 
	\end{tabular} 
	\caption{The cases to compare are indicated with the labels presented in Table \ref{tab:runs_ideal}. Here the reference case ($b$ for ``benchmark'') is $\Lambda\text{CDM} + N_{\mathrm{eff}}$ with all systematic parameters fixed to the ideal value, like in the simulated spectra. The cases under scrutiny ($s$ for ``systematics'') are $\Lambda\text{CDM}+ N_{\mathrm{eff}}$ with $\mathrm{Cal}^{\nu} = \mathrm{Cal}_{\rm E}^{\nu} = 1.01$ (introducing a mismatch) and $\Lambda\text{CDM}+ N_{\mathrm{eff}}$ with Gaussian priors on the calibrations per channel $\mathrm{Cal}^{\nu}$ and flat priors on the polarization efficiencies $\mathrm{Cal}_{\rm E}^{\nu}$, both using the smooth spectra.} \label{tab:shift/sigma_Neff_cals_smoothsim_neff}
\end{table}
\FloatBarrier

\subsection{Tables for the \lcdm~model}

\begin{table}[!htbp]
 \renewcommand\thetable{C.2.1}
\footnotesize
 \centering 
	\begin{tabular}{| P{1.3  cm} | P{1.55 cm}| P{1.55 cm} | P{1.5 cm}| P{1.5 cm} | P{1.6 cm}| P{1.5 cm} |} 
		\hline 
		Params & \multicolumn{2}{|c|}{\makecell{$s$: $\Delta$sys-smooth \\ $b$: fid-smooth}} &  \multicolumn{2}{|c|}{\makecell{$s$: $\Delta \mathcal{N}$1-smooth \\ $b$: fid-smooth}} &  \multicolumn{2}{|c|}{\makecell{$s$: $\Delta \mathcal{U}$-smooth \\$b$: fid-smooth}} \\
		\hline
    & $\frac{{\mu}_s - {\mu}_b}{{\sigma}_s}$ & $\frac{{\sigma}_s}{{\sigma}_b}$ & $\frac{{\mu}_s - {\mu}_b}{{\sigma}_s}$ & $\frac{{\sigma}_s}{{\sigma}_b}$ & $\frac{{\mu}_s - {\mu}_b}{{\sigma}_s}$ & $\frac{{\sigma}_s}{{\sigma}_b}$ \\
		\hline 
$A_\mathrm{s}$ & 0.701 & 1.003 & 0.010 & 1.006 & -0.027 & 1.002\\ 
$100 \theta_s$ & -0.314 & 0.992 & -0.018 & 0.996 & 0.005 & 1.003\\ 
$n_\mathrm{s}$ & 0.064 & 0.988 & -0.016 & 0.994 & 0.012 & 0.993\\ 
$\Omega_\mathrm{b}h^2$ & 0.627 & 1.012 & 0.009 & 1.008 & -0.026 & 1.009\\ 
$\Omega_\mathrm{c}h^2$ & 0.397 & 1.001 & 0.020 & 1.010 & -0.017 & 0.994\\ 
$\tau_\mathrm{reio}$ & 0.578 & 0.992 & 0.006 & 1.003 & -0.018 & 0.997\\ 
$\sigma_8$ & 1.012 & 1.005 & 0.020 & 1.022 & -0.037 & 1.014\\ 
$H_0$ & -0.298 & 0.997 & -0.019 & 1.007 & 0.012 & 0.994\\ 
\hline
$a_\mathrm{tSZ}$ & 6.909 & 0.874 & 0.007 & 1.330 & 0.400 & 2.385\\ 
$a_\mathrm{kSZ}$ & -10.301 & 0.936 & -0.032 & 1.737 & -0.449 & 4.496\\ 
$a_p$ & -1.850 & 1.011 & 0.097 & 3.858 & 0.500 & 19.707\\ 
$\beta_p$ & -5.675 & 0.999 & -0.032 & 4.096 & 0.117 & 5.128\\ 
$a_c$ & 14.586 & 1.231 & 0.073 & 2.512 & 0.493 & 7.913\\ 
$\beta_c$ & -21.587 & 0.982 & -0.024 & 3.259 & -0.142 & 3.981\\ 
$a_s$ & 13.754 & 1.028 & -0.059 & 5.697 & -0.409 & 21.708\\ 
$a_\mathrm{dust}^\mathrm{TT}$ & -1.213 & 0.987 & 0.026 & 1.087 & 0.440 & 2.289\\ 
$a_\mathrm{dust}^\mathrm{TE}$ & -0.170 & 0.965 & 0.005 & 1.010 & 0.331 & 1.376\\ 
$a_\mathrm{dust}^\mathrm{EE}$ & -0.867 & 0.999 & 0.047 & 1.413 & 0.487 & 5.783\\ 
$a_\mathrm{ps}^\mathrm{EE}$ & -0.183 & 0.914 & -0.031 & 0.978 & -0.058 & 0.973\\ 
$a_\mathrm{ps}^\mathrm{TE}$ & 0.053 & 1.022 & 0.004 & 1.001 & -0.027 & 0.970\\ 
$\xi$ & 10.595 & 0.985 & 0.056 & 1.991 & 0.086 & 2.321\\		
		\hline 
	\end{tabular} 
	\caption{The cases to compare are indicated with the labels presented in Table \ref{tab:runs_ideal_lcdm}. Here the reference case ($b$ for ``benchmark'') is $\Lambda\text{CDM}$ with bandpass shifts $\Delta^{\nu}$ and all the other systematic parameters fixed to the fiducial, ideal values. The cases under scrutiny ($s$ for ``systematics'') are $\Lambda\text{CDM}$ with $\Delta^{\nu} \neq 0$ GHz and with different priors on $\Delta^{\nu}$ (Gaussian prior centered on the fiducial value with $\sigma$ = 1 GHz and flat priors), using smooth spectra  without systematics.} \label{tab:shift/sigma_bsh_lcdm}
\end{table}
\FloatBarrier

\begin{table}[!htbp] 
 \renewcommand\thetable{C.2.2}
\footnotesize
 \centering 
	\begin{tabular}{| P{1.3  cm} | P{1.55 cm}| P{1.55 cm} | P{1.55 cm}| P{1.5 cm} | P{1.5 cm}| P{1.5 cm} |} 
		\hline 
		Params & \multicolumn{2}{|c|}{\makecell{$s$: $\alpha$sys-smooth \\ $b$: fid-smooth}} &  \multicolumn{2}{|c|}{\makecell{$s$: $\alpha \mathcal{N}$0.25-smooth \\ $b$: fid-smooth}} &  \multicolumn{2}{|c|}{\makecell{$s$: $\alpha \mathcal{U}$-smooth \\ $b$: fid-smooth}} \\
		\hline
		 & $\frac{{\mu}_s - {\mu}_b}{{\sigma}_s}$ & $\frac{{\sigma}_s}{{\sigma}_b}$ & $\frac{{\mu}_s - {\mu}_b}{{\sigma}_s}$ & $\frac{{\sigma}_s}{{\sigma}_b}$ & $\frac{{\mu}_s - {\mu}_b}{{\sigma}_s}$ & $\frac{{\sigma}_s}{{\sigma}_b}$ \\
		\hline 
$A_\mathrm{s}$ & 0.004 & 0.999 & 0.039 & 0.993 & 0.031 & 0.998\\ 
$100 \theta_s$ & 0.005 & 1.000 & -0.003 & 0.990 & 0.034 & 0.987\\ 
$n_\mathrm{s}$ & -0.003 & 0.994 & 0.012 & 0.987 & 0.010 & 0.979\\ 
$\Omega_\mathrm{b}h^2$ & 0.002 & 0.993 & 0.011 & 1.005 & -0.079 & 1.006\\ 
$\Omega_\mathrm{c}h^2$ & 0.004 & 1.002 & -0.017 & 0.996 & -0.092 & 0.990\\ 
$\tau_\mathrm{reio}$ & 0.001 & 0.997 & 0.039 & 0.990 & 0.008 & 0.992\\ 
$\sigma_8$ & 0.007 & 1.004 & 0.033 & 1.000 & -0.022 & 1.001\\ 
$H_0$ & -0.002 & 0.998 & 0.016 & 0.996 & 0.074 & 0.992\\ 
\hline
$a_\mathrm{tSZ}$ & -0.011 & 0.997 & -0.010 & 1.004 & -0.028 & 1.010\\ 
$a_\mathrm{kSZ}$ & 0.000 & 0.993 & -0.009 & 1.003 & 0.018 & 1.000\\ 
$a_p$ & 0.038 & 1.000 & 0.037 & 0.997 & -0.013 & 1.001\\ 
$\beta_p$ & -0.027 & 1.000 & -0.042 & 0.994 & 0.001 & 0.999\\ 
$a_c$ & -0.037 & 1.006 & -0.019 & 1.012 & 0.018 & 1.007\\ 
$\beta_c$ & 0.026 & 1.004 & 0.026 & 1.002 & -0.004 & 1.001\\ 
$a_s$ & 0.007 & 1.012 & 0.018 & 1.008 & 0.041 & 1.011\\ 
$a_\mathrm{dust}^\mathrm{TT}$ & 0.035 & 0.999 & 0.004 & 1.013 & -0.026 & 1.004\\ 
$a_\mathrm{dust}^\mathrm{TE}$ & 0.005 & 1.002 & -0.019 & 1.007 & 0.002 & 1.007\\ 
$a_\mathrm{dust}^\mathrm{EE}$ & -0.015 & 0.999 & -0.006 & 1.004 & 0.101 & 1.026\\ 
$a_\mathrm{ps}^\mathrm{EE}$ & -0.016 & 0.992 & -0.027 & 0.984 & 0.064 & 1.040\\ 
$a_\mathrm{ps}^\mathrm{TE}$ & -0.045 & 0.998 & -0.029 & 0.998 & -0.004 & 0.999\\ 
$\xi$ & -0.000 & 0.995 & 0.001 & 1.001 & 0.009 & 1.002\\ 		
		\hline 
	\end{tabular} 
	\caption{The cases to compare are indicated with the labels presented in Table \ref{tab:runs_ideal_lcdm}. Here the reference case ($b$ for ``benchmark'') is $\Lambda\text{CDM}$ with polarization angles $\alpha^{\nu}$ and all the other systematic parameters fixed to the fixed to the fiducial, ideal value. The cases under scrutiny ($s$ for ``systematics'') are $\Lambda\text{CDM}$ with $\alpha^{\nu} \neq 0^\circ$ and with different priors on $\alpha^{\nu}$ (Gaussian prior centered on the fiducial value with $\sigma = 0.25^\circ$ and flat, positive priors), on the smooth spectra without systematics.} \label{tab:shift/sigma_alfa_lcdm}
\end{table}
\FloatBarrier

\begin{table}[!htbp] 
 \renewcommand\thetable{C.2.3}
\footnotesize
 \centering 
	\begin{tabular}{| P{1.3  cm} | P{1.55 cm}| P{1.55 cm} | P{1.55 cm}| P{1.5 cm} | P{1.5 cm}| P{1.5 cm} |} 
		\hline 
		Params & \multicolumn{2}{|c|}{\makecell{$s$:Csys-smooth \\ $b$: fid-smooth}} &  \multicolumn{2}{|c|}{\makecell{$s$: C$\mathcal{N}$0.01-smooth \\ $b$: fid-smooth}} &  \multicolumn{2}{|c|}{\makecell{$s$: C$\mathcal{N}$0.01$\Delta \mathcal{N}$1--smooth \\ $b$: fid-smooth}} \\
		\hline
		 & $\frac{{\mu}_s - {\mu}_b}{{\sigma}_s}$ & $\frac{{\sigma}_s}{{\sigma}_b}$ & $\frac{{\mu}_s - {\mu}_b}{{\sigma}_s}$ & $\frac{{\sigma}_s}{{\sigma}_b}$ & $\frac{{\mu}_s - {\mu}_b}{{\sigma}_s}$ & $\frac{{\sigma}_s}{{\sigma}_b}$ \\
		\hline 
$A_\mathrm{s}$ & 1.833 & 0.994 & 0.064 & 1.162 & 0.064 & 1.165\\ 
$100 \theta_s$ & 1.561 & 0.997 & 0.049 & 1.010 & 0.028 & 0.997\\ 
$n_\mathrm{s}$ & 1.201 & 0.992 & 0.038 & 1.014 & 0.030 & 1.020\\ 
$\Omega_\mathrm{b}h^2$ & -1.650 & 1.005 & -0.031 & 1.022 & -0.054 & 1.018\\ 
$\Omega_\mathrm{c}h^2$ & -3.090 & 0.984 & -0.092 & 1.074 & -0.080 & 1.082\\ 
$\tau_\mathrm{reio}$ & -0.187 & 0.967 & -0.005 & 1.037 & 0.002 & 1.041\\ 
$\sigma_8$ & 0.160 & 0.999 & 0.018 & 1.107 & 0.025 & 1.099\\ 
$H_0$ & 2.691 & 1.012 & 0.086 & 1.068 & 0.069 & 1.074\\ 
\hline
$a_\mathrm{tSZ}$ & 0.668 & 1.036 & 0.036 & 1.222 & 0.047 & 1.488\\ 
$a_\mathrm{kSZ}$ & 0.522 & 1.034 & -0.012 & 1.109 & -0.037 & 1.786\\ 
$a_p$ & 2.624 & 1.031 & 0.082 & 1.863 & 0.090 & 4.174\\ 
$\beta_p$ & -0.053 & 1.009 & -0.028 & 1.044 & 0.005 & 4.199\\ 
$a_c$ & 0.929 & 1.043 & 0.014 & 1.142 & 0.078 & 2.635\\ 
$\beta_c$ & 0.054 & 1.016 & 0.022 & 1.038 & -0.007 & 3.366\\ 
$a_s$ & 5.875 & 1.030 & 0.085 & 3.233 & 0.004 & 6.519\\ 
$a_\mathrm{dust}^\mathrm{TT}$ & -0.113 & 1.023 & 0.016 & 1.103 & 0.020 & 1.193\\ 
$a_\mathrm{dust}^\mathrm{TE}$ & 0.166 & 1.022 & -0.003 & 1.011 & 0.008 & 1.023\\ 
$a_\mathrm{dust}^\mathrm{EE}$ & 2.262 & 1.054 & 0.050 & 1.149 & 0.069 & 1.517\\ 
$a_\mathrm{ps}^\mathrm{EE}$ & 1.984 & 1.689 & 0.041 & 1.043 & 0.049 & 1.040\\ 
$a_\mathrm{ps}^\mathrm{TE}$ & -0.169 & 1.023 & -0.020 & 1.001 & -0.016 & 0.989\\ 
$\xi$ & 0.084 & 1.012 & -0.006 & 1.099 & 0.025 & 2.136\\ 		
		\hline 
	\end{tabular} 
	\caption{The cases to compare are indicated with the labels presented in Table \ref{tab:runs_ideal_lcdm}. Here the reference case ($b$ for ``benchmark'') is $\Lambda\text{CDM}$ with polarization angles $\alpha^{\nu}$ and all the other systematic parameters fixed to the fiducial, ideal value. The cases under scrutiny ($s$ for ``systematics'') are $\Lambda\text{CDM}$ with $\mathrm{Cal}^{\nu}$ = $\mathrm{Cal}_{\rm E}^{\nu} = 1.01$ and two cases with marginalization over calibrations: one with Gaussian priors on $\mathrm{Cal}^{\nu}$ and flat priors on $\mathrm{Cal}_{\rm E}^{\nu}$ and one with the same priors on calibrations plus a gaussian prior on $\Delta^{\nu}$ ($\sigma =$ 1 GHz). Ideal smooth spectra are used in all cases.} \label{tab:shift/sigma_cal_lcdm}
\end{table}
\FloatBarrier

\FloatBarrier

\section{Correlation matrices for the cases with marginalization}
\begin{figure}[h!]
		\centering
		{\includegraphics[width=1.\textwidth]{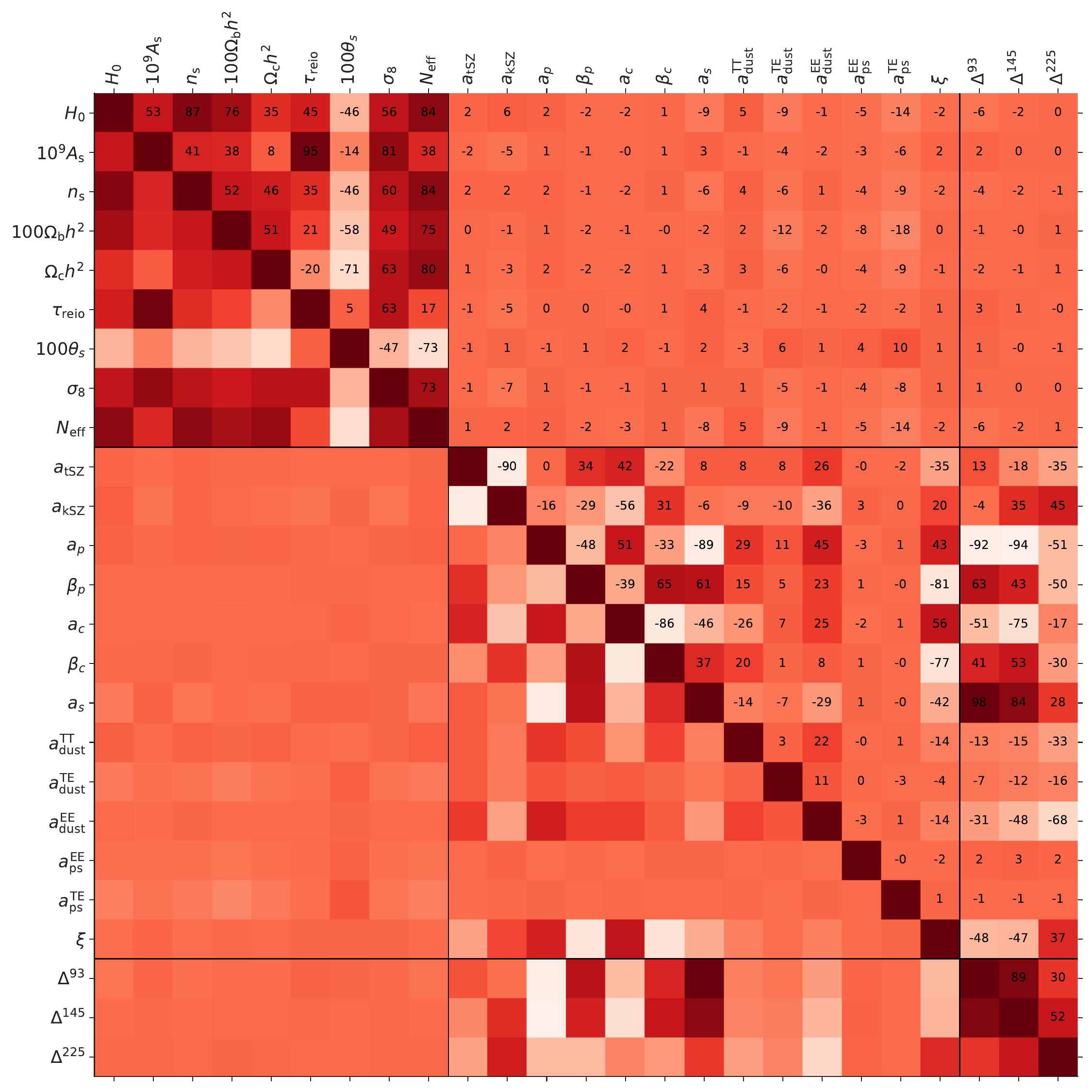}}
		{\caption{Correlation matrix for the case in which we marginalize over $\Delta^{\nu}$ with Gaussian priors $\mathcal{N}$(0,1) GHz (label: \texttt{$\Delta \mathcal{N}$1-smooth}). The correlation factors are multiplied by 100.} \label{fig:corr_delta}}
	\end{figure}

\begin{figure}[h!]
		\centering
		{\includegraphics[width=1.\textwidth]{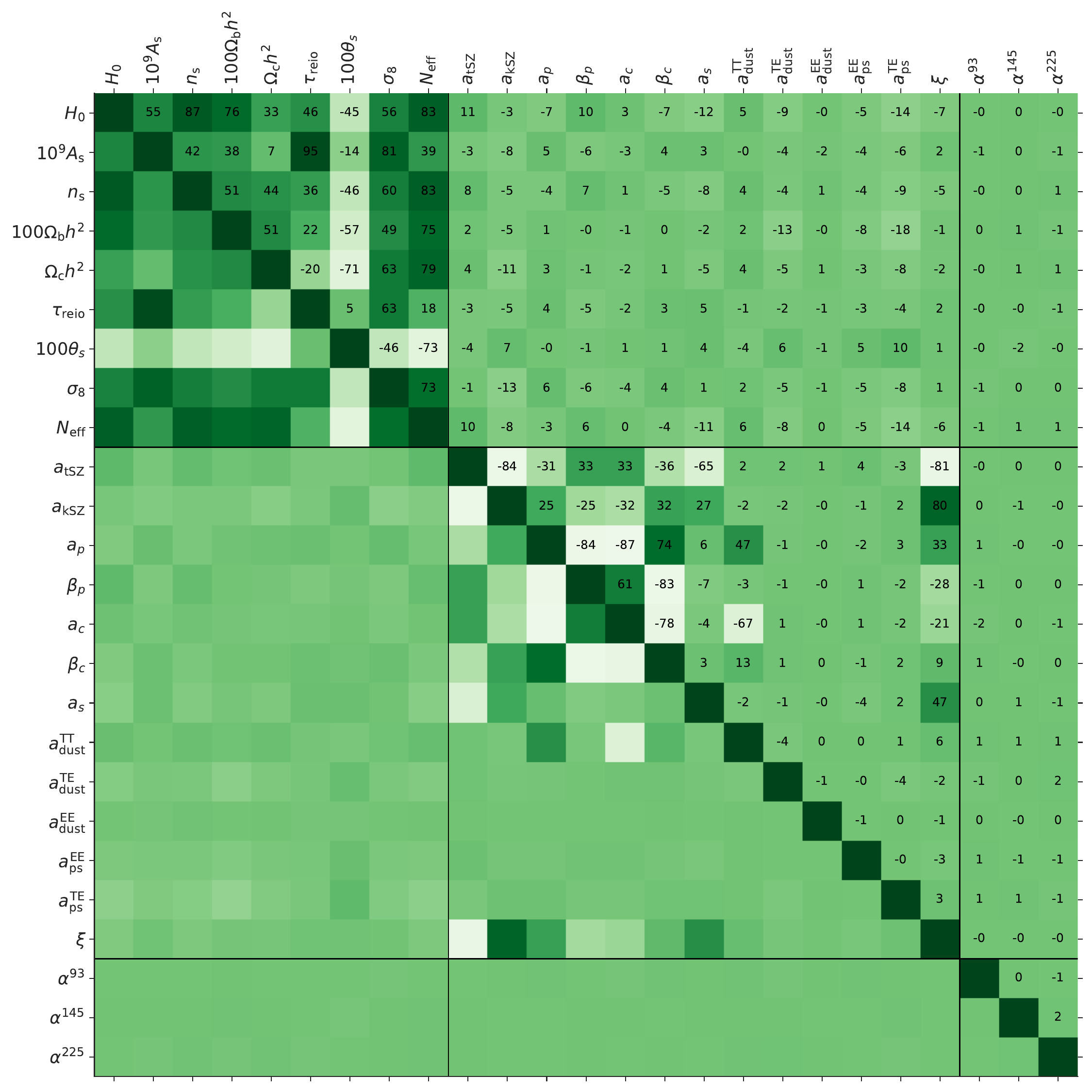}}
		{\caption{Correlation matrix for the case in which we marginalize over $\alpha^{\nu}$ with Gaussian priors $\mathcal{N}$(0,0.25)$^\circ$ (label: \texttt{$\alpha \mathcal{N}$0.25-smooth}). The correlation factors are multiplied by 100.} \label{fig:corr_alpha}}
	\end{figure}

\begin{figure}[h!]
		\centering
		{\includegraphics[width=1.\textwidth]{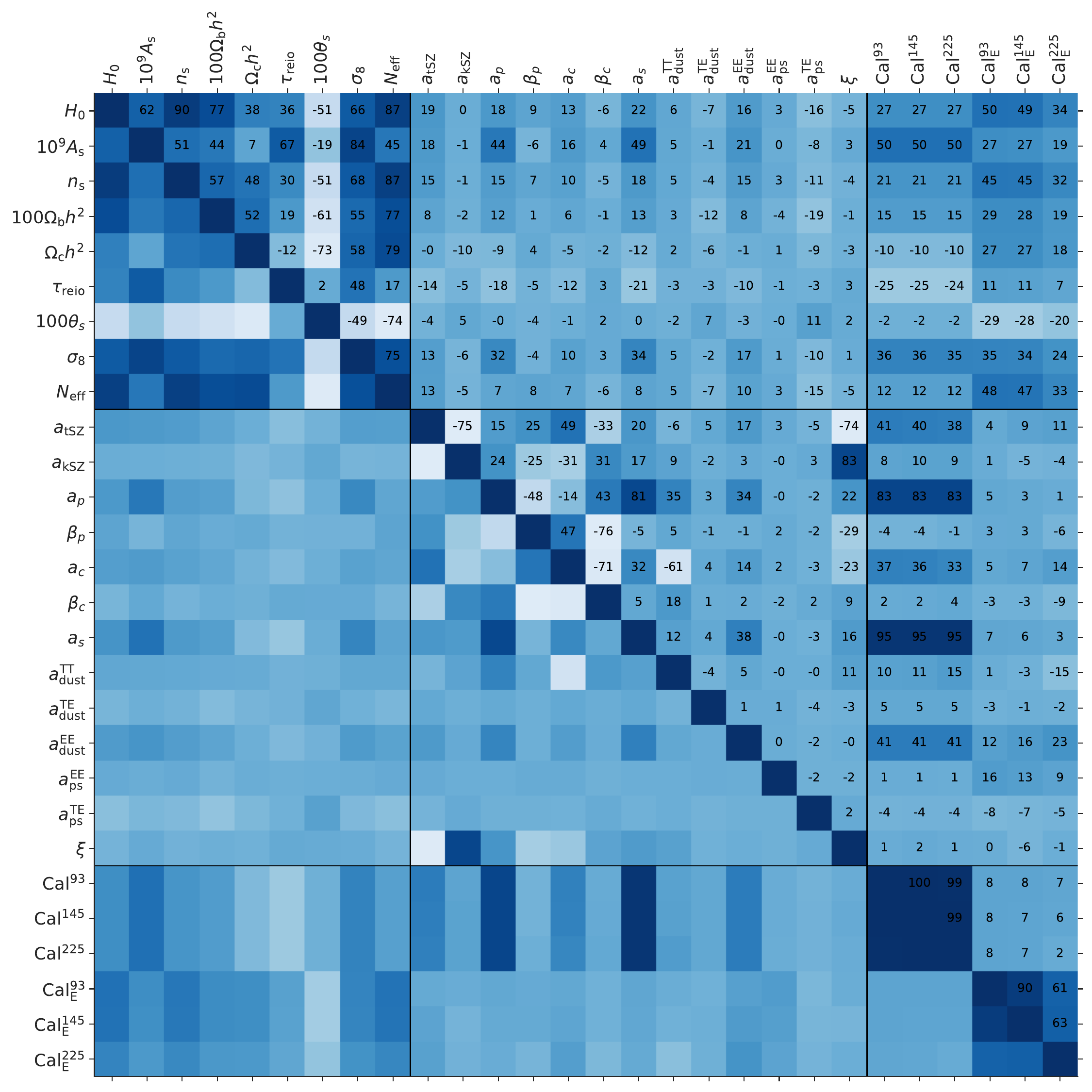}}
		{\caption{Correlation matrix for the case in which we marginalize over calibrations with Gaussian priors $\mathcal{N}$(1,0.01) GHz (label: \texttt{C$ \mathcal{N}$0.01-smooth}). The correlation factors are multiplied by 100.} \label{fig:corr_cals}}
	\end{figure}

In Figures \ref{fig:corr_delta}-\ref{fig:corr_cals}, we present the correlation matrices for the cases in which we marginalize over systematic parameters with Gaussian priors (\texttt{$\Delta \mathcal{N}$1-smooth}, \texttt{$\alpha \mathcal{N}$0.25-smooth} and \texttt{C$\mathcal{N}$0.01-smooth}).

\FloatBarrier

\FloatBarrier

\FloatBarrier

\end{document}